\documentclass[11pt,epsf]{article}
\usepackage{hyperref}
\usepackage{subfigure}
\usepackage{graphics}
\usepackage{amssymb}
\usepackage{epsf}
\usepackage{rotating}
\usepackage{axodraw}
\usepackage{pstricks}

\def\eps{\epsilon}
\def\a{\alpha}
\def\b{\beta}
\def\la{\lambda}
\def\g{\gamma}

\def\si{\sigma}
\def\lag{\langle}
\def\rag{\rangle}

\newcommand{\bea}{\begin{eqnarray}}
\newcommand{\eea}{\end{eqnarray}}

\newcommand{\psl}{p \! \! \!  /}
\newcommand{\qsl}{q \! \! \!  /}
\newcommand{\lsl}{l \! \! \!  /}

\newcommand{\beqa}{\begin{eqnarray}}
\newcommand{\eeqa}{\end{eqnarray}}
\def\beq{\begin{equation}}
\def\eeq{\end{equation}}
\def\nn{\nonumber}

\begin{document}

\begin{center}
\vspace{.5cm}
{\bf\large  Conformal Anomalies and the Gravitational Effective Action: \\}
{\bf The $TJJ$ Correlator for a Chiral Fermion \\}
\vspace{1.5cm}
{\bf\large $^{(a)}$Roberta Armillis, $^{(a,b)}$Claudio Corian\`{o} and  $^{(a)}$Luigi Delle Rose}
\footnote{roberta.armillis@le.infn.it, claudio.coriano@le.infn.it, luigi.dellerose@le.infn.it}
\vspace{1cm}

{\it  $^{(a)}$Dipartimento di Fisica, Universit\`{a} del Salento \\
and  INFN Sezione di Lecce,  Via Arnesano 73100 Lecce, Italy}\\
\vspace{.5cm}
{\it $^{(b)}$ Department of Physics, University of Crete\\ Heraklion, Crete, Greece}

\begin{abstract}
We compute in linearized gravity all the contributions to the gravitational effective action due to a virtual Dirac fermion, related to the conformal anomaly. This requires, in perturbation theory, the identification of the gauge-gauge-graviton vertex off mass shell, involving the correlator of the energy-momentum tensor and two vector currents ($TJJ$), which is responsible for the generation of the gauge contributions to the conformal anomaly in gravity. We also present the anomalous effective action in the inverse mass of the fermion as in the Euler-Heisenberg case. 
  
\end{abstract}
\end{center}
\newpage
\section{Introduction}
Investigations of conformal anomalies in gravity (see \cite{Duff:1993wm} for an historical overview and references) \cite{Deser:1976yx} and in gauge theories \cite{Adler:1976zt,Collins:1976yq,Freedman:1974ze} as well as in string theory, have been of remarkable significance along the years. In cosmology, for instance, \cite{Starobinsky:1980te} (see also \cite{Shapiro:2008sf} for an overview)  the study of the gravitational trace anomaly has been performed in an attempt to solve the problem of the ``graceful exit" (see for instance \cite{Fabris:2000gz, Shapiro:2001rd, Shapiro:2001rh, Pelinson:2002ef}). In other analysis it has been pointed out that the conformal anomaly may prevent the future singularity occurrence in various dark energy models \cite{Nojiri:2004pf,Nojiri:2005sx}. 

In the past the analysis of the formal structure of the effective action for gravity in four dimensions,  obtained by integration of the trace anomaly \cite{Riegert:1984kt, Fradkin:1983tg}, has received a special attention, showing that the variational solution of the anomaly equation, which is non-local, can be made local by the introduction of extra scalar fields. The gauge contributions to these anomalies are identified at 1-loop level from a set of diagrams - involving fermion loops with two external gauge lines and one graviton line - and are characterized, as shown recently by Giannotti and Mottola in \cite{Giannotti:2008cv}, by the presence of anomaly poles. Anomaly poles are familiar from the study of the chiral anomaly in gauge theories and describe the non-local structure of the effective action. In the case of global anomalies, as in QCD chiral dynamics, they signal the presence of a non-perturbative phase of the fundamental theory, with composite degrees of freedom (pions) which offer an equivalent description of the fundamental lagrangian, matching the anomaly, in agreement with 't Hooft's principle. Previous studies of the role of the conformal anomaly in cosmology concerning the production of massless gauge particles and the identification of the infrared anomaly pole are those of Dolgov \cite{Dolgov:1981nw,Dolgov:1971ri}, while a discussion of the infrared pole from a dispersive derivation 
is contained in \cite{Horejsi:1997yn}.\\

In a related work \cite{Armillis:2009sm}  we have shown that anomaly poles are typical of the perturbative description of the chiral anomaly not just in some special kinematical conditions, for instance in the collinear region, where the coupling of the anomalous gauge current to two (on-shell) vector currents (for the AVV diagram) involves a pseudoscalar intermediate state (with a collinear and massless fermion-antifermion pair) but under any kinematical conditions.   
They are the most direct - and probably also the most significant - manifestation of the anomaly in the 
perturbative diagrammatic expansion of the effective action. On a more speculative side, the interpretation of the pole in terms of composite degrees of freedom could probably have direct physical implications, including the  condensation of the composite fields, very much like Bose Einstein (BE) condensation of the pion field, under the action of gravity. Interestingly, in a recent paper, Sikivie and Yang have pointed out that Peccei-Quinn axions (PQ) may form BE condensates \cite{Sikivie:2009qn}. With these motivations in mind, in this work, which parallels a previous 
investigation of the chiral gauge anomaly \cite{Armillis:2009sm}, we study the perturbative structure of the off-shell effective action showing the appearance of similar singularities under general kinematic conditions. Our investigation is a first step towards the computation of the exact effective action describing the coupling of the Standard Model to gravity via the conformal anomaly, that we hope to discuss in the future.  

In our study we follow closely the work of \cite{Giannotti:2008cv}. There the authors have presented a complete off-shell classification of the invariant amplitudes of the relevant correlator responsible for the conformal anomaly, which involves the energy momentum tensor (T) and two vector currents  (J),  $TJJ$, and have thoroughly  investigated it in the QED case, drawing on the analogy with the case of the chiral anomaly. The analysis of
$\cite{Giannotti:2008cv}$ is based on the use of dispersion relations, which are sufficient to identify the anomaly poles of the amplitude from the spectral density of this correlator, but not to characterize completely the off-shell effective action of the theory and the remaining non-conformal contributions, which will be discussed in this paper. The poles that we extract from the complete effective action include both the usual poles derived from the 
spectral analysis of the diagrams, which are coupled in the infrared (IR) and other extra poles which account for the anomaly but are decoupled in the same limit. These extra poles appear under general kinematic configurations and are typical of the off-shell as well as of the on-shell effective action, both for massive and massless fermions.

We also show, in agreement with those analysis, that the pole terms which contribute to the conformal anomaly are indeed only obtained in the on-shell limit of the external gauge lines, and identify all the mass corrections to the correlator in the general case.  This analysis is obtained by working out all the relevant kinematical limits of the perturbative corrections. We present the complete anomalous off-shell effective action describing the interaction of gravity with the photons, written in a form in which we separate the non-local contribution due to the anomaly pole from the rest of the action (those which are conformally invariant in the massless fermion limit). Away from the conformal limit of the theory we present  a $1/m$ expansion of the effective action as in the Euler-Heisenberg approach. This expansion, naturally, does not convey the presence of non-localities in the effective action due to the appearance of massless poles.

The computation of similar diagrams, for the on-shell photon case, appears in older contributions
by Berends and Gastmans \cite{Berends:1975ah} using dimensional regularization, in their study of the gravitational scattering of photons and by Milton using Schwinger's methods \cite{Milton:1976jr}. The presence of an anomaly pole in the amplitude has not been investigated nor noticed in these previous analysis, since they do not appear explicitly in their results, nor the $1/m$ expansion of the three form factors of the on-shell vertex, contained in \cite{Berends:1975ah}, allows their identification in the S-matrix elements of the theory.
Two related analysis by Drummond and Hathrell in their investigation of the gravitational contribution to the self-energy of the photon \cite{Drummond:1978hh} and the renormalization of the trace anomaly \cite{Drummond:1979pp} included the same on-shell vertex. Later, this same vertex has provided the ground for several elaborations concerning a possible superluminal behaviour of the photon in the presence of an external gravitational field \cite{Shore:2003zc}.



 \section{The conformal anomaly and gravity}
In this section we briefly summarize some basic and well known aspects of the trace anomaly in quantum
gravity and, in particular, the identification of the non-local action whose variation generates a given trace anomaly.

We recall that the gravitational trace anomaly in 4 spacetime dimensions generated by quantum effects in a classical gravitational and electromagnetic background is given by the expression
\beq
T_\mu^\mu= -\frac{1}{8} \left[ 2 b \,C^2 + 2 b' \left( E - \frac{2}{3}\square R\right) + 2 c\, F^2\right]
\eeq
where $b$, $b'$ and $c$ are parameters that  for a single fermion in the theory result  $b = 1/320 \, \pi^2$,  $b' = - 11/5760 \, \pi^2$,
and $c= -e^2/24 \, \pi^2$; furthermore $C^2$ denotes the Weyl tensor squared and $E$ is the Euler density given by
\beqa
C^2 &=& C_{\lambda\mu\nu\rho}C^{\lambda\mu\nu\rho} = R_{\lambda\mu\nu\rho}R^{\lambda\mu\nu\rho}
-2 R_{\mu\nu}R^{\mu\nu}  + \frac{R^2}{3}  \\
E &=& ^*\hskip-.1cm R_{\lambda\mu\nu\rho}\,^*\hskip-.1cm R^{\lambda\mu\nu\rho} =
R_{\lambda\mu\nu\rho}R^{\lambda\mu\nu\rho} - 4R_{\mu\nu}R^{\mu\nu}+ R^2.
\eeqa

The effective action is identified by solving the following variational equation by inspection
\beq
-\frac{2}{\sqrt{g}}g_{\mu\nu} \frac{\delta \Gamma}{\delta g_{\mu\nu}}=T_\mu^\mu.
\label{traceq}
\eeq
Its solution is well known and is given by the non-local expression
\beqa
&& \hspace{-.6cm}S_{anom}[g,A] = \label{Tnonl}\\
&&\frac {1}{8}\int d^4x\sqrt{-g}\int d^4x'\sqrt{-g'} \left(E - \frac{2}{3} \square R\right)_x
 G_4(x,x')\left[ 2b\,C^2
 + b' \left(E - \frac{2}{3} \square R\right) + 2c\, F_{\mu\nu}F^{\mu\nu}\right]_{x'} \nonumber.
\eeqa
Notice that we are omitting $\sqrt{g}R^2$ terms which are not necessary at one loop level.
 The notation $G_4(x,x')$ denotes the Green's function of the
differential operator defined by
\beq
\Delta_4 \equiv  \nabla_\mu\left(\nabla^\mu\nabla^\nu + 2 R^{\mu\nu} - \frac{2}{3} R g^{\mu\nu}\right)
\nabla_\nu = \square^2 + 2 R^{\mu\nu}\nabla_\mu\nabla_\nu +\frac{1}{3} (\nabla^\mu R)
\nabla_\mu - \frac{2}{3} R \square\,
\label{operator4}
\eeq
and requires some boundary conditions to be specified.
This operator is conformally covariant, in fact under a rescaling of the metric one can show that
\beq
g_{\mu\nu}=e^\sigma\bar{g}_{\mu\nu} \rightarrow \Delta_4=e^{-2 \sigma}\bar{\Delta}_4.
\eeq
Notice that the general solution of (\ref{traceq}) involves, in principle, also a conformally invariant part that is not
identified by this method.
As in ref. \cite{Giannotti:2008cv}, we concentrate on the contribution proportional to $F^2$ and perform an expansion of this term for a weak gravitational field and drop from this action all the terms which are at least quadratic in the deviation of the metric from flat space
\beq
g_{\mu\nu}=\eta_{\mu\nu} +\kappa \, h_{\mu\nu} \hspace{1.0cm} \kappa^2=16\,  \pi \, G,
\label{metricex}
\eeq
with $G$ the gravitational constant. The non-local action reduces to
\beq
S_{anom}[g,A] = -\frac{c}{6}\int d^4x\sqrt{-g}\int d^4x'\sqrt{-g'}\, R^{(1)}_x
\, \square^{-1}_{x,x'}\, [F_{\alpha\beta}F^{\alpha\beta}]_{x'}\,,
\label{simplifies}
 \eeq
 valid for a weak gravitational field. In this case 
 \beq
 R^{(1)}_x\equiv \partial^x_\mu\, \partial^x_\nu \, h^{\mu\nu} - \square \,  h, \qquad h=\eta_{\mu\nu} \, h^{\mu\nu}.
 \eeq
 The presence of the Green's function of the $\square$ operator in Eq. (\ref{simplifies}) is the clear indication that the solution of the anomaly equation is characterized by an anomaly pole. In the next sections we are going to perform a direct diagrammatic computation of this action and reobtain from it the pole contribution identified in the dispersive analysis of \cite{Giannotti:2008cv} and the conformal invariant extra terms which are not present in (\ref{simplifies}). We start with an analysis of the
correlator following an approach which is close to that followed in ref. \cite{Giannotti:2008cv}. The crucial point of the derivation presented in that work is the imposition of the Ward identity for the $TJJ$ correlator (see
Eq. (\ref{TWard}) below) which allows to eliminate all the Schwinger (gradients) terms which otherwise plague any derivation based on the canonical formalism and are generated by the equal-time commutator of the energy momentum tensor with the vector currents. In reality, this approach can be bypassed by
just imposing at a diagrammatic level the validity of an operatorial relation for the 
trace anomaly, evaluated at a nonzero momentum transfer, together with the conservation of the
vector currents on the other two vector vertices of the correlator.

\section{The construction of the full amplitude $\Gamma^{\mu \nu \a \b} (p,q)$}
We consider the standard QED lagrangian
\beq
\mathcal{L}=-\frac{1}{4} F_{\mu\nu} F^{\mu\nu} + \, i \,  \bar{\psi} \gamma^\mu(\partial_\mu - i \, e \, A_\mu)\psi
- m \bar{\psi}\psi,
\eeq
with the energy momentum tensor split into the free fermionic part $T_f$,  the interacting fermion-photon part
$T_{fp}$ and the photon contribution $T_{ph}$ which are given by
\beq
T^{\mu\nu}_{f} = -i \bar\psi \gamma^{(\mu}\!\!
\stackrel{\leftrightarrow}{\partial}\!^{\nu)}\psi + g^{\mu\nu}
(i \bar\psi \gamma^{\lambda}\!\!\stackrel{\leftrightarrow}{\partial}\!\!_{\lambda}\psi
- m\bar\psi\psi),
\label{tfermionic}
\eeq
\beq
T^{\mu\nu}_{fp} = -\, e J^{(\mu}A^{\nu)} + e g^{\mu\nu}J^{\lambda}A_{\lambda}\,,
\eeq
and
\beq
T^{\mu\nu}_{ph} = F^{\mu\lambda}F^{\nu}_{\ \ \lambda} - \frac{1}{4} g^{\mu\nu}
F^{\lambda\rho}F_{\lambda\rho},
\label{tphoton}
\eeq
where the current is defined as
\beq
J^{\mu}(x) = \bar\psi (x) \gamma^{\mu} \psi (x)\,.\\
\label{vectorcurrent}
\eeq

\begin{figure}[t]
\begin{center}
\includegraphics[scale=0.9]{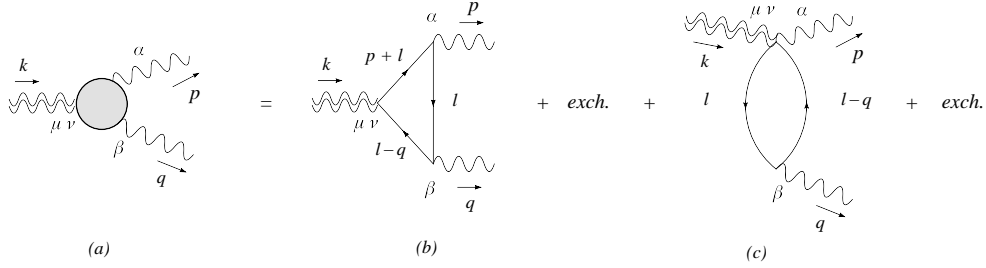}
\caption{\small The complete one-loop vertex (a) given by the sum of the 1PI contributions called $V^{\mu \nu \a \b} (p,q)$ (b) and $W^{\mu \nu \a \b} (p,q)$ (c). }
\label{vertex}
\end{center}
\end{figure}

In the coupling to gravity of the total energy momentum tensor
\beq
T^{\mu\nu}\equiv T_{f}^{ \mu\nu} +T_{fp}^{ \mu\nu} + T_{ph}^{ \mu\nu}
\eeq
we keep terms linear in the gravitational field, of the form  $h_{\mu\nu} T^{\mu\nu}$, and we have
introduced some standard notation for the symmetrization of the tensor indices and left-right derivatives $H^{(\mu\nu)} \equiv (H^{\mu\nu} + H^{\nu\mu})/2$ and
$\stackrel{\leftrightarrow}{\partial}\!\!_{\mu} \equiv
(\stackrel{\rightarrow}{\partial}\!\!_{\mu} - \stackrel{\leftarrow}{\partial}\!\!_{\mu})/2$.
It is also convenient to introduce a partial energy momentum tensor $T_p$, corresponding to the sum of the Dirac and interaction terms
\beq
T_p^{\mu\nu}\equiv T_{f}^{ \mu\nu} +T_{fp}^{ \mu\nu}
\eeq
which satisfies the inhomogeneous equation
\beq
\partial_\nu T_p^{\mu\nu}= -\partial_\nu T_{ph}^{ \mu\nu}.
\eeq
Using the equations of motion for the e.m. field  $\partial_\nu F^{\mu\nu}=J^{\mu}$, the inhomogeneous
equation becomes
\beq
\partial_\nu T_p^{\mu\nu}= F^{\mu\lambda} J_\lambda.
\eeq
There are two ways to identify the contributions of $T^{\mu\nu}$ and $T^{ \mu\nu}_p$  in the perturbative expansion of the effective action.
In the formalism of the background fields, the $T_pJJ$ correlator can be extracted from the defining functional integral
\beqa
\langle T_p^{\mu\nu}(z)\rangle_A &\equiv& \int D\psi D\bar{\psi} \,\,T^{\mu\nu}_p (z) \,\,e^{i \int d^4 x \mathcal{L} + \int J\cdot A(x) d^4 x}\nonumber \\
&=& \langle T^{\mu\nu}_p \, e^{i \int d^4 x \, J\cdot A(x)}\rangle
\eeqa
expanded through second order in the external field $A$. The relevant terms in this expansion are explicitly given by
\beq
\langle T_p^{\mu\nu}(z)\rangle_A = \frac{1}{2!}\langle T_{f}^{\mu\nu}(z) (J\cdot A)(J\cdot A) \rangle +
\langle T_{fp}^{\mu\nu}(J\cdot A)\rangle + ... \, ,
\eeq
with  $(J\cdot A)\equiv \int d^4 x J\cdot A(x)$.
The corresponding diagrams are extracted via two functional derivatives respect to the background field $A_\mu$
and are given by
\beq
\Gamma^{\mu\nu\alpha\beta} (z; x, y) \equiv \frac{ \delta^2 \lag T_p^{\mu\nu} (z) \rag_A}
{\delta A_{\alpha}(x)\delta A_{\beta}(y)} \bigg\vert_{A=0}
= V^{\mu\nu\alpha\beta} + W^{\mu\nu\alpha\beta}
\eeq

\beq
 V^{\mu\nu\alpha\beta}=(i \, e )^2 \, \lag T_{f}^{\mu\nu} (z) J^{\alpha} (x) J^{\beta} (y) \rag_{A=0}
\eeq
\beqa
W^{\mu\nu\alpha\beta} &=&\frac{ \delta^2 \lag T_{fp}^{\mu\nu} (z) (J\cdot A) \rag}
{\delta A_{\alpha}(x)\delta A_{\beta}(y)} \bigg\vert_{A=0} \nonumber \\
&=& \delta^4(x-z)g^{\alpha (\mu} \Pi^{\nu )\beta}(z, y)
+ \delta^4 (y-z)g^{\beta(\mu} \Pi^{\nu )\alpha}(z, x)
- g^{\mu\nu}[\delta^4(x-z) -  \delta^4(y-z) ]\Pi^{\alpha\beta}(x, y) \nonumber \\
\label{tjjcorrelator}
\eea
These two contributions are of $O(e^2)$.
Alternatively, one can directly compute the matrix element
\beq
\mathcal{M}^{\mu\nu} = \langle 0| T_p^{\mu\nu}(z) \int d^4 w d^4 w' J\cdot A(w) J\cdot A(w')|\gamma \gamma \rangle,
 \eeq
 which generates the diagrams (b) and (c) shown in Fig.\ref{vertex}, respectively called the ``triangle" and the
 ``t-p-bubble" (``t-" stays for tensor), together with the two ones obtained for the exchange of
 $p$ with $q$ and $\alpha$ with $\beta$.
 
 The conformal anomaly appears in the perturbative expansion of $T_p$ and involves these four diagrams.
The electromagnetic contribution is responsible for other two diagrams whose invariant amplitudes are
well-defined and will be used to fix the ill-defined amplitudes present in the tensor expansion of
$T_p^{\mu\nu}$ by using a Ward identity.

The lowest order contribution is obtained, in the background field formalism, from Maxwell's e.m. tensor, and is given by
\beqa
S^{\mu\nu\alpha\beta} &=&\frac{ \delta^2 \lag T_{ph}^{\mu\nu} (z) \rag}
{\delta A_{\alpha}(x)\delta A_{\beta}(y)} \bigg\vert_{A=0}.
\label{config}
\eeqa
Equivalently, it can be obtained from the matrix element
\beq
\langle 0| T_{ph}^{\mu\nu}| \gamma\gamma\rangle
\eeq
which allows to identify the vertex in momentum space.
Using (\ref{config}) we easily obtain
\beqa
S_{\mu\nu\alpha\beta}(z,x,y) &=& 2 g_{\alpha\beta} \partial_{\left( \right.\mu} \delta_{x z}\partial_{\nu\left.\right)}\delta_{yz} - 2 g_{\beta\left(\mu\right.}\partial_{\nu\left.\right)}\delta_{x z}\partial_\alpha\delta_{y z}  -
2 g_{\alpha\left(\nu\right.}\partial_{\left.\mu\right)}\delta_{yz}\partial_\beta\delta_{x z}\nonumber \\
&&+ g_{\alpha\mu}g_{\beta\nu} \partial_\lambda \delta_{yz} \partial^\lambda \delta_{x z}
 + g_{\alpha\nu}g_{\beta\mu} \partial_\lambda \delta_{yz} \partial^\lambda \delta_{x z}
 + g_{\mu\nu} \partial_{\beta}\delta_{x z} \partial_\alpha \delta_{y z}  
 - \partial_\rho \delta_{y z} \partial_\rho \delta_{x z} g_{\alpha\beta} g_{\mu\nu}
\nonumber \\
\eeqa
where $\partial_\mu\delta_{x z}\equiv \partial/\partial {x^\mu} \delta(x-z)$ and so on. In momentum space
this lowest order vertex is given by
\beqa
S^{\mu\nu\alpha\beta} &=&  
\big(p^{\mu} q^{\nu} + p^{\nu} q^{\mu}\big) \, g^{\alpha\beta}
+ p\cdot q\, \big(g^{\alpha\nu} g^{\beta\mu} + g^{\alpha\mu} g^{\beta\nu}\big) 
- g^{\mu\nu} \, (p\cdot q\,g^{\alpha\beta} - q^{\alpha} p^{\beta} ) \nn \\
&& - \, \big(g^{\beta\nu} p^{\mu} + g^{\beta\mu} p^{\nu} \big) \,q^{\alpha}
- \big (g^{\alpha\nu} q^{\mu} + g^{\alpha\mu} q^{\nu }\big) \, p^{\beta}.
\eeqa

The corresponding vertices which appear respectively in the triangle diagram and on the t-bubble
at $O(e^2)$ are given by
\beqa
 V^{\prime \, \mu\nu}(k_1, k_2)&=&\frac{1}{4} \left[\gamma^\mu (k_1 + k_2)^\nu
+\gamma^\nu (k_1 + k_2)^\mu \right] - \frac{1}{2} g^{\mu \nu}
[\gamma^{\lambda}(k_1 + k_2)_{\lambda} - 2 m]   \,,\\
 W^{\prime\,\mu\nu\alpha} &=& -\frac{1}{2} (\gamma^\mu g^{\nu\alpha}
+\gamma^\nu g^{\mu\alpha}) + g^{\mu \nu}\gamma^{\alpha}, 
\eeqa
where $k_1 (k_2)$ is outcoming (incoming).
Using the two vertices $V^{\prime \, \mu\nu}(k_1, k_2)$ and $W^{\prime\,\mu\nu\alpha} $ we obtain for the diagrams (b) and (c) of Fig.\ref{vertex}
\bea
&&V^{\mu\nu\alpha\beta}(p,q)=
- (-ie)^2 i^3  \int \frac{d^4 l}{(2 \pi)^4} 
\, \frac{{\rm tr}
\left\{V^{\prime\,\mu\nu}(l+p,l-q) (\lsl -\qsl +m)\gamma^{\beta}\,
(\lsl + m)\, \gamma^{\alpha}(\lsl +\psl + m)  \right\}}
{[l^2 - m^2] \, [(l-q)^2 - m^2] \, [(l+p)^2 - m^2] }\,,
\nn \\
\label{V}
\eea
and
\bea
W^{\mu\nu\alpha\beta}(p,q)   &=&
- (i e^2) i^2 \,  
\int \frac{d^4 l}{(2 \pi)^4} \, \frac{{\rm tr}\left\{W^{\prime \, \mu\nu\alpha}
\,(\lsl + m)\gamma^{\beta} (\lsl -\qsl +m) \right\}} {[l^2 - m^2][(l-q)^2 - m^2]},
\label{W}
\eea
%
%
%
so that the one-loop amplitude in Fig.~\ref{vertex} results
\bea
\Gamma^{\mu\nu\alpha\beta} (p,q) =
V^{\mu\nu\alpha\beta}(p,q) +\, V^{\mu\nu\b\a}(q,p)
+ W^{\mu\nu\alpha\beta}(p,q)+\, W^{\mu\nu\b\a}(q,p).
\label{Gamma}
\eea
The bare Ward identity which allows to define the divergent amplitudes that contribute to the anomaly in $\Gamma$ in terms of the remaining finite ones is obtained by re-expressing the classical equation
\beq
\partial_\nu T_{ph}^{\mu\nu} = - F^{\mu\nu} J_\nu
\eeq
as an equation of generating functionals in the background electromagnetic field
\beq
\partial_\nu \langle T_{ph}^{\mu\nu} \rangle_A= - F^{\mu\nu}\langle J_\nu\rangle_A,
\label{WIeq}
\eeq
which can be expanded perturbatively as
\beq
 \partial_\nu \langle T_{ph}^{\mu\nu} \rangle_A= - F^{\mu\nu}\langle J_\nu \int d^4 w (i e)J\cdot A (w)\rangle_
+ \,  ... \, .
\label{firstterm}
 \eeq
 Notice that we have omitted the first term in the Dyson's series of $\langle J_\nu\rangle_A$, shown on the r.h.s of
 (\ref{firstterm}) since $\langle J_\nu\rangle=0$.
The bare Ward identity then takes the form
\beq
 \partial_\nu  \Gamma^{\mu\nu\alpha\beta} =  \frac{ \delta^2 \left( F^{\mu\lambda}(z) \lag J_\lambda (z)\rag_A \right)}
{\delta A_{\alpha}(x)\delta A_{\beta}(y)} \bigg\vert_{A=0}
\eeq
which takes contribution only from the first term on the r.h.s of Eq. (\ref{firstterm}). This relation can be written in momentum space. For this we use the definition of the vacuum polarization
\beq
\Pi^{\alpha\beta}(x,y) \equiv  - i e^2 \langle J_\alpha(x) J_\beta(y)\rangle,
\eeq
or
\bea
\Pi^{\alpha\beta} (p) &=&  - i^2 \, (-ie)^2 
\int \frac{d^4 l}{(2\pi)^4} \frac{{\rm tr}\left\{\gamma^{\alpha}
\,(\lsl + m)\gamma^{\beta} (\lsl + \psl +m) \right\}} {[l^2 - m^2] \, [(l+p)^2 - m^2]} \nn \\
&=&    (p^2 g^{\alpha\beta} - p^{\alpha}p^{\beta}) \,\Pi (p^2,m^2)
\label{polmom}
\eea
with
\beq
\Pi (p^2,m^2) = \frac{ e^2}{36 \, \pi^2  \, p^2 }
\biggl[ 6  \, \mathcal A_0(m^2) +  p^2 - 6 \, m^2
   - \, 3 \, \mathcal B_0(p^2,m^2) \left(2 m^2 + p^2  \right)\biggr],
\label{vacuumpol}
\eeq
which obviously satisfies the Ward identity $p_{\alpha} \, \Pi^{\alpha\beta} (p) =0$. The expressions of the $\mathcal A_0$ and $\mathcal B_0$ contributions are given in Appendix \ref{scalars}.

Using these definitions, the unrenormalized Ward identity which allows to completely characterize the form of the correlator in momentum space becomes
\bea
k_\nu\,\Gamma^{\mu\nu\alpha\beta}(p,q)&=&  \,\,\,\,
\left(q^\mu p^\alpha p^\beta -q^\mu g^{\alpha\beta} p^2 +g^{\mu\beta}q^\alpha p^2
-g^{\mu\beta}p^\alpha p\cdot q \right)   \Pi(p^2) \nonumber\\
&&+ \left(p^\mu q^\alpha q^\beta - p^\mu g^{\alpha\beta} q^2
+g^{\mu\alpha}p^\beta q^2 - g^{\mu\alpha}q^\beta p\cdot q\right) \Pi(q^2) \,.
\label{TWard}
\eea

\subsection{Tensor expansion and invariant amplitudes of $\Gamma$}
\label{tensorexpansion}
\begin{table}
$$
\begin{array}
{|c 
 | 
 c 
 |
c 
 |
c 
|
c
|
c 
|}
\hline
& & & & & \\[-.5cm]
\begin{array}[t]{c}
p^{\mu} p^{\nu} p^{\alpha} p^{\beta}\\
q^{\mu} q^{\nu} q^{\alpha} q^{\beta}
\end{array}
&
\begin{array}[t]{c}
p^{\mu} p^{\nu} p^{\alpha} q^{\beta}\\
p^{\mu} p^{\nu} q^{\alpha} p^{\beta}\\
p^{\mu} q^{\nu} p^{\alpha} p^{\beta}\\
q^{\mu} p^{\nu} p^{\alpha} p^{\beta}
\end{array}
&
\begin{array}[t]{c}
p^{\mu} p^{\nu} q^{\alpha} q^{\beta}\\
p^{\mu} q^{\nu} p^{\alpha} q^{\beta}\\
q^{\mu} p^{\nu} p^{\alpha} q^{\beta}
\end{array}
&
\begin{array}[t]{c}
p^{\mu} q^{\nu} q^{\alpha} p^{\beta}\\
q^{\mu} p^{\nu} q^{\alpha} p^{\beta}\\
q^{\mu} q^{\nu} p^{\alpha} p^{\beta}
\end{array}
&
\begin{array}[t]{c}
p^{\mu} q^{\nu} q^{\alpha} q^{\beta}\\
q^{\mu} p^{\nu} q^{\alpha} q^{\beta}\\
q^{\mu} q^{\nu} p^{\alpha} q^{\beta}\\
q^{\mu} q^{\nu} q^{\alpha} p^{\beta}
\end{array}
&
\begin{array}[t]{c}
g^{\mu\nu}g^{\alpha\beta}\\
g^{\alpha\mu}g^{\beta\nu}\\
g^{\alpha\nu}g^{\beta\mu}
\end{array}
\\[2.1cm]
\hline
& & & & & \\[-.5cm]
\begin{array}{c}
p^{\mu} p^{\nu} g^{\alpha\beta}\\
p^{\mu} q^{\nu} g^{\alpha\beta}\\
q^{\mu} p^{\nu} g^{\alpha\beta}\\
q^{\mu} q^{\nu} g^{\alpha\beta}
\end{array}
&
\begin{array}{c}
p^{\beta} p^{\nu} g^{\alpha\mu}\\
p^{\beta} q^{\nu} g^{\alpha\mu}\\
q^{\beta} p^{\nu} g^{\alpha\mu}\\
q^{\beta} q^{\nu} g^{\alpha\mu}
\end{array}
&
\begin{array}{c}
p^{\beta} p^{\mu} g^{\alpha\nu}\\
p^{\beta} q^{\mu} g^{\alpha\nu}\\
q^{\beta} p^{\mu} g^{\alpha\nu}\\
q^{\beta} q^{\mu} g^{\alpha\nu}
\end{array}
&
\begin{array}{c}
p^{\alpha} p^{\nu} g^{\beta\mu}\\
p^{\alpha} q^{\nu} g^{\beta\mu}\\
q^{\alpha} p^{\nu} g^{\beta\mu}\\
q^{\alpha} q^{\nu} g^{\beta\mu}
\end{array}
&
\begin{array}{c}
p^{\mu} p^{\alpha} g^{\beta\nu}\\
p^{\mu} q^{\alpha} g^{\beta\nu}\\
q^{\mu} p^{\alpha} g^{\beta\nu}\\
q^{\mu} q^{\alpha} g^{\beta\nu}
\end{array}
&
\begin{array}{c}
p^{\alpha} p^{\beta} g^{\mu\nu}\\
p^{\alpha} q^{\beta} g^{\mu\nu}\\
q^{\alpha} p^{\beta} g^{\mu\nu}\\
q^{\alpha} q^{\beta} g^{\mu\nu}
\end{array}
\\ [1.2cm]\hline
\label{WI}
\end{array}
$$
\caption{\small The 43 tensor monomials built up from the metric tensor and the two independent momenta $p$ and $q$ into which a general fourth rank tensor can be expanded.
\label{monomials}}
\end{table}
The full one-loop amplitude $\Gamma$ can be expanded on the basis provided by the 43 monomial tensors listed in Tab.\ref{monomials}
\bea
\Gamma^{\mu\nu\alpha\beta} (p,q) = \, \sum_{i=1}^{43} \, A_i(k^2,p^2,q^2) \, l_i^{\mu \nu \a \b} (p,q).
\eea
Since the amplitude $\Gamma^{\mu\nu\alpha\beta} (p,q) $ has total mass dimension equal to $2$ it is obvious that not all the coefficients $A_i$ are convergent. They can be divided into $3$ groups:
\begin{itemize}
\item[a)]  $A_1 \leq A_i \leq A_{16}$ - multiplied by a product of four momenta, they have mass dimension $-2$ and therefore are UV finite;
\item [b)] $A_{17} \leq A_i \leq A_{19}$ - these have mass dimension $2$ since the four Lorentz indices of the amplitude are carried by two metric tensors
\item [c)] $A_{20} \leq A_i \leq A_{43}$ - they appear next to a metric tensor and two momenta, have mass dimension $0$ and are divergent.
\end{itemize}
The way in which the $43$ invariant amplitudes will be managed in order to reduce them to the $13$ named $F_i (k^2,p^2,q^2)$ is the subject of this section.
The reduction is accomplished in 4 different steps and has as a guiding principle the elimination of the divergent amplitudes $A_i$ in terms of the convergent ones after imposing some conditions on the whole amplitude.
We require
\begin{itemize}
\item[a)] the symmetry in the two indices $	\mu$ and $\nu$  of the symmetric energy-momentum tensor $T^{\mu\nu}$;
\item[b)] the conservation of the two vector currents on $p^{\a}$ and $q^{\b}$;
\item[c)] the Ward identity on the vertex with the incoming momentum $k$ defined above in Eq. (\ref{WI}).
\end{itemize}
Condition a) becomes
\bea
\Gamma^{\mu\nu\alpha\beta} (p,q) = \Gamma^{\nu\mu\alpha\beta} (p,q),
\eea
giving a linear system of $43$ equations; $15$ of them being identically satisfied because the tensorial structures are already symmetric in the exchange of $\mu$ and $\nu$, while the remaining $14$ conditions are
\bea
&& \hspace{0.2cm} A_5=A_6, \hspace{1.7cm} A_8=A_9,  \hspace{1.7cm} A_{10}=A_{11},
 \hspace{1.5cm}  A_{13}=A_{14},  \hspace{1.5cm} A_{18}=A_{19}, \nn \\
&& A_{21}=A_{22},   \hspace{1.5cm} A_{24}=A_{28},  \hspace{1.5cm} A_{25}=A_{29},
 \hspace{1.5cm} A_{26} =  A_{30},  \hspace{1.5cm} A_{27} = A_{31}, \nn \\
&& A_{32} = A_{36},  \hspace{1.5cm}  A_{34} = A_{37},  \hspace{1.5cm} A_{33} = A_{38},
 \hspace{1.5cm}  A_{35} =  A_{39},
 \label{conda}
\eea
where all $A_i$ are thought as functions of the invariants $k^2, p^2, q^2$.
After substituting (\ref{conda}) into $\Gamma^{\mu\nu\alpha\beta} (p,q)$ the $43$ invariant tensors of the decomposition are multiplied by only $29$ invariant amplitudes. Condition b), which is vector current conservation on the two vertices with indices $\a$ and $\b,$ allows to re-express some divergent $A_i$ in terms of other finite ones
\bea
p_\a \, \Gamma^{\mu\nu\alpha\beta} (p,q) = q_{\b} \, \Gamma^{\mu\nu\alpha\beta} (p,q) =0.
\label{WI12}
\eea
This constraint generates two sets of $14$ independent tensor structures each, so that in order to fulfill (\ref{WI12}) each coefficient is separately set to vanish. The first Ward identity leads to a linear system composed of $10$ equations
\bea
p_\a \, \Gamma^{\mu\nu\alpha\beta} (p,q) = 0 	\Rightarrow
\left \{ \begin {array}{l}
A_{19} + A_{36} \, p \cdot p + A_{37} \, p \cdot q = 0, \\
A_{38} \,  p \cdot p+A_{39}\,  p \cdot q=0, \\
A_{17} +A_{40} \,  p \cdot p+A_{42}  \,   p \cdot q=0,\\
A_{41} \, p \cdot p+A_{43}  \,   p \cdot q=0, \\
A_{20}+2 A_{28}+A_{1} \, p \cdot p+A_{4}  \,  p \cdot q=0, \\
2 \,  A_{30}+A_{3} \,  p \cdot p+A_{7}  \,  p \cdot q=0, \\
A_{22}+A_{29}+A_{6} \,  p \cdot p+A_{11} \,   p \cdot q=0, \\
A_{31}+A_{9}\,  p \cdot p+A_{14}  \,  p \cdot q=0,\\
A_{23}+A_{12} \, p \cdot p+A_{16} \,   p \cdot q=0, \\
A_{15} \, p \cdot p+A_{2} \, p \cdot q=0;
\end{array} \right.
\eea
 we choose to solve it for the set $\{A_{15}, A_{17}, A_{19}, A_{23}, A_{28}, A_{29}, A_{30}, A_{31}, A_{39}, A_{43}\}$ in which only the first one is convergent and the others are UV divergent. The set would not include
 all the divergent $A_i$ since in the last equations appear two convergent coefficients, $A_{15} $ and $A_{2}$. \\Following our choice the result is
\bea
&& A_{15} = -A_{2}  \, \frac{ p \cdot q}{p \cdot p},  \hspace{5.3cm}
A_{17}= -A_{40} \,    p \cdot p - A_{42} \, p \cdot q, \\
&&A_{19} = -A_{36} \,  p \cdot p-A_{37} \, p \cdot q, \hspace{3.4cm}
A_{23}= -A_{12}\,   p \cdot p - A_{16} \, p \cdot q, \\
&&A_{28}= \frac{1}{2} \, \Big[-A_{20} - A_{1} \, p \cdot p - A_{4} \, p \cdot q\Big] , \hspace{1.8cm}
A_{29}= -A_{22}-A_{6} \,   p \cdot p-A_{11} \, p \cdot q, \\
&&A_{30}=   - \frac{1}{2} \Big[A_{3} p \cdot p + A_{7} \,   p \cdot q\Big] , \hspace{3.2cm}
A_{31}= -A_{9}\,    p \cdot p-A_{14} \, p \cdot q, \\
&& A_{39}= - A_{38} \frac{ p \cdot p}{p \cdot q}, \hspace{5.3cm}
A_{43} = -  A_{41} \, \frac{ p \cdot p}{p \cdot q}.
\eea
In an analogous way we go on with the second Ward identity (WI) after replacing the solution of the previous system in the original amplitude. The new one is indicated by $\Gamma^{\mu\nu\alpha\beta}_{b} (p,q)$, where the subscript $b$ is there to indicate that we have applied condition $b)$ on $\Gamma$. The constraint gives
\bea
q_\b \, \Gamma^{\mu\nu\alpha\beta}_b (p,q) = 0 	\Rightarrow
\left \{ \begin {array}{l}
 A_{40}  \, p \cdot q+A_{41} \, q \cdot q \,=0, \\
 A_{1} \, p \cdot q+A_{3}  \, q \cdot q \,=0, \\
 A_{20}+A_{4} \, p \cdot q+A_{7} \, q \cdot q \,=0, \\
 A_{36}+A_{6} \, p \cdot q+A_{9}  \, q \cdot q \,=0,\\
 A_{22}+A_{37}+A_{11} \, p \cdot q+A_{14}  \, q \cdot q \,=0, \\
 2 A_{38}+A_{12} \, p \cdot q- A_{2}  \frac{ \, p \cdot q \, q\cdot q} {p \cdot p }=0.
 \end{array} \right.
\eea
We solve these equations determining the amplitudes in the set $\{ A_1, A_{20}, A_{22}, A_{36}, A_{38}, A_{40}\}$ in terms of the remaining ones, obtaining
\bea
&& A_{38}= -\frac{A_{12} \, p \cdot p \,    \, p \cdot q \, -A_{2} \, p \cdot q \,  \, q \cdot q \, } {2   \, p \cdot p \, },
\hspace{2.3cm} A_{40}= -\frac{A_{41}   \, q \cdot q \, }{\, p \cdot q \, },\\
&& \hspace{0.2cm} A_{1}= -\frac{A_{3}   \, q \cdot q \, }{\, p \cdot q \, }, \hspace{5.8cm}
A_{20}= -A_{4}   \, p \cdot q \, -A_{7} \, q \cdot q \, , \\
&& A_{22}= -A_{37}-A_{11}   \, p \cdot q \, -A_{14} \, q \cdot q \, , \hspace{2.8cm}
A_{36}= -A_{6}    \, p \cdot q \, -A_{9} \, q \cdot q.
\eea
The manipulations above have allowed a reduction of the number of invariant amplitudes from the initial $43$ to $13$ using the  $\{\mu,\nu\}$ symmetry ($14$ equations), the first WI on $p_\a$ ($10$ equations) and the second WI on $q_\b$ ($6$ equations). The surviving invariant amplitudes in which the amplitude $\Gamma^{\mu\nu\alpha\beta}_c (p,q)$ can be expanded using the form factors are
$\{ A_2, A_3, A_4, A_6, A_7, A_9, A_{11}, A_{12}, A_{14}, A_{16}, A_{37}, A_{41}, A_{42} \}$. This set still 
contains $3$ divergent amplitudes, $(A_{37}, A_{41}, A_{42} )$. The amplitude $\Gamma^{\mu\nu\alpha\beta}_c (p,q) $ is indeed ill-defined until we impose on it condition c), that is Eq. (\ref{TWard}). This condition gives
\bea
\textrm{Eq.} (\ref{TWard})	\Rightarrow
\left \{ \begin {array}{l}
 \vspace{0.2cm} 
 -A_{3} \, \,  \Big[1 + \frac{ \, p \cdot p \,  }{2 \, p \cdot q \, }\Big] \, \,  +A_{6}  +\frac{1}{2}  \,  A_{7}   \,  -A_{9} \,  -  \,  \frac{A_{41}    }{\, p \cdot q \, } =0, \\
 \vspace{0.2cm}
A_{37}+A_{42}+A_{4}\, \,  [ p \cdot p \, + p \cdot q] \,\,  + \,  A_{11}   \, p \cdot q \, +\frac{1}{2}  \,   \, A_{7} \,  \, q \cdot q \, + \nn \\
\vspace{0.2cm} 
\hspace{3cm} 
 +A_{11}  \,   \, q \cdot q \, +\frac{1}{2} \, A_3 \, \, \frac{ \, p \cdot p \,  \, q \cdot q \, }{   \, p \cdot q \, }=0, \\
\vspace{0.2cm} 
\frac{1}{2} A_2  \frac{\, p \cdot q   \, q \cdot q } {  \, p \cdot p}
- A_{41}  \frac{ \, p \cdot p + \, q \cdot q }{ \, p \cdot q} -
\frac{1}{2} A_3  \, p  \cdot p + A_7 ( \, p \cdot p  + \frac{1}{2} \, p \cdot q ) 
+A_6  \, p  \cdot q  \nn \\
\vspace{0.2cm} 
\hspace{3cm} 
+ A_{12} (\frac{1}{2} \, p  \cdot q +  q \cdot q ) +A_{14}
 Ê( \, p  \cdot q  +2 \, q \cdot q ) +
2 Ê A_{37} -\Pi (p^2)-\Pi (q^2)=0
  \end {array} \right.
\eea
From this condition we obtain
\bea
&& A_{37} =  -\frac{A_2}{4} \frac{ p \cdot q  \,  q \cdot q}{ p \cdot p}
+\frac{1}{4} \, A_3 \, p \cdot p -\frac{1}{4} A_7 \left(Ê2 \, p \cdot p + p \cdot q \right) 
+ \frac{1}{2} \, A_{41} \left ( \frac{p \cdot p + q \cdot q} { p \cdot q} \right ) 
  \nn \\
&&
\hspace{1cm}
- \frac{1}{2} \, A_6 \, Ê p\cdot  q
- \frac{1}{4} A_{12} \left(Ê p \cdot q  + 2 \, q \cdot q \right) 
- \frac{1}{2} A_{14} \left(Ê p \cdot  q + 2 
 Ê q \cdot  q \right ) + \frac{1}{2} \left[ \Pi (p^2)+ \Pi
 Ê (q^2) \right]  \nn \\ \\
&& A_{41} = - \frac{A_3}{2} \, p \cdot p -  \, \left(A_3 - A_6 - A_7 + A_9 \right) \, p \cdot q  \\
&& A_{42} =  \frac{A_3}{2} \, p \cdot p \, \left( \frac{p \cdot p } {p \cdot q } + 1 - \frac{ q \cdot q} {p \cdot q} \right)
+ \frac{1}{2} \, A_7 \left( p \cdot p + p \cdot q - q \cdot q\right) - A_4 \, \left( p \cdot p + p \cdot q \right) \nn \\
&& \hspace{3cm} - (A_6 - A_9 )\,  p \cdot p
+ (A_{14} - A_{11}) (q \cdot q + p \cdot q).
\eea
After these steps we end up with an expression for $\Gamma$ written in terms of only $10$  invariant amplitudes, that are $\mathcal{X}\equiv\{A_2, A_3,A_4,A_6,A_7,A_9,A_{11},A_{12},A_{14},A_{16} \}$,  significantly reduced respect to the original $43$. Further reductions are possible (down to $8$ independent invariant amplitudes), however, since these reductions just add to the complexity of the related tensor structures, it is convenient to select an appropriate set of reducible (but finite) components characterized by a simpler tensor structure and present the result in that form. The 13 amplitudes introduced in the final decomposition are, in this respect, a good choice since the corresponding tensor structures are rather simple. These tensors are combinations of the $43$ monomials listed in Tab.\ref{monomials}.

The set $\mathcal{X}$ is very useful for the actual computation of the tensor integrals and for the study of their reduction to scalar form. To compare with the previous study of Giannotti and Mottola \cite{Giannotti:2008cv} we have mapped the computation of the components of the set $\mathcal{X}$ into their structures $F_i$ $(i=1,2,..,13) $. Also in this case, the truly independent amplitudes are 8. One can extract, out of the 13 reducible amplitudes, a consistent subset of 8 invariant amplitudes. The remaining amplitudes in the 13 tensor structures are, in principle, obtainable from this subset.

\subsection{Reorganization of the amplitude}
Before obtaining the mapping between the amplitudes in $\mathcal{X}$ and the structures $F_i$, we briefly describe the tensor decomposition introduced in \cite{Giannotti:2008cv} which defines these 13 structures.
We define the rank-2 tensors
\bea
&&u^{\alpha\beta}(p,q) \equiv (p\cdot q) \,  g^{\alpha\beta} - q^{\alpha} \, p^{\beta}\,,\\
&&w^{\alpha\beta}(p,q) \equiv p^2 \, q^2 \, g^{\alpha\beta} + (p\cdot q) \, p^{\alpha} \, q^{\beta}
- q^2 \,  p^{\alpha} \, p^{\beta} - p^2 \, q^{\alpha} \, q^{\beta}\,,
\label{uwdef}
\eea
which are Bose symmetric,
\bea
&&u^{\alpha\beta}(p,q) = u^{\beta\alpha}(q,p)\,,\\
&&w^{\alpha\beta}(p,q) = w^{\beta\alpha}(q,p)\,,
\eea
and conserve vector current,
\bea
&&p_{\alpha} \, u^{\alpha\beta}(p,q)  = q_{\beta} \, u^{\alpha\beta}(p,q) = 0\,,\\
&&p_{\alpha} \, w^{\alpha\beta}(p,q)  = q_{\beta} \, w^{\alpha\beta}(p,q) = 0\,.
\eea
\begin{table}
$$
\begin{array}{|c|c|}\hline
i & t_i^{\mu\nu\alpha\beta}(p,q)\\ \hline\hline
1 &
\left(k^2 g^{\mu\nu} - k^{\mu } k^{\nu}\right) u^{\alpha\beta}(p.q)\\ \hline
2 &
\left(k^2g^{\mu\nu} - k^{\mu} k^{\nu}\right) w^{\alpha\beta}(p.q)  \\ \hline
3 & \left(p^2 g^{\mu\nu} - 4 p^{\mu}  p^{\nu}\right)
u^{\alpha\beta}(p.q)\\ \hline
4 & \left(p^2 g^{\mu\nu} - 4 p^{\mu} p^{\nu}\right)
w^{\alpha\beta}(p.q)\\ \hline
5 & \left(q^2 g^{\mu\nu} - 4 q^{\mu} q^{\nu}\right)
u^{\alpha\beta}(p.q)\\ \hline
6 & \left(q^2 g^{\mu\nu} - 4 q^{\mu} q^{\nu}\right)
w^{\alpha\beta}(p.q) \\ \hline
7 & \left[p\cdot q\, g^{\mu\nu}
-2 (q^{\mu} p^{\nu} + p^{\mu} q^{\nu})\right] u^{\alpha\beta}(p.q) \\ \hline
8 & \left[p\cdot q\, g^{\mu\nu}
-2 (q^{\mu} p^{\nu} + p^{\mu} q^{\nu})\right] w^{\alpha\beta}(p.q)\\ \hline
9 & \left(p\cdot q \,p^{\alpha}  - p^2 q^{\alpha}\right)
\big[p^{\beta} \left(q^{\mu} p^{\nu} + p^{\mu} q^{\nu} \right) - p\cdot q\,
(g^{\beta\nu} p^{\mu} + g^{\beta\mu} p^{\nu})\big]  \\ \hline
10 & \big(p\cdot q \,q^{\beta} - q^2 p^{\beta}\big)\,
\big[q^{\alpha} \left(q^{\mu} p^{\nu} + p^{\mu} q^{\nu} \right) - p\cdot q\,
(g^{\alpha\nu} q^{\mu} + g^{\alpha\mu} q^{\nu})\big]  \\ \hline
11 & \left(p\cdot q \,p^{\alpha} - p^2 q^{\alpha}\right)
\big[2\, q^{\beta} q^{\mu} q^{\nu} - q^2 (g^{\beta\nu} q^ {\mu}
+ g^{\beta\mu} q^{\nu})\big]  \\ \hline
12 & \big(p\cdot q \,q^{\beta} - q^2 p^{\beta}\big)\,
\big[2 \, p^{\alpha} p^{\mu} p^{\nu} - p^2 (g^{\alpha\nu} p^ {\mu}
+ g^{\alpha\mu} p^{\nu})\big] \\ \hline
13 & \big(p^{\mu} q^{\nu} + p^{\nu} q^{\mu}\big)g^{\alpha\beta}
+ p\cdot q\, \big(g^{\alpha\nu} g^{\beta\mu}
+ g^{\alpha\mu} g^{\beta\nu}\big) - g^{\mu\nu} u^{\alpha\beta} \\
& -\big(g^{\beta\nu} p^{\mu}
+ g^{\beta\mu} p^{\nu}\big)q^{\alpha}
- \big (g^{\alpha\nu} q^{\mu}
+ g^{\alpha\mu} q^{\nu }\big)p^{\beta}  \\ \hline
\end{array}
$$
\caption{Basis of 13 fourth rank tensors satisfying the vector current conservation on the external lines with momenta $p$ and $q$. \label{genbasis}}
\end{table}
These two tensors are used to build the  set of $13$ tensors catalogued in Table \ref{genbasis}. They are linearly independent for generic $k^2, p^2, q^2$
different from zero. Five of the $13$ tensors are Bose symmetric, namely,
\bea
t_i^{\mu\nu\alpha\beta}(p,q) = t_i^{\mu\nu\beta\alpha}(q,p)\,,\qquad i=1,2,7,8,13\,,
\eea
while the remaining eight tensors form four pairs which are overall related by Bose symmetry
\bea
&&t_3^{\mu\nu\alpha\beta}(p,q) = t_5^{\mu\nu\beta\alpha}(q,p)\,,\\
&&t_4^{\mu\nu\alpha\beta}(p,q) = t_6^{\mu\nu\beta\alpha}(q,p)\,,\\
&&t_9^{\mu\nu\alpha\beta}(p,q) = t_{10}^{\mu\nu\beta\alpha}(q,p)\,,\\
&&t_{11}^{\mu\nu\alpha\beta}(p,q) = t_{12}^{\mu\nu\beta\alpha}(q,p)\,.
\label{tpairs}
\eea
The amplitude in (\ref{Gamma}) can be expanded in this basis composed as
\bea
\Gamma^{\mu\nu\alpha\beta}(p,q) =  \, \sum_{i=1}^{13} F_i (s; s_1, s_2,m^2)\ t_i^{\mu\nu\alpha\beta}(p,q)\,,
\label{Gamt}
\eea
where the invariant amplitudes $F_i$ are functions of the kinematical invariants $s=k^2=(p+q)^2$, $s_1=p^2$, $s_2=q^2$ and of the internal mass $m$. In \cite{Giannotti:2008cv} the authors use the
Feynman parameterization and momentum shifts in order to identify the expressions of these amplitudes in terms of parametric integrals, which was the approach followed also by Rosenberg in his original identification of the 6 invariant amplitudes of the AVV anomaly diagram.
If we choose to reorganize all the monomials into the simpler set of $13$ tensor groups shown in Tab.\ref{genbasis}, then we need to map the $A_i$  in $\mathcal{\chi}$ and the $F_i$'s. The mapping is given by
\bea
F_1  &=&    \frac{1} {3 \,   k^2 \, }
   \Big[ A_4 (4  \, p \cdot q + 3  \, p \cdot p) \, + 2 A_{11} ( \, p \cdot q \,  + \, 2 \, q \cdot q )
    \, + 2 A_6 \, p \cdot p \,  \, \Big. \nn \\
   && \Big. + 2 A_7 \, q \cdot q \,
    \,   - 2 A_{14} \, q \cdot q \,
    - A_{16} \, q \cdot q \,   + 2 A_3 \, \frac{p \cdot p \,   \, q \cdot q }{p \cdot q}\, \Big]  , 
 \label{mapping1}   \\
F_2  &=&    \frac{1} {3 \, \, k^2 \, } \left[ - 2 A_3 \left(\, \frac{p \cdot p}{p \cdot q } \, + 2    \, \right) + 4 A_6 \, \, + A_7 \,  \, - 2 A_9   \, -A_{12} \,  \right]  , \\
F_3  &=&    \frac{1}  {12 \, k^2 \, }
\left[  A_4 ( 2 \, p \cdot q \, + 3  \, q \cdot q \,)  - 2 A_{11}  ( \, p \cdot q \, + 2  \, q \cdot q \, ) - 2 A_6 \, p \cdot p  \,  \right.\nn \\
&&  \left. - 2 A_7 \, q \cdot q  \,
    + 2 A_{14} \, q \cdot q \,   \, + A_{16} \, q \cdot q \,  - 2 A_3 \, \frac{ p \cdot p \,  \, q \cdot q }{p \cdot q}\, \right]  \\
F_4  &=&   \frac{A_7}{4  \, p \cdot p \, } +
\frac{1}{12 \,  k^2 }
\left[ 2 A_3 \left (\, \frac{p \cdot p}{p \cdot q \,} \, + 2  \right) \, -4 A_6  \,
- A_7  \, + 2 A_9 \,  \, + A_{12}  \, \right]  \\
F_5  &=&
   \frac{A_{16}}{4} +
   \frac{1}   {12 \,\, k^2}
   \left[ -2 \, A_{6} \, p \cdot p \,
   - 2 \, A_3\,  \frac{ q \cdot q \,  p \cdot p \, } { p \cdot q \, }
   +A_4 \, (-3 \, p \cdot p \, -4  \,  p \cdot q \,)\right.  \nn \\
   && \left. - 2 \, A_{11} \,( p \cdot q \, + 2 q \cdot q \, ) - 2 \, A_7 \,
   q \cdot q \,+2\,  A_{14} \, q \cdot q \, + A_{16}\, q \cdot q \, \right],\\
F_6  &=&   \frac{A_{12}}{4   \, q \cdot q \, } +
\frac{1}   {12 \,   k^2 }
\left[ -4 \, A_6 - A_7 + \, 2 \,  A_9 + A_{12} \, + \, 2 \, A_3
\left(\frac{   p \cdot p}{p \cdot q} +\, 2 \right)   \right], \\
F_7 &=&
   \frac{A_{11}}{2} + \frac{1}{ \, p \cdot q \, ^2}  \left( A_9 \, q \cdot q \, p \cdot p \, + \frac{A_6}{2} \, p \cdot p \, p \cdot q \,
   + \frac{A_{14}}{2} \, q \cdot q \, p \cdot q \, \right)
    +\frac{1} {6 \,   k^2}  \Big[  \, A_4 (- 4   \, p \cdot q \,- 3 \, p \cdot p \, )  \Big. \nn\\
&& \Big.
  - 2 \, A_{11} (\, p \cdot q \,  +2 \, \, q \cdot q \,)
    - 2 \, A_6   \, p \cdot p \, -\, 2 \, A_7 \, q \cdot q \,
     + \, 2 \, A_{14} \, q \cdot q \,
  + \,A_{16} \, q \cdot q \,  -\, 2 \, A_3 \frac{ \, p \cdot p \,   \, q \cdot q \,}{ \, p \cdot q}  \Big] ,\nn \\ \\
F_8  &=&   \frac{1}{6 \,   k^2} \, \Big[  2 \,  A_3 \left( \frac{ \, p \cdot p }{p \cdot q}\,  + 2 \right)
-3   \, \frac{A_9} {p \cdot q}\,\left(  \, p \cdot p + q \cdot q \right)  - \,4 \, A_6 \,
-\, A_7   \,  - \, 4 \, A_9  + \, A_{12}  \, \Big]  \\
F_9  &=&
   \frac{A_6 }{p \cdot q} \,  + \, A_9  \frac{\, q \cdot q \, }{\, p \cdot q \, ^2},  \\
F_{10}  &=&   A_9 \, \frac{p \cdot p } {\, p \cdot q \, ^2} + \frac{ A_{14}}{\, p \cdot q \,}, \\
F_{11}  &=&   \frac{A_{12}}{2
   \, q \cdot q \, }-\frac{A_2}{2 \, p \cdot p \, }, \\
F_{12}  &=&   \frac{A_3}{ 2 \, p \cdot q \, }
    + \frac{A_7} {2 \, p \cdot p \, }, \\
F_{13}  &=&
\frac{1}{2} A_6  \, \left( p \cdot p \, + p \cdot q \,- q \cdot q \,\right)
+\frac{1}{4} A_7  \,  \left(p \cdot p \, + p \cdot q \, - q \cdot q \,\right )
+\frac{A_2 \, p \cdot q \, q \cdot q \,} {4 \, p \cdot p \,}
+A_{14} \,  \left(\frac{p \cdot q \,}{2}+q \cdot q \,\right) \nn \\
&&
+\frac{1}{4}   A_{12} \,  \left(p \cdot q \, + 2 \, q \cdot q \,\right)
+\frac{A_3}{4 \, p \cdot q \,}
   \left(p \cdot p \,^2 + (p \cdot q \, +q \cdot q \,) \, p \cdot p \,+2   p \cdot q \, q \cdot q \,\right)    \nn\\
   &&
   +\frac{1}{2}
   A_9 \left[q \cdot q \,+ p \cdot p \, \left(\frac{2
  q \cdot q \,}{p \cdot q \,} +1\right) \right]
  -\frac{1}{2} \, [\Pi (p) +\Pi (q)]. 
  \label{mapping13}
\eea
We have shown how to obtain the $13$ $F_i$' s, starting from our derivation of the one-loop full amplitude $\Gamma^{\mu\nu\a\b} (p,q)$ leading to the ten invariant amplitudes of the set $\mathcal{X}$. Since we know the analytical expression of the $A_i$ involved, we can go one step further and give all the $F_i$' s in their analytical form in the most general kinematical configuration.
\section{Trace condition in the non-conformal case} 
Similarly to the chiral case, we can fix the correlator by requiring the validity of a trace condition on the 
amplitude, besides the two Ward identities on the 
conserved vector currents and the Bose symmetry in their indices. This approach is alternative to the imposition of the Ward identity (\ref{TWard}) but nevertheless equivalent to it. At a diagrammatic level we obtain 
\bea
g_{\mu \nu} \Gamma^{\mu\nu\alpha\beta}(p,q) = \Lambda^{\alpha \beta}(p,q) - \, \frac{e^2}{6 \pi^2}  \, u^{\alpha\beta}(p,q).
\label{betaf}
\eea

We comment below on this equation, in relation to the scales present in the perturbative expansion of the correlator, which are, besides the fermion mass $m$, the energy at which we probe the correlator ($s$) and the subtraction point after 
renormalization ($\mu$ or $M$). We have also defined 
\bea
\Lambda^{\alpha \beta}(p,q) &=& -m \, (i e)^2 \int d^4x \, d^4y \, e^{i p\cdot x+iq \cdot y} \langle \bar \psi \psi J^{\alpha}(x) J^{\beta}(y) \rangle \nn \\
&=& - m \, e^2 \,  \int \frac{d^4 l}{(2 \pi)^4} \, tr \left\{\frac{i}{\lsl+\psl - m} \g^{\alpha} \frac{i}{\lsl-m} \g^{\beta} \frac{i}{\lsl-\qsl-m} \right\} + \textrm{exch.}
\eea

 A direct computation gives
\bea
\Lambda^{\alpha \beta}(p,q) = 
G_1(s,s_1,s_2,m^2) \, u^{\alpha \beta}(p,q) 
+ G_2(s,s_1,s_2,m^2) \, w^{\alpha \beta}(p,q),
\eea
where
\bea
3 \, s \, F_1(s,s_1,s_2,m^2) &=& G_1(s,s_1,s_2,m^2) - \frac{e^2}{6 \pi^2}  \\
3 \, s \, F_2(s,s_1,s_2,m^2) &=& G_2(s,s_1,s_2,m^2)
\eea
and 
\bea
G_1(s,s_1,s_2,m^2) &=& 
Ê\frac{ e ^2 \gamma Ê m^2}{ \pi^2 \sigma } +\frac{e^2\, \mathcal D_2(s,s_2,m^2)\, Ês_2 m^2}{ \pi^2 \sigma ^2} 
\left[s^2+4 s_1 s-2 s_2 s-5 s_1^2+s_2^2+4 s_1 Ê s_2\right]  Ê \nn \\
 Ê %
&& Ê \hspace{-3cm} 
- Ê \frac{e^2 \, \mathcal D_1 (s,s_1,m^2)\, s_1 m^2}{ \pi^2 \sigma ^2}
 Ê \left[-\left(s-s_1\right){}^2+5 s_2^2-4 \left(s+s_1\right)
 Ê s_2\right]  Ê\nn \\
 Ê %
 Ê&& \hspace{-3cm} Ê
 - e^2 \, \mathcal C_0 (s,s_1,s_2,m^2)\, 
 \left[ 
\frac{ m^2 \gamma}{2 \pi^2  \sigma^2} Ê \left[ \left(s-s_1\right){}^3-s_2^3+\left(3 s+s_1\right)Ê s_2^2+\left(-3 s^2-10 s_1 s+s_1^2\right) s_2 \right] -\frac{2 m^4 \gamma }{ \pi^2 \sigma }\right], \nn \\ \\ 
G_2(s,s_1,s_2,m^2) &=&
 - \frac{2 e^2 m^2}{ \pi^2 \sigma } 
 - \frac{2 e^2 \, \mathcal D_2(s,s_2,m^2) m^2}{\pi^2 \sigma Ê ^2} \, Ê\left[\left(s-s_1\right){}^2-2
 Ê s_2^2+\left(s+s_1\right) s_2\right]  Ê Ê Ê\nn \\
 Ê %
 Ê && \hspace{-3cm}
 Ê -\, \frac{2 \, e^2 \mathcal D_1 (s,s_1,m^2) \, m^2}{\pi^2 \sigma ^2}
 Ê \left[s^2+\left(s_1-2 s_2\right) s-2 s_1^2+s_2^2+s_1 s_2\right] \nn \\
 Ê %
 Ê Ê&& \hspace{-3cm}
 - \, e^2 \mathcal C_0 (s,s_1,s_2,m^2) \,
 Ê Ê\biggl[\frac{4 m^4}{ \pi^2 \sigma}
 Ê+\frac{m^2}{\pi^2 \sigma ^2} \, Ê\left[ s^3-\left(s_1+s_2\right) s^2
 Ê- \left(s_1^2-6 Ês_2 s_1+s_2^2\right) \right. s \nn \\
 Ê%
 Ê&& \hspace{3 cm} +\left. \left(s_1-s_2\right){}^2
 Ê \left(s_1+s_2\right)\right] \biggr],
\eea
where $\gamma \equiv s -s_1 - s_2$, $\si \equiv s^2 - 2 (s_1+s_2)\, s + (s_1-s_2)^2$ and the scalar integrals $ \mathcal D_1 (s,s_1,m^2)$, $ \mathcal D_2 (s,s_1,m^2)$, $ \mathcal C_0 (s,s_1,s_2,m^2)$ for generic virtualities and masses are defined in Appendix \ref{scalars}. \\
We have checked that the final expressions of the form factors in the most general case, obtained either by imposing this condition on the 
energy momentum tensor or the Ward identity in the form given by Eq. (\ref{WIeq}) exactly coincide. In Appendix \ref{alternative}  we discuss this relation in the simpler case of a massless fermion in the loop.

\section{ The off-shell massive $\lag TJJ \rag $ correlator}
To obtain the explicit expression of the parametric integrals which describe the form factors, we follow an approach similar to that of \cite{Armillis:2009sm}, for the case of the chiral gauge anomaly. These 
have been obtained by re-computing the anomaly diagrams by dimensional reduction together with the tensor-to-scalar decomposition of the Feynman amplitudes. For instance, in  \cite{Armillis:2009sm} we have given the explicit expressions of the parametric integrals of Rosenberg using this trick. The correctness of the result can be 
checked numerically by comparing the parametric forms to the explicit computation. In this case the procedure is identical, though the computations are very involved.
By comparing the two approaches we extract, indirectly, an explicit expression of the parametric forms of these integrals, introduced in
\cite{Giannotti:2008cv}. We have checked that indeed there is perfect numerical agreement between our computation and the parametric result, as discussed in Appendix \ref{numerical}.

We introduce in this section the main results of our computation which will be used in the next sections for further analysis. The complete expressions of the form factors $F_i$ ($i=1, \dots,13$) in the massive and then in the massless case are contained in Appendix \ref{Fimassive} and \ref{Fimassless} respectively, whereas  the master integrals are collected in Appendix \ref{scalars}. In both cases the virtualities of the external lines are generic and denoted by $s_1,s_2$. After presenting the complete expressions, we discuss several kinematical limits of the result, in particular the on-shell limit for the two vector lines ($s_1 \rightarrow 0$, $s_2  \rightarrow 0$) in order to better understand the structure of the whole correlator. The appearance of generalized anomaly poles in the correlator and their IR decoupling under the most general conditions will be discussed thoroughly.

Notice that $F_{13}$ contains two vacuum polarization diagrams with the two photon momenta which are divergent and we are bond to define a suitable renormalization of the 2-point function which will affect the running of the coupling. In the next section we will address the explicit relation between renormalization schemes and running of the coupling in the context of the renormalization of the correlator.

 \subsection{Anomaly poles and their UV/IR significance}
 There are close similarities between the effective action in the case of the chiral gauge anomaly and the conformal case, due to the presence of massless poles. In \cite{Armillis:2009sm} we have analyzed the fact that in the chiral case the anomaly is entirely generated by the longitudinal component $w_L$, which is indeed isolated for {\em any} configuration of the photon momenta. This is somehow unexpected since the dispersive analysis shows that the pole in $w_L$ is coupled only under a specific kinematic condition, and is usually 
 interpreted as an infrared effect. Nevertheless there is a complete equivalence between the representation of the anomaly diagram in the Rosenberg representation - where the pole is not extracted as an independent component - and the L/T representation in which the pole is isolated under any kinematical configuration (and even in the massive case). This is apparent from the broken anomalous Ward identities satisfied by the AVV diagram where the mass corrections and the anomaly term can be separately identified \cite{Armillis:2009sm}.

To illustrate the emergence of a similar behaviour in the case of the conformal anomaly, it is sufficient to notice in the expression of 
$F_1$ given in Eq. (\ref{Fone}) the presence of the isolated contribution $( F_{1\,pole}\equiv - e^2/(18 \pi^2 s))$ which survives in the massless limit but is present also in the massive case. This component, indeed, is responsible for the trace anomaly also in the massive case, even though there appear extra corrections with mass-dependent terms. Obviously also in this case, which is generic from the kinematical point of view, one can clearly show that the pole does not couple in the infrared if we compute the residue of the 
entire amplitude. The anomaly pole, in fact, appears in the spectral function only in a special kinematic configuration when the fermion-antifermion pair of the anomaly diagram is collinear. However both in the case of the AVV diagram and in the conformal case, as evident from the expression of $F_1$, it reappears as 
an extra contribution and is responsible for the trace anomaly. It is rather easy to show 
the pole dominance of the anomaly away from the conformal point (massive case) at high energy, since the non anomalous terms present in $F_1$ and $F_2$ are subleading at large $s$. We are entitled to separate the pole contribution, which describes the non-local contribution to the trace anomaly, from the rest, and rewrite the $F_1$ form factor and effective action, respectively, as 
\beq
F_1=F_{1\, pole} + \tilde{F}_1
\eeq
and
\beq
\mathcal{S}=\mathcal{S}_{pole} +\tilde\mathcal{S}.
\eeq
The reminder $(\mathcal{S})$ includes all the remaining contributions coming from the several form 
factors of the expansion, while the pole part gives 

\beq
\mathcal{S}_{pole}= - \frac{e^2}{ 36 \pi^2}\int d^4 x d^4 y \left(\square h(x) - \partial_\mu\partial_\nu h^{\mu\nu}(x)\right)  \square^{-1}_{x\, y} F_{\alpha\beta}(x)F^{\alpha\beta}(y).
\eeq
As we have just mentioned, it is not difficult to show that the anomaly pole in $F_1$, in the general kinematical case (e.g. for off-shell photons and a massive fermion in the loop) decouples in the infrared (i.e. its residue vanishes) while it remains coupled in the massless on-shell limit. In other configurations (for any of the two photons off-shell) is also decoupled. This behaviour is in perfect analogy with the chiral case \cite{Armillis:2009sm}.  
 
\subsection{Massive and massless contributions to anomalous Ward identities and the trace anomaly} 
Anomalous effects are associated with massless fermions, and for this reason, when we analyze the contribution to the anomaly for a massive correlator, we need to justify the distinction between massless and massive contributions.  The latter  contribute to the anomalous Ward identity, in our approach, via terms of $O(m^2/s^2)$, where $s=k^2$ is the virtuality of the graviton vertex. At nonzero momentum transfer $(k\neq 0)$ the second term on the right-hand side of Eq. (\ref{betaf}) is interpreted as
an anomalous contribution, proportional to an asymptotic $\beta$ function $(\beta_{as})$ of the theory, coming from the residue of the anomaly pole which appears in the form factor $F_1$. While the appearance of the asymptotic $\beta$ function of the theory 
(which coincides with the $\beta$ function of the $\overline{MS}$ scheme) is expected at large $s$, where all the remaining scales 
of the theory $(s_1,s_2, m)$ can be dropped, corrections to the asymptotic description in the ultraviolet (UV) are expected. At the same time, in the far infrared (IR) region, below the fermion mass, the anomalous contribution should approach zero in a certain fashion, which will be specified below.

A complete quantitative understanding of this point for a general kinematics (e.g. for $s\neq 0$) remains, in a way, an open issue, 
 but much more can be said for the simpler case of zero momentum transfer, where a consistent pattern of separation between massless and massive contributions to the correlator emerge in the UV region. In this case the virtuality of the 
 two photons and the fermion mass $m$ (plus a renormalization scale $\mu$ or $M$) are the scales which appear in the renormalized perturbative expansion.  Related analysis have been presented in 
 \cite{Goncalves:2009sk} and  \cite{Giannotti:2008cv} and our conclusions do not differ from these previous investigations. We summarize the main points. 

Respect to the case of the chiral anomaly, the trace anomaly is connected with the regularization procedure involved in the computation of the diagrams. In our analysis we have used dimensional regularization (DR) and we have imposed conservation of the vector currents, the symmetry requirements on the correlator and the conservation of the energy momentum tensor. As we move from 4 to $d$ spacetime dimensions (before that we renormalize the theory), the anomaly pole term appears quite naturally in the expression of the correlator. This is not surprising, since QED in $d\neq 4$  dimensions is not even classically 
conformal invariant and the trace of the energy momentum tensor in the classical theory 
involves both a $F^2$ term ($\sim (d-4) F^2$) beside, for a massive correlator, a $\bar{\psi}\psi$ contribution. 
Let's summarize the basic features concerning the renormalization property of the correlator as they emerge from our direct computation. 

1) The anomalous Ward identity obtained by tracing the correlator ($\Gamma^{\mu\nu\alpha\beta}$) with $g_{\mu\nu}$ involves only the $F_1$ and $F_2$ 
form factors in the massive case; in the massless case the scale breaking appears uniquely due to $F_1$ via the term $e^3/(12\pi^2)u^{\alpha\beta}(p,q)$, as pointed out before. The finiteness of the two form factors involved in the trace of the correlator is indeed evident.
2) The residue of the pole term $(e^3/(12\pi^2))$ in $F_1$ is affected by the renormalization of the entire correlator  (the form factor $F_{13}$ is the only one requiring renormalization) only by the re-definition of the bare coupling $(e^2)$ in terms of the renormalized coupling $(e_R^2)$ through the renormalization factor $Z_3$. 
At this point, the interpretation of the residue at the pole as a contribution proportional to the $\beta$ function of the theory is, in a way, ambiguous \cite{Manohar:1996cq}, since the $\beta$ function is related to a given renormalization scheme. We stress once more that  Eq. (\ref{betaf}) does not involve a renormalization scheme - which at this point has not yet been defined - but just a regularization. 
We have regulated the infinities of the theory but we have not specified a subtraction of the infinities. For this reason, the substitution 
\beq
(e^3/(12\pi^2))\to 2 \beta_{as}(e)/e
\label{asym}
\eeq
which attributes the mass-independent term in $F_1$ to a specific $\beta$ function, the asymptotic one $(\beta_{as})$, as we are going to elaborate below, requires some clarification. 

To fully appreciate this point, it is convenient to go back to the unrenormalized Ward identity 
(\ref{TWard}) and differentiate it with respect to the momentum $q$ and then set $p=-q$ $(k=0)$ by going to zero momentum transfer. One obtains the derivative Ward identity 

\beq
g_{\mu\nu}\Gamma^{\mu\nu\alpha\beta}(p,-p)= 2 p^2 \frac{d\Pi}{d p^2}(p^2)(p^2 g^{\alpha\beta}- p^{\alpha}p^{\beta}).
\label{unren}
\eeq
The appearance of the derivative of the scalar self-energy of the photon on the right-had side of the previous equation is particularly illuminating, since it allows to relate this expression to a particular $\beta$ function of the theory, which is not the asymptotic 
$\beta_{as}$ considered in Eq.~(\ref{asym}). This $\beta$ function is useful for describing the IR running of the coupling. 

To illustrate this point we start from the expression of the scalar amplitude appearing in the photon self-energy in DR
\beq
\Pi(p^2,m)=\frac{e^2}{2 \pi^2}\left( \frac{1}{6 \epsilon} - \frac{\gamma}{6}- 
\int_0^1 dx x (1-x) \log\frac{m^2 - p^2 x(1-x)}{4 \pi \mu^2}\right)
\eeq
whose renormalization at zero momentum gives 
\beq
\Pi_R(p^2,m)=\Pi(p^2,m) - \Pi(0,m)=-\frac{e^2}{2 \pi^2}\int_0^1 x(1-x) 
\log\frac{ m^2 - p^2 x (1-x)}{m^2}.
\eeq

Using this expression, we can easily compute 
\beq
2 p^2\frac{d\Pi}{d p^2}=2 p^2\frac{d\Pi_R}{d p^2}= - \frac{e^2}{6\pi^2} + \frac{e^2\ m^2}{\pi^2}\int_0^1\,dx\ \frac{x(1-x)}
{m^2 - p^2 x(1-x) }.
\label{separated}
\eeq
Notice that this result does not depend on the renormalization scheme due to the presence of the derivative respect to $p^2$. 
Notice also that the $\beta$ function of the theory evaluated in the zero momentum subtraction scheme is exactly given by the right-hand side of the previous expression 
\beq
2 p^2\frac{d\Pi_R}{d p^2}= - \frac{\beta (e^2,p^2)}{e^2},
\label{defin}
\eeq

(where $\beta(e^2,p^2)= 2 e \beta(e, p^2)$), but this result does not hold, generically, in any scheme. The identification of anomalous (massless) 
effects in the theory, as exemplified by these simpler Ward identity, should then be obtained by extracting the appropriate 
$\beta$ function of the theory, whose running should be driven by the effective massless degrees of freedoms (fermions, in our case) at the relevant observation scale  ($p^2$).  

Clearly, in the case of Eq. (\ref{defin}) all the mass contributions have been absorbed into the very definition of the $\beta$ function.
Notice that if $ p^2 \ll m^2$ this $\beta$ function, after a rearrangement gives
\beq
- \frac{\beta (e^2,p^2)}{e^2}=\frac{e^2}{\pi^2}\int_0^1 dx \frac{p^2 x^2 (1-x)^2}{m^2 - p^2 x (1-x)}
\eeq
and therefore it vanishes as $\beta\sim O(p^2/m^2)$ for $p^2\to 0$. Equivalently, by taking the $m\to \infty$ limit we recover the expected decoupling of the fermion (due to a vanishing $\beta$)
since we are probing the correlator at a scale ($p^2$) which is not sufficient to resolve the contribution of the fermion loop. On the contrary, as $p^2\to \infty$, with $m$ fixed, the running of the $\beta$ function is the usual asymptotic one 
$\beta_{as}(e^2)\sim e^4/(6 \pi^2)$ modified by corrections $O(m^2/p^2)$. The UV limit is characterized by the same running typical of the massless case, as expected. 

Notice that the right-hand side of Eq. (\ref{unren}), as we have already remarked, does not depend on the renormalization scheme, while the $\beta$ function does and Eq. (\ref{defin}) should be understood as a definition. For this reason, $\beta(e^2,p^2)$ correctly describes the IR running of the coupling as $p^2 \ll m^2$, and in this case it is obvious that massless anomalous effects of scale breaking are not present in this specific limit.

In the case of regularization scheme different from zero momentum subtraction, there are some differences which should be taken into consideration. For instance, 
in a mass-dependent scheme one subtracts the value of the graph at a Euclidean momentum point $p^2=-M^2$, redefining the scalar self-energy as
\beq
\Pi^R(p^2,m,M)= \Pi(p^2,m) - \Pi(p^2=-M^2, m) = \frac{e^3}{2 \pi^2}\left[\int_0^1 dx\ x(1-x)\ \log{m^2-p^2 x(1-x)\over m^2+M^2 x(1-x)}\right]
\eeq
which gives, respect to the previous ($M=0$) scheme, a $\beta$ function now of the form 
\beqa
\beta(e) &=& - \frac{e}{2} M \frac{d}{d M} \frac{e^2}{2 \pi^2} \int_0^1 dx x (1-x) \log 
\frac{m^2 - p^2 x (1-x)}{m^2 + M^2 x (1-x)}\nonumber \\
&=&\frac{e^3}{2 \pi^2}\int_0^1 dx x (1-x) \frac{M^2 x (1-x)}{m^2 + M^2 x(1-x)}.
\eeqa
For large values of $M$, this $\beta$ function describes the usual UV running since
\beq
\beta(e)\sim \frac{e^3}{2 \pi^2}\int_0^1 dx x(1-x)=\beta(e)_{as}=\frac{e^3}{12 \pi^2}.
\eeq
In this second scheme, the (regularization independent) right-hand side of Eq. (\ref{unren}) can be interpreted as due to an anomalous contribution coming from the pole plus some explicit mass corrections, as obvious from  the first and second term of (\ref{separated}). We conclude with some considerations on a third (mass-independent) scheme. 

In the $\overline{MS}$ scheme, the renormalization of the photon self-energy is performed via the subtraction 
\beq
\Pi_R(p^2, m, \mu)= \Pi(p^2, m,\mu)- \frac{e^2}{12 \pi^2}\left( \frac{1}{\epsilon} + \gamma - \log 4 \pi\right)
\eeq
which gives directly an asymptotic  $\beta$ function since
\beqa
\beta(e) &=& \frac{e}{2}\mu \frac{d}{d\mu}\Pi_R(p^2, m,\mu)\nonumber \\
&=& \frac{e^3}{2 \pi^2}\int_0^1 dx x (1-x)=\frac{e^3}{12 \pi^2}.
\eeqa
It is clear, from these considerations, that a judicious definition of the $\beta$ function allows a correct interpretation of 
the right-hand side of (\ref{unren}) and (\ref{separated}). In the $\overline{MS}$ scheme, the breaking of scale 
invariance can be attributed to a UV running of the coupling (for $p^2\gg m^2$) plus mass corrections which are suppressed as 
$O(m^2/p^2)$. Notice that in this case the renormalization scale $(\mu^2)$ should be $O(p^2)$, since we should not allow large logarithms to be present in the perturbative expansion. In this sense, the extrapolation of the $\overline{MS}$ result to 
$p^2\sim\mu^2 \ll m^2$ should be forbidden by the same criterion, since large logs of the relevant scales ($\log(m/\mu)$) would  
otherwise be generated. In the far infrared region $p^2\ll m^2$ the use of the same $\beta$ function is indeed not appropriate, 
since the same scheme does not correctly describe the decoupling of the anomaly, which instead should occur, since there is no massless fermion in the theory.

To conclude this discussion we just mention that the $\overline{MS}$ scheme can be used, obviously, both to describe the far IR and the far UV regions of the theory, with the condition that we are bound to choose a vanishing $\beta$ function at $p^2\ll m^2$ and an asymptotic one 
for $p^2\gg m^2$ and assuring continuity of the gauge coupling across the fermion mass scale though the $\beta$-function is discontinuous. This is the standard procedure followed in the $\overline{MS}$ scheme as, for instance, in QCD factorization, 
improved with the inclusion of threshold effects at the crossing scales (see for instance \cite{Cafarella:2005zj}, \cite{Cafarella:2008du}) where the number of massless flavours change.

\subsection{The off-shell massless $\lag TJJ \rag $ correlator}
Clearly, as we perform the massless limit on the amplitude, the residue of the same anomaly pole - identified above 
in the contribution $F_{1\, pole}$ -  is still present, but will now be decoupled in the infrared. 
 
In the massless case the scalar functions $F_i$ depend only on the kinematic invariants $s,s_1,s_2$ but we still retain the last entry of these functions and set it equal to $0$  for clarity, using the notation $F_i \equiv F_i (s;s_1,s_2,0)$. These new functions are computed starting from the massive ones and letting $m\rightarrow 0$ and $\mathcal A_0 (m^2) \rightarrow 0$, i.e. eliminating all the massless tadpoles generated in the zero fermion mass limit.

 The off-shell massless invariant amplitudes $F_i (s; s_1, s_2,0)$ are here given in terms of a new set of master integrals listed in Appendix \ref{scalars}. We give here only the simplest invariant amplitudes, leaving the remaining ones to the appendix 	\ref{Fimassless}. 
The anomaly pole is clearly present in $F_1$, which is given by
\bea
%
 %
 \underline{ \bf \underline{F_{1} (s;\,s_1,\,s_2,\,0)}}&=&
- \frac{e^2 }{18 \pi^2 s}, 
 \eea
while
  %
  \bea
 \underline{ \bf \underline{F_{2} (s;\,s_1,\,s_2,\,0)}} &=& 0. 
  \eea
 The complete $\langle TJJ \rangle$ correlator is very complicated in this case as the long expressions of the form factors show, but a deeper analysis of its poles by computing the residue in $s=0$ can be useful to draw some conclusions.  The single pieces of $\Gamma^{\mu\nu\a\b}(s;\,s_1,\,s_2,\,0)$ indeed contribute as
\bea
\lim_{s\rightarrow0} \, s F_1 (s;\,s_1,\,s_2,\,0) \, t^{\mu\nu\a\b}_1 &=&
 - \frac{e^2}{18 \, \pi^2 } \, t^{\mu\nu\a\b}_1\big\vert_{s=0},\\
%
%
\lim_{s\rightarrow0} \, s F_3 (s;\,s_1,\,s_2,\,0) \, t^{\mu\nu\a\b}_3 &=&
\,\frac{e^2}{72 \, \pi^2 } \, t^{\mu\nu\a\b}_3 \big\vert_{s=0},\\
%
%
\lim_{s\rightarrow0} \, s F_5 (s;\,s_1,\,s_2,\,0) \, t^{\mu\nu\a\b}_5 &=&
\,\frac{e^2 }{72\, \pi^2} \, t^{\mu\nu\a\b}_5 \big\vert_{s=0}, \\
%
%
\lim_{s\rightarrow0} \, s F_7 (s;\,s_1,\,s_2,\,0) \, t^{\mu\nu\a\b}_7 &=&
\frac{e^2}{36\, \pi^2} \, t^{\mu\nu\a\b}_7 \big\vert_{s=0},
\eea
while $F_2$ is absent in the massless case. The residues of the $F_i(s;\,s_1,\,s_2,\,0) $ not included in the equation above are all vanishing. Combining the results given above one can easily check that the entire correlator is completely free from anomaly poles as
\bea
\lim_{s\rightarrow0} \, s \, \Gamma^{\mu\nu\a\b}(s;\,s_1,\,s_2,\,0) =0
\eea
in this rather general configuration. A similar result holds for the correlator responsible for the chiral anomaly and shows the decoupling of polar contributions in the infrared. 

\subsection{ The on-shell massive $\lag TJJ \rag $ correlator}
A particular  case of the $\lag TJJ \rag $ correlator is represented by its on-shell version with a
massive fermion in the loop.  If we contract $u^{	\a \b} (p,q)$ and $w^{	\a \b} (p,q)$ with the polarization tensors $\eps_\a (p)$ and $\eps_\b (q)$ requiring $\eps_\a (p) \, p^\a=0$, \mbox{${\eps_\b (p) \, p^\b=0}$ }, the first  tensor remains unchanged while  $w^{\a \b} (p,q)$ becomes $\widetilde w^{	\a \b} (p,q)= s_1 \, s_2 \, g^{\a \b}$. This will be carefully taken into account when computing the $s_1 \rightarrow 0$, $s_2 \rightarrow 0$ limit of the product of the invariant amplitudes $F_i$ with their corresponding tensors $t_i^{\,\mu\nu\a\b}$ ($i=1,\dots,13$).\\
The invariant amplitudes reported below describe $F_i(s;0,0,m^2)$ whose tensors $t_i^{\,\mu\nu\a\b}$  are also finite and non-vanishing. They are
\bea
\underline{ \bf \underline{F_1 (s;\,0,\,0,\,m^2)}} &=&
- \frac{e^2 }{18 \pi^2  s} \,  + \, \frac{e^2  m^2}{3 \pi^2 s^2} \, - \, \frac{e^2\, m^2} {3 \pi^2 s}\mathcal C_0 (s, 0, 0, m^2) 
\bigg[\frac{1}{2 \,  }-\frac{2 m^2}{ s}\bigg],  \nn \\
\underline{ \bf \underline{F_3 (s;\,0,\,0,\,m^2)}}  &=&
- \frac{e ^2}{144 \pi^2 s} - \frac{ e ^2 m^2}{12 \pi^2  s^2}
- \, \frac{e^2 \, m^2}{4 \pi^2  s^2} \mathcal D (s, 0, 0, m^2) \, 
\nn \\
&&
- \, \frac{e^2 \, m^2}{6 \pi^2 s } \mathcal C_0(s, 0, 0, m^2 )\, \left[ \frac{1}{2} + \frac{m^2}{s}\right], \nn \\
\underline{ \bf \underline{F_5 (s;\,0,\,0,\,m^2)}}  &=& \underline{ \bf \underline{F_3 (s;\,0,\,0,\,m^2)}} , \nn \\
\underline{ \bf \underline{F_7 (s;\,0,\,0,\,m^2)}}  &=&  - 4 \, \underline{ \bf \underline{F_3 (s;\,0,\,0,\,m^2)}} \nn \\
\underline{ \bf \underline{F_{13\, R} (s;\,0,\,0,\,m^2)}}  &=&
 \frac{11  e ^2}{144  \pi^2 }  +   \frac{ e ^2 m^2}{4 \pi^2 s}
 + \, e^2\mathcal C_0 (s, 0,0,m^2) \,\left[ \frac{m^4}{2 \pi^2 s}+\frac{ m^2}{4 \pi^2 }\right] \nn \\
 && + \,e^2  \mathcal D (s, 0, 0, m^2) \, \left[ \frac{5 m^2}{12 \pi^2  s} + \frac{1}{12}\right],
\label{masslesslimit}
\eea
where the on-shell scalar integrals $\mathcal D (s, 0, 0, m^2)$ and $\mathcal C_0 (s, 0,0,m^2)$ are computed in Appendix \ref{scalars}; here
$F_{13\, R}$ denotes the renormalized amplitude, obtained by first removing the UV pole present in the photon self-energy by the usual renormalization of the photon wavefunction and then taking the on-shell limit. The remaining  invariant amplitudes $F_i (s, 0,0,m^2)$ are zero or multiply vanishing  tensors in this kinematical configuration so they do not contribute to the correlator. \\
The limit from the massive on-shell form factors to the massless ones is clearer by looking at the series expansion of the scalar integrals around $m=0$
\bea
C_0 (s,0,0,m^2) = \frac{1}{2 \, s} \, \left[ \log\left(-\frac{s}{m^2}\right)\right]^2 - \frac{ 2  \, m^2}{s^2} \log \left(- \frac{s}{m^2} \right) + O (m^3)
\eea
and from this we obtain for $F'_1$
\bea
F_1'(s,0,0,m^2) =   
\frac{e^2 \, m^2}{3 \pi^2 \, s^2} \left\{
 1 -  \frac{1}{4} \left[ \log\left(-\frac{s}{m^2}\right)\right]^2 \right\},
\eea
where the notation  $F'_1$ denotes the first form factor after the subtraction of the pole in $1/s$.

Using the results given above, the full massive on-shell amplitude is given by
\bea
\Gamma^{\mu \nu \a \b} (s;0,0,m^2)\,  = \,
&& \, F_1 \,(s; 0,0,m^2) \, \widetilde t_1^{\, \mu \nu \a \b}  \, +\,
F_3 \, (s; 0,0,m^2) \, (\widetilde t_3^{\, \mu \nu \a \b} \, + \,  \widetilde t_5^{\, \mu \nu \a \b}
- \, 4 \, \widetilde t_7^{\, \mu \nu \a \b}) \nn \\
 && + \, F_{\, 13,\, R\,} (s; 0,0,m^2) \, \widetilde t_{13}^{\, \mu \nu \a \b}  ,
\eea
so that the invariant amplitudes reduce from $13$ to $3$ and the three linear combinations of the tensors can be taken as a new basis
\bea
 \widetilde t_1^{\, \mu \nu \a \b} &=& \lim_{s_1,s_2 \rightarrow 0} \,  t_1^{\, \mu \nu \a \b} =
 (s \, g^{\mu\nu} - k^{\mu}k^{\nu}) \, u^{\a \b} (p,q)  
 \label{widetilde1}\\
 \widetilde t_3^{\, \mu \nu \a \b} \, + \,  \widetilde t_5^{\, \mu \nu \a \b}
- \, 4 \, \widetilde t_7^{\, \mu \nu \a \b} &=&  \lim_{s_1,s_2 \rightarrow 0}  \, (t_3^{\, \mu \nu \a \b} \, + \,  t_5^{\, \mu \nu \a \b}
- \, 4 \,  t_7^{\, \mu \nu \a \b}) = \nn  \\
&& \hspace{-0.5cm} - 2 \, u^{\a \b} (p,q) \left( s \, g^{\mu \nu} + 2 (p^\mu \, p^\nu + q^\mu \, q^\nu )
- 4 \, (p^\mu \, q^\nu + q^\mu \, p^\nu) \right)  \\
\widetilde{t}^{\, \mu \nu \alpha \beta}_{13} &=&  \lim_{s_1,s_2 \rightarrow 0} \,  t_{13}^{\, \mu \nu \a \b} =
\big(p^{\mu} q^{\nu} + p^{\nu} q^{\mu}\big)g^{\alpha\beta}
+ \frac{s}{2} \big(g^{\alpha\nu} g^{\beta\mu} + g^{\alpha\mu} g^{\beta\nu}\big) \nn \\
&&  \hspace{-0.5cm} - g^{\mu\nu} (\frac{s}{2} g^{\alpha \beta}- q^{\alpha} p^{\beta})
-\big(g^{\beta\nu} p^{\mu}
+ g^{\beta\mu} p^{\nu}\big)q^{\alpha}
 - \big (g^{\alpha\nu} q^{\mu}
+ g^{\alpha\mu} q^{\nu }\big)p^{\beta}, 
\label{widetilde13}
\nn \\
\eea
 as  previously done in the literature \cite{Berends:1975ah}.
If we extract the residue of the full amplitude we realize that even though some functions $F_i (s, 0,0,m^2) $ have kinematical singularities in $1/s$ this polar structure is no longer present in the complete massive correlator
\bea
\lim_{s \rightarrow 0} s \, \Gamma^{\mu \nu \alpha \beta} = 0
\eea
showing that in the massive case the $\lag TJJ \rag $ correlator exhibits no poles. In a following section we will comment on the interpretation of these massless poles exploiting the analogy with a similar situation
encountered in the case of the gauge anomaly.

 \section{The general effective action and its various limits }
 In this section we present results for the correlator in various kinematical limits. We start from its expression
 in the on-shell massive case and then perform its expansion in $1/m$ which will be used in a next section
 to extract the corresponding effective action. As a final step we show the on-shell structure of the invariant amplitudes in the conformal limit.
 
 It is possible to identify from them the structure of the effective action in its most general form. 
If we denote by $\mathcal{S}_i$ the contribution to the effective action due to each form factor $F_i$, then we can write it in the form 
\beq
\mathcal{S}_i= \int d^4 x \,d^4 y \,d^4 z\, \hat{t}_i^{\mu\nu\alpha\beta}(z,x,y) 
h_{\mu\nu}(z)A_\alpha(x) A_\beta(y)\int \frac{d^4 p\, d^4 q }{(2 \pi)^8}e^{-i p\cdot(x-z) -i q\cdot(y-z)} 
F_i(k,p,q)
\eeq
where $k\equiv p +q$. We have introduced the operatorial version of the tensor structures $t_i^{\mu\nu\alpha\beta}$, denoted by $\hat{t}_i$ that will be characterized below. Defining 
\beq
\hat{p}_x^\alpha\equiv i \frac{\partial}{\partial x_\alpha}, \qquad \hat{q}_y^\alpha\equiv i \frac{\partial}{\partial y_\alpha}, \qquad \hat{k}_z^\alpha\equiv - i \frac{\partial}{\partial z_\alpha}
\eeq
and using the identity
\beq
\hat{F}_i ( \hat{k}_z, \hat{p}_x, \hat{q}_y) \delta^4(x-z)\delta^4(y-z)
=\int \frac{d^4 p\, d^4 q }{(2 \pi)^8}e^{-i p\cdot(x-z) -i q\cdot(y-z)} 
F_i(k,p,q)
\eeq
where formally $\hat{F}_i$ is the operatorial version of $F_i$, 
we can arrange the anomalous effective action also in the form 
\beq
\mathcal{S}_i=\int d^4 x d^4 y d^4 z  \hat{F}_i ( \hat{k}_z, \hat{p}_x, \hat{q}_y)\left[\delta^4(x-z)\delta^4(y-z)\right]\hat{t}_i^{\mu\nu\alpha\beta}(z,x,y) h_{\mu\nu}A_\alpha(x)A_\beta(y).
\eeq
For instance we get
\bea
\hat{t}_1^{\mu\nu\alpha\beta}(z,x,y) h_{\mu\nu}A_\alpha(x)A_\beta(y)&=&
\frac{1}{2}\left(\square_z h(z) - 
\partial^z_\mu \partial^z_\nu h^{\mu\nu}(z)\right) F_{\alpha\beta}(x)F^{\alpha\beta}(y),\\
\hat{t}_2^{\mu\nu\alpha\beta}(z,x,y) h_{\mu\nu}A_\alpha(x)A_\beta(y) &=&
\left(\square_z h(z) - 
\partial^z_\mu \partial^z_\nu h^{\mu\nu}(z)\right) \partial_\mu F^{\mu}_\lambda(x)\partial_\nu F^{\nu\lambda}(y), \\
\hat{t}_3^{\mu\nu\alpha\beta}(z,x,y) h_{\mu\nu}A_\alpha(x)A_\beta(y) &=&
\frac{1}{2} h^{\mu\nu}(z)\left(\square_x g_{\mu\nu} - 
4 \partial^x_\mu \partial^x_\nu \right) F_{\alpha\beta}(x)F^{\alpha\beta}(y), \\
\hat{t}_4^{\mu\nu\alpha\beta}(z,x,y) h_{\mu\nu}A_\alpha(x)A_\beta(y) &=&
h^{\mu\nu}(z)\left(\square_x g_{\mu\nu}- 
4 \partial^x_\mu \partial^x_\nu \right) \partial_\mu F^{\mu}_\lambda(x)\partial_\nu F^{\nu\lambda}(y), \\
\hat{t}_5^{\mu\nu\alpha\beta}(z,x,y) h_{\mu\nu}A_\alpha(x)A_\beta(y) &=&
\frac{1}{2} h^{\mu\nu}(z)\left(\square_y 
g_{\mu\nu} - 
4 \partial^y_\mu \partial^y_\nu \right) F_{\alpha\beta}(x)F^{\alpha\beta}(y), \\
\hat{t}_6^{\mu\nu\alpha\beta}(z,x,y) h_{\mu\nu}A_\alpha(x)A_\beta(y) &=& 
h^{\mu\nu}(z)\left(\square_y g_{\mu\nu}- 
4 \partial^y_\mu \partial^y_\nu \right) \partial_\mu F^{\mu}_\lambda(x)\partial_\nu F^{\nu\lambda}(y), \\
\hat{t}_7^{\mu\nu\alpha\beta}(z,x,y) h_{\mu\nu}A_\alpha(x)A_\beta(y) &=&
 \frac{1}{2} h^{\mu\nu}(z)\left(
\partial^{x\,\lambda} \partial^y_\lambda g_{\mu\nu} - 
2 (\partial^y_\mu \partial^x_\nu  +  \partial^y_\nu \partial^x_\mu)\right)  F_{\alpha\beta}(x)F^{\alpha\beta}(y), \\
\hat{t}_8^{\mu\nu\alpha\beta}(z,x,y) h_{\mu\nu}A_\alpha(x)A_\beta(y) &=&
h^{\mu\nu}(z)\left(  \partial^{x\,\lambda} \partial^y_\lambda g_{\mu\nu} - 
2 (\partial^y_\mu \partial^x_\nu  +  \partial^y_\nu \partial^x_\mu)\right) \partial_\mu F^{\mu}_\lambda(x)\partial_\nu F^{\nu\lambda}(y)
\eea
and similar expressions for the remaining tensor structures. However, the most useful forms of the effective action involve an expansion in the fermions mass, as in the $1/m$ formulation (the Euler-Heisenberg form) or for small $m$. In this second case the non-local contributions obtained from the anomaly poles appear separated from the massive terms, 
showing the full-fledged implications of the anomaly. This second formulation allows a smooth massless limit, where the breaking of the conformal anomaly is entirely due to the massless fermion loops.

In the ${1}/{m}$ case, for on-shell gauge bosons, the result turns out to be particularly simple. We obtain  
\bea
F_{1} (s, 0, 0, m^2) &=& 
\frac{7 e ^2 }{2160 \pi^2 } \frac{1}{m^2} + \frac{e ^2 s}{3024 \, \pi^2 }  \frac{1}{m^4} 
+O\left(\frac{1}{m^6}\right), \\
F_{3} (s, 0, 0, m^2) &=& F_{5} (s, 0, 0, m^2) = 
\frac{e^2 }{4320 \, \pi^2 }  \frac{1}{m^2} 
+ \frac{e^2 s}{60480 \, \pi^2} \frac{1}{m^4} +O\left(\frac{1}{m^6}\right), \\
F_{7} (s, 0, 0, m^2) &=&  - 4 \, F_{3} (s, 0, 0, m^2) \\
F_{13, R} (s, 0, 0, m^2) &=& 
\frac{11 e^2 s}{1440 \, \pi^2 } \frac{1}{m^2} + \frac{11 e^2 s^2}{20160 \, \pi^2 } \frac{1}{m^4} 
+ O\left(\frac{1}{m^6}\right),
\eea
which can be rearranged in terms of three independent tensor structures. Going to configuration space, the linearized expression of the contribution to the gravitational effective action due to the $TJJ$ vertex, in this case, can be easily obtained in the form
\bea
S_{TJJ} &=& \int d^4 x d^4 y d^4 z \, \Gamma^{\mu \nu \alpha\beta}(x,y,z) A_{\alpha}(x) \, A_{\beta}(y) \, h_{\mu \nu}(z) \nn \\
&=& \frac{ 7 \, e^2}{4320 \, \pi^2 \, m^2} \int d^4 x \left( \Box h - \partial^{\mu} \partial^{\nu} h_{\mu \nu}\right) F^2 \nn \\
&-& \frac{e^2}{4320 \, \pi^2 \, m^2} \int d^4 x \left( \Box h F^2 - 8 \partial^{\mu} F^{\alpha \beta} \partial^{\nu} F_{\alpha \beta} h_{\mu \nu} + 4 (\partial^{\mu} \partial^{\nu} F_{\alpha \beta})F^{\alpha \beta} h_{\mu \nu} \right) \nn \\
&+& \frac{11 \, e^2}{1440 \, \pi^2 \, m^2} \int d^4 x T^{\mu \nu}_{ph} \Box h_{\mu \nu}.
\eea
which shows three independent contributions linear in the (weak) gravitational field.

\section{ The massless (on-shell) $\lag TJJ \rag $ correlator}
The non-local structure of the effective action, as we have pointed out in the previous sections, is not apparent 
within an expansion in $1/m$, nor this expansion has a smooth match with the massless case. 

The computation of the correlator $\Gamma^{\mu\nu\alpha\beta}  (s; 0,0,0) $ hides some subtleties in the  massless fermion limit  (with on-shell external photons), as the form factors $F_i$ and the tensorial structures $t_i$ both contain the kinematical invariants $s_1, s_2$.  For this reason the limit of both factors 
(form factor and corresponding tensor structure)  $F_i \, t_i^{\mu\nu\a\b}$ has to be taken carefully, starting from the expression of the massless $F_i (s; s_1,s_2,0) $ listed in Appendix \ref{Fimassless} and from the tensors $t_i^{\mu\nu\a\b}$ contracted with the physical polarization tensors. In this case only few form factors survive and in particular  
\bea
F_{1} (s, 0, 0, 0) &=& - \frac{e^2}{18 \pi^2  s}, \\
F_{3} (s, 0, 0, 0) &=&  F_{5} (s, 0, 0, 0) = - \frac{e^2}{144 \pi^2 \, s}, \\
F_{7} (s, 0, 0, 0) &=& -4 \, F_{3} (s, 0, 0, 0), \\
F_{13, R} (s, 0, 0, 0) &=& - \frac{e^2}{144 \pi^2} \, \left[ 12 \log \left(-\frac{s}{\mu^2}\right) - 35\right],
\eea
and hence the whole correlator with two onshell photons on the external lines is 
\bea
\Gamma^{\mu\nu\alpha\beta}  (s; 0,0,0) &=& 
F_{1} (s, 0, 0, 0) \, \widetilde{t}^{\, \mu \nu \alpha \beta}_{1} 
+ F_{3} (s, 0, 0, 0) \, \left( \widetilde{t}^{\, \mu \nu \alpha \beta}_{3}  + \widetilde{t}^{\, \mu \nu \alpha \beta}_{5}
- 4 	\, \widetilde{t}^{\, \mu \nu \alpha \beta}_{7}  \right) + 
F_{13, R} \, \widetilde{t}^{\, \mu \nu \alpha \beta}_{13}  \nn \\
&=&
- \frac{e^2}{48 \pi^2 \, s} \left[ 
\left(2 \, p^\b \, q^\a- s \, g^{\a\b}\right) \left( 2 \, p^\mu \, p^\nu + 2 \, q^\mu \, q^\nu - s \, g^{\mu\nu} \right)\right]
+ \, F_{13, R} \, \widetilde{t}^{\, \mu \nu \alpha \beta}_{13},
\label{gamma00}
\eea
where $\widetilde{t}^{\, \mu \nu \alpha \beta}_{i}$ are the tensors defined in eqs.~(\ref{widetilde1}-\ref{widetilde13}).

The study of the singularities in $1/s$ for this correlator requires a different analysis for $F_1$ and the remaining form factors, as explicitly shown in eq.~\ref{gamma00}, where $F_1$ has been kept aside from the others, even if it is proportional to $F_3$. Indeed $F_1$ is the only form factor multiplying a non zero trace tensor, $\widetilde{t}^{\, \mu \nu \alpha \beta}_{1}$, and responsible for the trace anomaly. If we take the residue of the onshell correlator for physical polarizations of the photons in the final state we see how the $4$ form factors and their tensors combine in such a way that the result is different from zero as 
\bea
\lim_{s\rightarrow 0 } \, s \, \Gamma^{\mu\nu\alpha\beta}  (s; 0,0,0)  = 
- \frac{e^2}{12 \pi^2 }  \, p^\b \, q^\a (p^\mu \, p^\nu + q^\mu \, q^\nu),
\eea
 where clearly each singular part in $1/s$ present in $F_1,F_3, F_5,F_7$ added up and the logarithmic behaviour in $s$ of $F_{13}$ has been regulated by the factor $s$ in front when taking the  limit. The result shows that the pole, in this case, is coupled in the IR, as shown by the dispersive analysis.

 \section{Conclusions}
We have presented a computation of the $TJJ$ correlator, responsible for the appearance of gauge contributions to the conformal anomaly in the effective action of gravity. We have used our results to present the general form of the gauge contributions to this action, in the limit of a weak gravitational field.   One interesting feature of this correlator is the presence of an anomaly pole \cite{Giannotti:2008cv}.

 Usually anomaly poles are interpreted as affecting the infrared region of the correlator and appear only in one special kinematical configuration, which requires massless fermions in the loop and on-shell conditions for the external gauge lines. In general, however, the anomaly pole affects the UV region even if it is not coupled in the infrared.  This surprising feature of the anomaly is present both in the case of the chiral anomaly \cite{Armillis:2009sm} {\em and} in the conformal anomaly. Here we have extracted explicitly this behaviour by a general analysis of the correlator, extending our previous study of the chiral gauge anomaly. 

Indeed anomaly poles are the most interesting feature, at perturbative level, of the anomaly, being it conformal or chiral, and are described by mixed diagrams involving either a scalar (gravitational case) \cite{Giannotti:2008cv} or a pseudoscalar (chiral case) \cite{Armillis:2009sm,Coriano:2008pg}. The connection between the infrared and the ultraviolet, signalled by the presence of these contributions, should not be too surprising in an anomalous context. The pole-like behaviour of an anomalous correlator is usually ``captured" by a variational solution of a given anomaly equation, which implicitly assumes the presence of a pole term in the integrated 
functional \cite{Armillis:2009im}. By rediscovering the pole in perturbation theory, obviously, one can clearly conclude that variational solutions of the anomaly equations are indeed correct, although they miss homogeneous solutions  to the Ward identity, that indeed must necessarily be identified by an off-shell perturbative analysis of the correlators. This is the approach followed here and in \cite{Armillis:2009sm}.

We have also seen that  the identification of the massless anomaly pole allows to provide a  ``mixed" formulation of the effective action in which the pole is isolated from the remaining mass terms, 
extracted  in the $\tilde\mathcal{S}_{pole}$ part of the anomalous action, which could be used for further studies. 
We have also emphasized that a typical $1/m$ expansion of the anomalous effective action fails to convey fully the presence of scaleless contributions. 

There are various applications of our analysis which can be of interest for further studies. The first concerns the possible implication of these types of effective actions in cosmology, especially in inflationary scenarios where the coupling of gravity to matter via gauge interactions and the conformal anomaly plays an interesting phenomenological role. As we have stressed in the introduction, 
the local description of an anomalous effective action involves additional degrees of freedom which can be identified in the case of the gauge anomaly \cite{Coriano:2008pg} as well as in the conformal case \cite{Giannotti:2008cv}. In \cite{Giannotti:2008cv} the authors describe the role of the corresponding scalar degrees of freedom 
in the effective action emphasizing their meaning as possible composite. In the case of an anomalous gauge theory the derivative coupling of the anomaly pole to the anomalous gauge current indicates that its nature is that of a quasi Nambu-Goldstone mode. If the parallel with chiral theories holds, one should be able to characterize the physical property of this state, including its BE condensation under the action of gravity. Similar features could be shared by the scalar state(s) described by the conformal anomaly.  Of particular interest are the extensions of these analysis to the case of supersymmetric theories, in particular to $N=1$ superconformal theories, where the R-current, the supersymmetry current and the energy momentum tensor belong to the same supermultiplet, as are their corresponding anomalies. Clearly our computation is the first step in this direction, and can be extended with the inclusion of other types of fields in the perturbative expansion, reaching, as a starting point, all the relevant fields of the Standard Model.  In general, one could also use our approach to come with a complete 
description of the interplay between supersymmetry and the conformal anomaly, acting as a mediator of the gravitational interaction, which is of phenomenological interest. Finally, we mention that our analysis could be useful in order to test, in a specific realization perturbative results of conformal field theories in four spacetime dimensions. These studies of the $TJJ$ correlator have been performed on rather general grounds, using conformal invariance as a unique assumption in order to infer the structure of the operator product expansion \cite{Osborn:1993cr}.  We hope to return to these points in the near future. 

 \centerline{\bf Acknowledgements}
We thank  Shun-Pei Miao, A. Petkou, T. Tomaras and R. Woodard for discussions. C.C. thanks the Theory Group at Crete for hospitality and partial financial support. This work is supported in part  by the European Union through the Marie Curie Research and Training Network ``Universenet'' (MRTN-CT-2006-035863).

\begin{appendix}

\section{Appendix. Definitions and conventions for the scalar integrals}
\label{scalars}
We collect in this appendix all the scalar integrals involved in this computation. To set all our conventions, we start with the definition of the one-point function, or massive tadpole $\mathcal  A_0 (m^2)$, the massive bubble $\mathcal B_0 (s, m^2) $  and  the massive three-point function $\mathcal C_0 (s, s_1, s_2, m^2)$
\bea
\mathcal A_0 (m^2) &=& \frac{1}{i \pi^2}\int d^n l \, \frac{1}{l^2 - m^2} 
= m^2 \left [ \frac{1}{\bar \eps} + 1 - \log \left( \frac{m^2}{\mu^2} \right )\right],\\
 \mathcal B_0 (k^2, m^2) &=&  \frac{1}{i \pi^2} \int d^n l \, \frac{1}{(l^2 - m^2) \, ((l - k )^2 - m^2 )} \nn \\
 &=& \frac{1}{\bar \eps} + 2 - \log \left( \frac{m^2}{\mu^2} \right ) - a_3 \log \left( \frac{a_3+1}{a_3-1}\right), \\
\mathcal C_0 (s, s_1, s_2, m^2) &=&
 \frac{1}{i \pi^2} \int d^n l \, \frac{1}{(l^2 - m^2) \, ((l -q )^2 - m^2 ) \, ((l + p )^2 - m^2 )} \nn \\
&=&- \frac{1}{ \sqrt \sigma} \sum_{i=1}^3 \left[Li_2 \frac{b_i -1}{a_i + b_i}   - Li_2 \frac{- b_i -1}{a_i - b_i} + Li_2 \frac{-b_i +1}{a_i - b_i}  - Li_2 \frac{b_i +1}{a_i + b_i}
   \right], 
\label{C0polylog}
\eea
with
\bea
a_i = \sqrt {1- \frac{4 m^2}{s_i }} \qquad \qquad
b_i = \frac{- s_i + s_j + s_k }{\sqrt{ \sigma}},
\eea
where $s_3=s$ and in the last equation $i=1,2,3$ and $j, k\neq i$. \\
The one-point and two-point functions  written before in  $n=4 - 2 \, \eps$ are divergent in dimensional regularization with the singular parts given by
\bea
\mathcal A_0 (m^2) ^{sing.}  \rightarrow  \frac{1}{\bar \eps} \, m^2,  \qquad \qquad
\mathcal B_0 (s, m^2) ^{sing.}  \rightarrow  \frac{1}{\bar \eps} ,
\eea
with 
\bea
\frac{1}{\bar \eps} = \frac{1}{\eps} - \g - \ln \pi
\label{bareps}
\eea
We use two finite combinations of scalar functions given by
\bea
&&  \mathcal B_0 (s, m^2) \, m^2 - \mathcal A_0 (m^2) =  m^2 \left[ 1 - a_3 \log \frac{a_3 +1}{a_3 - 1}  \right] , \\
&& \mathcal D_i \equiv \mathcal D_i (s, s_i,  m^2) =
\mathcal B_0 (s, m^2) - \mathcal B_0 (s_i, m^2) =  \left[ a_i \log\frac{a_i +1}{a_i - 1}
- a_3 \log \frac{a_3 +1}{a_3 - 1}  \right] \qquad i=1,2.
\label{D_i}
\nn \\
\eea
The scalar integrals $ \mathcal C_0 (s, 0,0,m^2) $ and $\mathcal D (s, 0, 0, m^2)$ are the $ \{ s_1 \rightarrow 0$, $s_2 \rightarrow 0 \} $ limits of the generic functions $\mathcal C_0(s,s_1,s_2,m^2)$ and $\mathcal D_1(s,s_1,m^2)$
\bea
\mathcal C_0 (s, 0,0,m^2) &=& \frac{1}{2 s} \log^2 \frac{a_3+1}{a_3-1}, \\
\mathcal D (s, 0, 0, m^2) &=& \mathcal D_1 (s,0,m^2)= \mathcal D_2(s,0,m^2) =
  \left[ 2 - a_3 \log \frac{a_3+1}{a_3-1}\right].
\eea
The master integrals denoted by $\mathcal B_0 (s,0)$, $\mathcal D_i (s,s_i,0)$ ($i=1,2$) and $\mathcal C_0 (s,s_1,s_2,0)$ are consistently redefined for $m=0$ (and $s<0$) as
 \bea
 && \mathcal B_0 (s,0 )= \biggl[ \frac{1}{\bar \eps} - \log \left(-\frac{s}{\mu^2} \right) + 2\biggr], \\
 &&\mathcal \mathcal D_i (s, s_i,0)=  \mathcal B_0 (s,0 ) -  \mathcal B_0 (s_i,0 ) =
 \log \left(\frac{s_i}{s} \right), \qquad i=1,2 \\
 && \mathcal C_0 (s, s_1, s_2, 0) = \frac{ 1}{s} \Phi (x,y),
 \eea
where $\mu$ is the renormalization scale and the function $\Phi (x, y)$  is defined as
\cite{Usyukina:1993ch}
\bea
\Phi( x, y) &=& \frac{1}{\la} \biggl\{ 2 [Li_2(-\rho  x) + Li_2(- \rho y)]  +
\ln \frac{y}{ x}\ln \frac{1+ \rho y }{1 + \rho x}+ \ln (\rho x) \ln (\rho  y) + \frac{\pi^2}{3} \biggr\},
\label{Phi}
\eea
with
\bea
 \la(x,y) = \sqrt {\Delta},
 \qquad  \qquad \Delta=(1-  x- y)^2 - 4  x  y,
\label{lambda} \\
\rho( x,y) = 2 (1-  x-  y+\la)^{-1},
  \qquad  \qquad x=\frac{s_1}{s} \, ,\qquad \qquad y= \frac {s_2}{s}\, .
\eea
The singularities in $1/\bar \eps$ and the dependence on the renormalization scale $\mu$ cancel out when taking into account the difference of two functions $\mathcal B_0$, so that the $\mathcal D_i$'s  are well-defined; the three-point master integral is convergent.

\section{Appendix. Alternative conditions on the correlator in the massless case}
\label{alternative}
As we have mentioned, one can follow an entirely different approach in order to fix the expression of the correlator. This is based on the requirement that the trace anomaly satisfies a well known 
operatorial relation which is imposed on the matrix elements at nonzero momentum. Specifically we proceed as follows, and illustrate this point in the massless limit.  We impose the value of the trace anomaly as a defining condition on the whole amplitude, so that the (new) request $c')$ will be 

\begin{itemize}
\item[$c'$)] the non-zero anomaly trace in the massless limit.
\end{itemize}
As the first two conditions $a)$ and $b)$, respectively the $\{ \mu \leftrightarrow \nu\} $ symmetry and the vector current conservation, remain the same as before, we continue illustrating the modifications due to this approach from this point on. The third condition is given by
\bea
g_{\mu\nu} \, \Gamma^{\mu\nu\a\b}(p,q) = c \, u^{\a\b}(p,q),
\label{trace}
\eea
 where c is related to the  usual QED $\b$-function as $c = - \frac{2 \, \b}{e}$. The resulting system  is
 \bea
Eq. (\ref{trace})	\Rightarrow
\left \{ \begin {array}{l}
 \vspace{0.2cm} 4 \, \frac{A_{41}}{p\cdot q} \,- A_7   + 2 \, A_9 - A_{12}  = 0, \\
 \vspace{0.2cm}
c  + 4 \, A_{37}+4 \, A_{42} + A_4 \,  p \cdot p- 2 \, A_6 \, p \cdot p + 
2  \,  A_{11} \, p \cdot  q+2 \, A_{14} \,  q \cdot q + A_{16} \,  q \cdot q =0,
\end {array} \right.
\eea
whose solutions for $A_{41}$ and $A_{37}$ read as
\bea
A_{41} &=& \frac{p \cdot q}{4} \left( A_7 - 2 \, A_9 + A_{12} \right)  \\
A_{37} + A_{42} &=& \frac{1}{4} \left [ - c   -A_4 \, 
  p \cdot p +2 \, A_6  \, p \cdot p  -2 \, A_{11} \, p \cdot q -2 \,  A_{14}
 q \cdot q - A_{16} \,  q \cdot q )\right].
\label{sumc}
\eea
As seen from the last equation, the second solution returns the sum of two UV divergent amplitudes, $A_{37}$ and $A_{42}$. However, an explicit computation shows that in the explicit mapping between the two sets of $A_i$  and  $F_i$ these two amplitudes appear in such a way that their divergences cancel. 
Therefore, reinserting the expressions of $A_{41}$ and $A_{37}$ extracted from Eq. (\ref{sumc}) into the expression of $\Gamma^{\mu\nu\a\b}(p,q)$  one finds another mapping between the form factors $A_i$ and  $F_i$, as previously done in Eqs.~(\ref{mapping1}-\ref{mapping13})
 \bea
 F_1 &=& \frac{ c }{3 k \cdot k}, \\
  F_2 &=& 0, \\
 F_3 &=& \frac{A_4}{4} -\frac{c }{12 \, k \cdot k }, \\
 F_4 &=& \frac{A_7}{4  \, p \cdot p}, \\
 F_5 &=&  \frac{A_{16}}{4} - \frac{ c}{12 k\cdot k}, \\
 F_6 &=& \frac{A_{12}} {4  \, q \cdot q}, \\
 F_7 &=& -\frac{c }{6  k\cdot k} + \frac{A_{11}}{2} 
 + \frac{(A_9   \, p \cdot p + A_{14}  \, p \cdot q)  \, q \cdot q}{2   \, p \cdot q^2}
 + \frac{ \, p \cdot p (A_6  \, p \cdot q + A_9    \, q \cdot q)}{2  \, p \cdot q^2}, \\
 F_8 &=& -\frac{A_9}{2  \, p \cdot q}, \\
  F_9 &=& \frac{A_6}{ \, p \cdot q}   + A_9 \frac{ q \cdot q}{ \, p \cdot q^2}, \\
  F_{10} &=& A_9 \frac{ \, p \cdot p} { \, p \cdot q^2} + \frac{A_{14}} {   \, p \cdot q}, \\
 F_{11} &=& \frac{A_{12}} {2\, q \cdot q} - \frac{A_2}{2  \, p \cdot p}, \\
 F_{12} &=& \frac{A_3}{2  \, p \cdot q} + \frac{A_7 }{2 \, p \cdot p} , \\
 F_{13} &=& \frac{1} {4    \, p \cdot q}\left[ 2 A_{11}  \, p \cdot q^2 + c    \, p \cdot q + 4 \, A_{42}   \, p \cdot q
 + A_4  \, p \cdot p  \, p \cdot q + 2 \, A_6  \, p \cdot p    \, p \cdot q  \right.\nn \\
 && \hspace{1cm} \left. 
 + \, 2 \, A_{14}  \, q \cdot q  \, p \cdot q
 + A_{16}  \, q \cdot q  \, p \cdot q+ 4 \, A_9  \, p \cdot p  \, q \cdot q\right];
\eea
 This new mapping leaves the invariant amplitudes from $F_9$ to $F_{12}$ the same as before, so the condition $c)$, i.e. the WI derived from Eq. (\ref{WIeq}) and $c')$ are perfectly equivalent in determining these $4$ form factors.
 
 \section{Appendix. Comparison with the parametric approach and numerical checks} 
\label{numerical}
The parametric approach of \cite{Giannotti:2008cv} allows, by combining the denominators of the various tensor amplitudes, to give parametric expressions for the form factors $F_i$ starting from a set scalar parametric integrals. Our results correspond to an explicit computation of these integrals. We will not give each integral separately, since they are rather lengthy. The mapping between the $F_i$'s in the parametric form and our 
expressions allow to perform numerical checks of our result. We have perfect agreement between the parametric forms derived in  \cite{Giannotti:2008cv}, computed numerically, and our explicit expressions in all the euclidean regions of the external momenta. We briefly clarify this point.
\begin{table}
$$
\begin{array}{|c|c|c|}\hline
j & C_j = {\rm coefficient\ of} & c_j(x,y) \\ \hline\hline
1 & p^{\mu}p^{\nu} p^{\alpha}p^{\beta} & - 4x^2 (1-x) (1-2 x) \\ \hline 
2 &\ \ (p^{\mu}q^{\nu} + q^{\mu}p^{\nu}) p^{\alpha}p^{\beta}\ \ & - x (1-x)(1-4x+8xy) + xy \\ \hline
3 & q^{\mu}q^{\nu} p^{\alpha}p^{\beta} &  2x(1-2y)(1-x-y + 2 x y )\\ \hline
4 & p^{\mu}p^{\nu} p^{\alpha}q^{\beta} & -2x(1-x)(1-2x)(1-2y)\\ \hline
5 &\ \ (p^{\mu}q^{\nu} + q^{\mu}p^{\nu}) p^{\alpha}q^{\beta}\ \ &\ \ \ x(1-x)(1-2y)^2 + y(1-y)(1-2x)^2\ \ \ \\ \hline
6 & q^{\mu}q^{\nu} p^{\alpha}q^{\beta} & - 2y(1-y)(1-2x)(1-2y)\\ \hline
7 & p^{\mu}p^{\nu} q^{\alpha}p^{\beta} &  2 x y(1-2x)^2 \\ \hline
8 &\ \ (p^{\mu}q^{\nu} + q^{\mu} p^{\nu}) q^{\alpha}p^{\beta}\ \ & - 2 x y(1-2x) (1-2y)\\ \hline
9 & q^{\mu}q^{\nu} q^{\alpha}p^{\beta} & 2 x y (1-2y)^2\\ \hline
10 & p^{\mu}p^{\nu} q^{\alpha}q^{\beta} & 2 y(1-2x)(1-x-y + 2xy)\\ \hline
11 &\ \ (p^{\mu}q^{\nu} + q^{\mu}p^{\nu}) q^{\alpha}q^{\beta}\ \ & - y (1-y)(1-4y+8xy) + xy\\ \hline
12 & q^{\mu}q^{\nu} q^{\alpha}q^{\beta} & - 4  y^2\,(1-2 y) (1-y)\\ \hline
\end{array}
$$
\caption{The twelve tensors with four free indices 
$(\mu\nu\alpha\beta)$ on $p,q$ used in ref.  \cite{Giannotti:2008cv} ) for the construction of the form factors 
$F_i$. At each coefficient functions $C_j(k^2;p^2,q^2)$ 
correspond a polynomial $c_j$ in the Feynman parameterized form, as given in Eq. (\ref{Cj}).}
\label{tensorcoeff}
\end{table}

Explicit formulae for all twelve 
finite coefficient functions may be given in the Feynman parameterized form,
\beq
C_j(k^2; p^2,q^2) = \frac{e^2}{4\pi^2}
\int_0^1 dx\int_0^{1-x} dy \ \frac{c_j(x,y)}{p^2\, x(1-x)+q^2\, y(1-y)+
2\,xy\,p\cdot q+m^2}\,,
\label{Cj}
\eeq
where the polynomials $c_i(x,y)$ for $i=1, \dots, 12$ are listed in Table \ref{tensorcoeff}. 

\bea
&& \hspace{-.5cm} F_1 = \frac{C_7 + C_8 + C_9}{3} + \frac{p^2}{3k^2}\,
(-C_1 + C_3 + C_8 - C_9) + \frac{q^2}{3k^2}\, (-C_7 + C_8 + C_{10} - C_{12})\,, \label{F1}\\
&& \hspace{-.5cm} F_3 = \frac{2C_7 -C_8 -C_9}{12} + \frac{p^2}{12k^2}\, 
(C_1 - C_3 -C_8 + C_9) + \frac{q^2}{12k^2}\, (C_7 - C_8 - C_{10} + C_{12})\,,\\
&& \hspace{-.5cm} F_5 = \frac{-C_7 -C_8 + 2 C_9}{12} +\frac{p^2}{12k^2}\,
(C_1 - C_3 - C_8 + C_9) + \frac{q^2}{12k^2}\, (C_7 - C_8 - C_{10} + C_{12})\,,\\
&& \hspace{-.5cm} F_7 = \frac{-C_7 + 2C_8 -C_9}{6} + \frac{p^2}{6k^2}\,
(C_1 - C_3 - C_8 + C_9) + \frac{q^2}{6k^2}\,(C_7 - C_8 - C_{10} + C_{12})\nonumber\\
&& \qquad\qquad\qquad + \frac{\ p^2\,q^2}{(p\cdot q)^2}\ C_5 + \frac{p^2 C_2 + q^2 C_{11}}
{2\,(p\cdot q)} \,. \label{F1357}
\eea

\bea
&& F_2 = \frac{C_1}{3q^2} + \frac{C_{12}}{3p^2} +
\frac{-C_1 + 2C_2 - 2 C_5 + 2 C_{11} - C_{12}}{3k^2}\,, \label{F2}\\
&& F_4 = -\frac{C_1}{12q^2} + \frac{3 C_{10} - C_ {12}}{12p^2} 
+ \frac{C_1 - 2 C_2 + 2 C_5 - 2 C_{11} + C_{12}}{12k^2}\,,\\
&& F_6 = \frac{-C_1 + 3 C_3}{12q^2} - \frac{C_{12}}{12p^2} 
+ \frac{C_1 - 2 C_2 + 2 C_5 - 2 C_{11} + C_{12}}{12k^2}\,,\\
&& F_8 = -\frac{C_5}{2 p\cdot q}  - \frac{C_1}{6q^2} - \frac{C_{12}}{6p^2} 
+ \frac {C_1 - 2 C_2 + 2 C_5 - 2 C_{11} + C_{12}}{6k^2}\,.
\label{F2468}
\eea
\bea
&& F_9 = \frac{C_2}{p\cdot q} + \frac{q^2\,C_5}{(p\cdot q)^2} \,,\\
&& F_{10}= \frac{p^2\,C_5}{(p\cdot q)^2} + \frac{C_{11}}{p\cdot q}\,,\\
&& F_{11} = \frac{C_3}{2q^2}-\frac{C_{12}}{2p^2}\,,\\
&& F_{12} = \frac{C_{10}}{2p^2}-\frac{C_1}{2q^2}. 
\eea \label{tenrel3}
\noindent

Finally, numerical checks on $F_{13}$ are performed on the UV convergent amplitude
\bea
&&F_{13} =\frac{\Pi_R(p^2) + \Pi_R(q^2)}{2} + \frac{\ p^2\,q^2}{p\cdot q}\ C_5 
+ \frac{p^4 C_4 + q^4 C_6}{4 p\cdot q}
+\frac{p\cdot q}{4}\, (2C_2 + C_3 + C_{10} + 2 C_{11})\nonumber\\
&&\qquad+ \frac{p^2}{4}\,(2C_2 + C_4 + 2C_5 + C_{10}) 
+ \frac{q^2}{4}\, (C_3 + 2 C_5 + C_6+ 2 C_{11})\,
\label{F13} 
\eea
where the scalar two-point functions have been renormalized by the subtraction of the UV $1/\epsilon$ pole.

\section{Appendix. The massive invariant amplitudes}
\label{Fimassive}
The off-shell massive form factors $F_i$, with
\begin{itemize}
\item \underline{\bf $s \neq 0$  \qquad $s_1 \neq 0$ \qquad $s_2 \neq 0$ \qquad $m \neq 0$}
\end {itemize}
and with $\gamma \equiv s -s_1 - s_2$, $\si \equiv s^2 - 2 (s_1+s_2)\, s + (s_1-s_2)^2$
are given by \footnote{We use boldfaced notation to facilitate their identification in the lengthier expressions}
\bea
 %
\underline{ \bf \underline{F_1 (s;\,s_1,\,s_2,\,m^2)}} &=&
  \frac{ e^2 \gamma   m^2}{3 \pi ^2 s \sigma } 
  + \frac{ e^2 \, \mathcal D_2(s,s_2,m^2)\,  s_2 \left[s^2+4 s_1 s-2 s_2 s-5 s_1^2+s_2^2+4 s_1
   s_2\right] m^2}{3 \pi^2 s \sigma ^2}   \nn \\
&&   \hspace{-3cm} 
-\, \frac{ e^2 }{18 \pi ^2 s}  
- \frac{ e^2 \, \mathcal D_1 (s,s_1,m^2)\, s_1
   \left[-\left(s-s_1\right){}^2+5 s_2^2-4 \left(s+s_1\right)
   s_2\right] m^2}{3 \pi^2 s \sigma ^2}  \nn \\
  && \hspace{-3cm}  
  - \,  \, e^2 \,  \mathcal \,C_0 (s,s_1,s_2,m^2)\, 
  \left[\frac{ m^2 \gamma
   \left[\left(s-s_1\right){}^3-s_2^3+\left(3 s+s_1\right)
   s_2^2+\left(-3 s^2-10 s_1 s+s_1^2\right) s_2\right]}
   {6 \pi^2  s \sigma ^2}-\frac{2 m^4 \gamma }{3  \pi^2 s \sigma }\right], \nn \\ \\
  \label{Fone}
  %
\underline{ \bf \underline{F_2 (s;\,s_1,\,s_2,\,m^2)}}&=&
 -\frac{2 e^2 m^2}{3 \pi^2 s \sigma } 
 - \frac{2 \, e^2 \, \mathcal D_2(s,s_2,m^2) \,  \left[\left(s-s_1\right){}^2-2
   s_2^2+\left(s+s_1\right) s_2\right] m^2}{3 \pi^2 \, s \, \sigma   ^2}      \nn \\
    && \hspace{-3cm}   
   - \, \frac{2  e^2 \, \mathcal D_1 (s,s_1,m^2) \, m^2}{3 \pi^2 s \sigma ^2}
   \left[s^2+\left(s_1-2 s_2\right) s-2 s_1^2+s_2^2+s_1 s_2\right] \nn \\
    && \hspace{-3cm} 
    \,- \, e^2 \, \mathcal C_0 (s,s_1,s_2,m^2) \, 
    \biggl[\frac{4 m^4}{3 \pi^2 s \sigma}
  +\frac{ m^2}{3 \pi^2 s \sigma ^2} \,  \left[ s^3-\left(s_1+s_2\right) s^2 
  - \left(s_1^2-6  s_2 s_1+s_2^2\right) \right. s \nn \\
  && \hspace{3 cm} +\left. \left(s_1-s_2\right){}^2
   \left(s_1+s_2\right)\right] \biggr], \nn \\ \\
  %
  %
\underline{ \bf \underline{F_3 (s;\,s_1,\,s_2,\,m^2)}} &=&
 \underline{ \bf \underline{F_5 (s;\,s_2,\,s_1,\,m^2)}} =
 - \frac{  e^2} {144 \pi^2 s  \sigma ^3} \biggl[
 s^6-3 \left(s_1-4 s_2\right) s^5+6 \left(3  s_1-7 s_2\right) s_2 s^4  \nn \\
   && \hspace{-2cm}
 +2 \left(5 s_1^3-69 s_2 s_1^2+117   s_2^2 s_1+23 s_2^3\right) s^3
 -3 \left(5 s_1^4-62 s_2  s_1^3+72 s_2^2 s_1^2+50 s_2^3 s_1+7 s_2^4\right) s^2 \nn \\
 && \hspace{-2cm}
 +3   \left(s_1-s_2\right){}^2 \left(3 s_1^3-24 s_2 s_1^2-33 s_2^2
   s_1+2 s_2^3\right) s   -2 \left(s_1-s_2\right){}^6\biggr] \nn \\
   && \hspace{-2cm}
   - \frac{ e^2 \gamma  m^2}{6 \pi^2 s \sigma   ^2}
\biggl[  s^2 -2 \left(s_1-3 s_2\right)  s+\left(s_1-s_2\right){}^2 \,\biggr] \nn \\
&&  \hspace{-2cm}
 - \frac{ \, e^2 \, \gamma}{12\pi^2 s \sigma ^2} \,  \biggl[ s^2+\left(5 s_2-2 s_1\right)  s+\left(s_1-s_2\right){}^2\biggr]
\left[ \mathcal B_0(s,m^2) m^2- \mathcal A_0(m^2) \right] \nn \\
   && \hspace{- 2cm}
   - \frac{ \,e^2 \, m^2 } {12 \pi^2 \, s \, \si^3}  \, \mathcal D_1 (s,s_1,m^2) \,
   \biggl[ \left(2 s+s_1\right)
   \left(s-s_1\right){}^4-12 \left(s+s_1\right) s_2^2
   \left(s-s_1\right){}^2   \nn \\
   &&\hspace{-1cm}  
   +s_1 \left(41 s+2 s_1\right) s_2   \left(s-s_1\right){}^2
   -  \left(6 s+5 s_1\right) s_2^4+\left(16
   s^2-41 s_1 s+14 s_1^2\right) s_2^3\biggr] \nn \\
  && \hspace{-3cm}
  - \frac{ \, e^2 \, s_1 } {48 \pi^2 \, \si^4 } \, \mathcal D_1 (s,s_1,m^2) \,  
  \Big[ \left(s-s_1\right){}^6+2 \left(14 s+11 s_1\right) s_2 \left(s-s_1\right){}^4 \nn \\
  && \hspace{-1cm}
  -\left(23 s^2-214 s_1  s+19 s_1^2\right) s_2^2 \left(s-s_1\right){}^2   +2
   -21 s_2^6+ \left(5 s_1-2 s\right) s_2^5 \Big. \nn \\
   && \hspace{-1cm} \Big. +\left(107 s^2-318 s_1 s+71
   s_1^2\right) s_2^4+8 \left(-11 s^3+18 s_1 s^2+17 s_1^2 s-8
   s_1^3\right) s_2^3\Big]  \nn \\
  &&  \hspace{-3cm} 
  - \frac{  \, e^2 \, s_2 \, \, m^2 } {12 \pi^2 \, s \, \si^3} \, \mathcal D_2 (s,s_2,m^2) \,
  \Big[s_2^4+\left(19 s+2 s_1\right)
   s_2^3-2 \left(12 s^2-23 s_1 s + 6 s_1^2\right)   s_2^2 \nn \\
  &&  \hspace{-2cm} -\left(s-s_1\right) \left(13 s^2-49 s_1 s+14
   s_1^2\right) s_2+\left(s-s_1\right){}^3 \left(17 s+5
   s_1\right)\Big] \nn \\
   &&  \hspace{-3cm} 
   -  \frac{ \, e^2 \, s_2 } {48 \pi^2 \, \si^4 } \, \mathcal D_2 (s,s_2,m^2) \,
   \Big[ s_2^6-2 \left(s-14 s_1\right) s_2^5 + \left(s^2+120 s_1 s-37 s_1^2\right) s_2^4 \nn \\
   && \hspace{-1cm}
    - 4 \left(s^3+49 s_1 s^2-69 s_1^2 s+13  s_1^3\right) s_2^3  +\left(s-s_1\right) \left(11 s^3-69 s_1
   s^2+309 s_1^2 s-83 s_1^3\right) s_2^2\nn \\
   && \hspace{-1cm} -2
   \left(s-s_1\right){}^3 \left(5 s^2-49 s_1 s-4 s_1^2\right) s_2 
   +3 \left(s-s_1\right){}^5 \left(s+5 s_1\right)\Big] \nn \\
   && \hspace{- 1cm}
   -\,  \, e^2 \, \mathcal C_0 (s,s_1,s_2,m^2)\, 
   \bigg[ \frac{\gamma  \, m^4}{3 \pi^2 \, s \, \sigma  ^2}
   \left[ s^2+\left(7 s_2-2 s_1\right)   s+\left(s_1-s_2\right){}^2\right]
    \nn \\
   && \hspace{-2cm}
   +\frac{m^2}{24 \pi^2  \, s \, \sigma ^3}   \left[  -s_2^6+\left(2 s_1-9 s\right) s_2^5    \right.
   +\left(12 s^2-65 s_1 s+s_1^2\right) s_2^4
       \nn \\
   &&  \hspace{-1cm}
   +2 \left(13 s^3-54 s_1 s^2+55 s_1^2 s-2 s_1^3\right)   s_2^3
   -\left(s-s_1\right) \left(45 s^3-133 s_1 s^2+15 s_1^2
   s+s_1^3\right) s_2^2 \nn \\
   && \hspace{-1cm}
    \left.
    +\left(s-s_1\right){}^3 \left(15 s^2+47
   s_1 s-2 s_1^2\right) s_2 
   +\left(s-s_1\right){}^5 \left(2   s+s_1\right)\right]
   \nn \\
  && \hspace{-2cm}
   + \frac { s_1 s_2} {8 \pi^2 \sigma ^4}
   \left[ 2 s^6+3 \left(s_2-3 s_1\right) s^5+\left(15 s_1^2+6  s_2 s_1-13 s_2^2\right) s^4  \right.
  \nn \\
    &&  \hspace{-1cm}
     +2 \left(-5 s_1^3-19 s_2 s_1^2+29  s_2^2 s_1+s_2^3\right) s^3
     +12 s_2 \left(4 s_1^3-4 s_2 s_1^2-3 s_2^2 s_1+s_2^3\right) s^2  \nn \\
 && \hspace{-1cm}
 \left.
   +\left(s_1-s_2\right){}^2
   \left(3 s_1^3-15 s_2 s_1^2-31 s_2^2 s_1-5 s_2^3\right)
   s-\left(s_1-s_2\right){}^4
   \left(s_1+s_2\right){}^2\right]\bigg], \nn \\ \\
  %
  %
 \underline{ \bf \underline{F_4 (s;\,s_1,\,s_2,\,m^2)}} &=&
 \underline{ \bf \underline{F_6 (s;\,s_2,\,s_1,\,m^2)}}  = \nn \\
 && \hspace{-3cm} 
 \frac{  e^2 m^2}{6 \pi^2 s \sigma ^2   s_1}  \biggl[ 3 s^3-2 \left(2 s_1+3 s_2\right)
   s^2+\left(-s_1^2+6 s_2 s_1+3 s_2^2\right) s+2 s_1
   \left(s_1-s_2\right){}^2\biggr] \nn \\
   && \hspace{-3cm}  
  + \frac{ e^2}{12 \pi^2  s \sigma ^2  s_1} \left[ \mathcal B_0(s,m^2) \,m^2    - \mathcal A_0 (m^2 )\right]
     \bigg[3 s^3-2 \left(2 s_1+3 s_2\right)  s^2+ \nn \\
     && \hspace{2cm}
     \left(-s_1^2+4 s_2 s_1+3 s_2^2\right) s
  +2 s_1 \left(s_1-s_2\right){}^2\bigg]  \nn \\
   && \hspace{-3cm} 
   + \frac{ e ^2 }{24 \pi^2 \sigma ^3 s_1}
   \bigg[
      -s_2^5+\left(6 s+11 s_1\right)  s_2^4-\left(14 s^2+s_1 s+5 s_1^2\right) s_2^3
      +\left(16 s^3-35 s_1 s^2+46 s_1^2 s-15 s_1^3\right)   s_2^2   \bigg. \nn \\
   && \hspace{-2cm} \bigg.   -\left(s-s_1\right){}^2 \left(9 s^2-11 s_1 s-6
   s_1^2\right) s_2+2 \left(s-s_1\right){}^4 \left(s+2 s_1\right)
   \bigg] \nn \\
  && \hspace{-3cm} 
  - \, e^2 \mathcal D_2 (s, s_2,m^2)
   \bigg[\frac{ m^2}{12 \pi^2  s   \sigma ^3 s_1}   \bigg(
   -2 \left(2 s+s_1\right) \left(s-s_1\right){}^4+\left(3 s^2-43 s_1 s+2
   s_1^2\right) s_2 \left(s-s_1\right){}^2  \bigg. \bigg.\nn \\
  && \hspace{-1cm} \bigg.  +\left(9 s+4 s_1\right) s_2^4 +\left(-23 s^2+29 s_1 s-10 s_1^2\right) s_2^3+\left(15
   s^3+2 s_1 s^2+5 s_1^2 s+6 s_1^3\right) s_2^2\bigg) 	\nn \\
   && \hspace{-2cm} 
   - \frac{1}{48 \pi^2 \sigma ^4 s_1} \bigg(
   3 \left(s+s_1\right)  \left(s-s_1\right){}^6-4 \left(4 s^2-14 s_1 s-5 s_1^2\right)
   s_2 \left(s-s_1\right){}^4 \bigg. \nn \\
   &&  \hspace{-2cm} +\left(35 s^3-119 s_1 s^2+169 s_1^2
   s-13 s_1^3\right) s_2^2 \left(s-s_1\right){}^2+\left(s-3
   s_1\right) s_2^6 - 8 \left(s^2+9 s_1 s+7 s_1^2\right)   s_2^5 \nn \\
   &&  \hspace{-2cm}  \bigg. +\left(25 s^3+159 s_1 s^2-197 s_1^2 s+157 s_1^3\right)
   s_2^4+4 \left(-10 s^4+21 s_1^2 s^2+28 s_1^3 s-27 s_1^4\right)
   s_2^3 \bigg) \bigg] \nn \\
    &&  \hspace{-3cm} 
    + \, e^2 \mathcal D_1 (s, s_1,m^2)\, \bigg[
    \frac{m^2}{12 \pi^2   s \sigma ^3 s_1}
    \bigg(
    2  s^5+\left(15 s_1-8 s_2\right) s^4+\left(-53 s_1^2-5 s_2 s_1+12 s_2^2\right) s^3 \bigg. \bigg. \nn \\
  && \hspace{-2cm}  +\left(49 s_1^3+46 s_2 s_1^2-33 s_2^2 s_1-8  s_2^3\right) s^2
    -\left(s_1-s_2\right) \left(9 s_1^3+52 s_2  s_1^2+23 s_2^2 s_1+2 s_2^3\right) s \nn \\
    && 	\hspace{-2cm}
    -2 s_1\left(s_1-s_2\right){}^3 \left(2 s_1+s_2\right)\bigg) 
    +\frac{1}{48 \pi^2 \sigma ^4}
   \bigg(
   s_2^6+4 \left(6 s+11 s_1\right)  s_2^5
   +\left(-87 s^2+106 s_1 s-91 s_1^2\right) s_2^4 \nn \\
   && \hspace{-2cm}
   +4 \left(22 s^3-69 s_1 s^2+40 s_1^2 s+s_1^3\right)  s_2^3
   +\left(s-s_1\right) \left(3 s^3-29 s_1 s^2+209 s_1^2 s-79   s_1^3\right) s_2^2 \nn \\
   && \hspace{-2cm}
   -8 \left(s-s_1\right){}^3 \left(6 s^2-13 s_1 s - 4 s_1^2\right) s_2+\left(s-s_1\right){}^5
   \left(19 s+5  s_1\right)\bigg)
   \bigg]  \nn \\
   && \hspace{-3cm} 
   + \, e^2 \mathcal C_0 (s,s_1,s_2,m^2)\, \bigg\{
    \frac{ m^4}{6 \pi^2 \sigma ^2} \left[ \frac{\sigma  \left(3 s+2 s_1\right)}{s 	\, s_1} +18 s_2\right]
   -\frac{m^2}{24 \pi^2 \sigma ^3}	 
   \bigg[-\frac{ \sigma ^2}{s s_1} \left(9  s^2+\left(59 s_1+3 s_2\right) s \right. \nn \\
   && \hspace{-1cm}
   \left. +2 s_1
   \left(s_1+s_2\right)\right)   +12 \left(3 s^2-3
   \left(22 s_1+7 s_2\right) s+s_1 \left(3 s_1-17
   s_2\right)\right) \sigma  \nn \\
   && \hspace{-1cm}
   +720 \, s \, s_1
   \left(\left(s-s_1\right){}^2-2 \left(s+s_1\right) s_2\right)\bigg]  
   -\frac{1}{16 \pi^2 \sigma ^4} 
   \bigg[-2 s \left(s-s_1\right){}^6 \nn \\
   && \hspace{-1cm}
   -2 \left(s^2+7 s_1 s+2 s_1^2\right) s_2 \left(s-s_1\right){}^4
   + 2 \left(7 s^3+6 s_1 s^2+11 s_1^2 s-4 s_1^3\right) s_2^4
    \nn \\
   && \hspace{-1cm}
   +12 \left(2 s^3-3 s_1 s^2-2 s_1^2 s+s_1^3\right) s_2^2 \left(s-s_1\right){}^2
   -4 \left(s+s_1\right) s_2^6
    + 6 \left(s^2-5 s_1 s+2 s_1^2\right)   s_2^5 \nn \\
    && \hspace{-1cm}
   - 4 \left(9 s^4-25 s_1 s^3+33 s_1^2 s^2-15 s_1^3 s+2 s_1^4\right) s_2^3 \bigg]
   \bigg\}, \nn \\ \\
 %
 %
 \underline{ \bf \underline{F_7 (s;\,s_1,\,s_2,\,m^2)}} &=&
\frac{  e ^2 m^2}{3\pi^2  s \sigma ^2} \bigg[\left(s^2+12 s_2 s-s_2^2\right)
  s_1+s_1^3-\left(2 s+s_2\right) s_1^2+\left(s-s_2\right){}^2 s_2 \bigg] \nn \\
  && \hspace{-3cm}
  +\frac{e^2}{72 \pi^2 }  \bigg[\frac{840 s \left( 2 \left(s+s_1\right)
  s_2-\left(s-s_1\right){}^2\right) s_1^2}{\sigma ^3} + \frac{6 \left(-13 s^2+166 s_1 s-13
  s_1^2+39 \left(s+s_1\right) s_2\right) s_1}{\sigma ^2} \nn \\
  && \hspace{-2cm}
  +\frac{3 \left(-s+27 s_1+s_2\right)}{\sigma}
  + \frac{2}{s} + \frac{9 s}{\gamma^2} -\frac{6}{\gamma} \bigg] \nn \\
  && \hspace{-3cm}
  +\frac{e^2}{6 \pi^2 } \,  \left[ \mathcal B_0(s,m^2) \, m^2 - \mathcal A_0(m^2) \right] \, \bigg[ \frac{14 s_1 s_2}{\sigma
  ^2}+\frac{s+s_1+s_2}{s \sigma }-\frac{3}{\gamma ^2}\bigg] \nn \\
  && \hspace{-3cm}
  - \frac{e^2}{16 \pi^2} \, \mathcal C_0 (s,s_1,s_2,m^2)\,  \bigg\{
  m^4 \left[ \frac{16 \gamma }{3 s \sigma }
  -\frac{96 s_1 s_2}{\sigma ^2}-\frac{16}{\gamma ^2}\right]
  +m^2 \bigg[ \frac{960 s
  \left(\left(s-s_1\right){}^2-2 \left(s+s_1\right) s_2\right) s_1^2}{\sigma^3} -\frac{4}{3s} \nn \\ %
  && \hspace{-1.5cm}
  +\frac{4}{\gamma } + \frac{16 \left(7 s_1^2-\left(74 s+21 s_2\right) s_1+s \left(7 s-19s_2\right)\right) s_1}{\sigma ^2}
  -\frac{4 \left(3 s \left(2s+s_2\right)+s_1 \left(87 s+4 s_2\right)\right)}{3 s \sigma }
  \bigg]\nn \\
  && \hspace{-2cm}
  -\frac{4 s_1 s_2}{\gamma ^2 \sigma ^4}
  \bigg[ \left(-9 s^2+22 s_2 s-4
  s_2^2\right) s_1^6 + 2 s \left(20 s^2-39 s_2 s+21 s_2^2\right) s_1^5 + \left(s-s_2\right){}^6 s_2 \left(4 s+s_2\right) \nn \\
  && \hspace{-2cm}
  + 2 s \left(s-s_2\right){}^4 \left(2 s^2+5 s_2 s+11 s_2^2\right) s_1
  +\left(-65 s^4+96 s_2 s^3+33 s_2^2 s^2-62 s_2^3 s+6 s_2^4\right) s_1^4 \nn \\
  && \hspace{-2cm}
  +2 s \left(27 s^4-22 s_2 s^3-108 s_2^2 s^2+102 s_2^3 s-31 s_2^4\right) s_1^3
  -\left(s-s_2\right){}^2 \big( 23 s^4+40 s_2 s^3-105 s_2^2 s^2 \nn \\
  && \hspace{-2.2cm}
  -34 s_2^3 s+4 s_2^4 \big) s_1^2+ s_1^8-2 s s_1^7 \bigg] \bigg\} \nn \\
  &&\hspace{-3cm}
  - \frac{e^2}{16 \pi^2} \, \mathcal D_1  (s,s_1,m^2)\, \bigg[
  \frac{2}{3} m^2 \bigg( \frac{3}{\gamma^2}(1-\frac{3 s_2}{s-s_1})
  + \frac{20 s_1^2-37 s s_1+s
  \left(9 s_2-19 s\right)}{\sigma s \left(s-s_1\right)} \nn \\
  && \hspace{-2cm}
  +\frac{8 s_1 \left(3 s_1^2-\left(61 s+3 s_2\right) s_1+s \left(3 s-19
  s_2\right)\right)}{ \sigma^2 s}
  +\frac{440 s_1^2
  \left(\left(s-s_1\right){}^2-\left(3 s+s_1\right) s_2\right)}{\sigma^3}\bigg) \nn \\
  && \hspace{-2cm}
  +\frac{ 2 s_1}{3 \gamma ^2 \sigma ^4} \bigg(-\left(317 s^2+227 s_1 s+64 s_1^2\right)
  s_2^6+\left(s_1-s\right){}^5 \left(-7 s^2+39 s_1 s+32 s_1^2\right)s_2 \nn \\
  && \hspace{-2cm}
  +\left(397 s^3+846 s_1 s^2-539 s_1^2 s+312 s_1^3\right)
  s_2^5-\left(s-s_1\right){}^3 \left(23 s^3+114 s_1 s^2+463 s_1^2 s-16 s_1^3\right) s_2^2 \nn \\
  && \hspace{-2cm}
  -\left(275 s^4+s_1 \left(1181 s^3+s_1 \left(3 s_1
  \left(93 s+94 s_1\right)-1441 s^2\right)\right)\right)
  s_2^4+\left(s-s_1\right) \big(103 s^4+767 s_1 s^3 \nn \\
  && \hspace{-2cm}
  -79 s_1^2 s^2-563 s_1^3
  s-36 s_1^4\big) s_2^3-23 s_2^8+\left(133 s+4 s_1\right)
  s_2^7+\left(s_1-s\right){}^7 \left(2 s+s_1\right) \bigg)
  \bigg] \nn \\
  && \hspace{-3cm}
  - \, e^2 \mathcal D_2(s,s_2,m^2) \, \bigg[
  \frac{ m^2}{6 \pi^2  s \gamma ^2 \sigma ^3} \bigg( \left(-6 s^2+49 s_1 s-7 s_1^2\right)s_2^5 +\left(s-s_1\right){}^4 \left(16 s^2-5 s_1 \left(s+s_1\right)\right)s_2  \nn \\
  && \hspace{-2cm}
  + s_2^7-\left(s-4 s_1\right) s_2^6 -4 s \left(s-s_1\right){}^6  + 2 \left(5 s^3-87 s_1 s^2+56 s_1^2 s-4 s_1^3\right) s_2^4
   \nn \\
  && \hspace{-2.5cm}
  -\left(s-s_1\right){}^2 \big(21 s^3+40 s_1 s^2+147 s_1^2 s -4 s_1^3\big) s_2^2
  + \left(5 s^4+164 s_1 s^3-68 s_1^2 s^2-16 s_1^3 s+11 s_1^4\right) s_2^3
  \bigg) \nn \\
  && \hspace{-2cm}
  +\frac{ s_2}{24 \pi^2 \gamma ^2 \sigma ^4} \bigg(
  -23 s_1^8+\left(133 s+4 s_2\right) s_1^7  -\left(317 s^2+227 s_2 s+64 s_2^2\right)s_1^6 \nn \\
  && \hspace{-2cm}
  +\left(s-s_2\right){}^5 \left(7 s^2-39 s_2 s-32 s_2^2\right)s_1
  -\left(s-s_2\right){}^7 \left(2 s+s_2\right) \nn \\
  && \hspace{-2cm}
  +\left(397 s^3+846 s_2 s^2-539 s_2^2 s+312 s_2^3\right)s_1^5
  -\left(s-s_2\right){}^3 \left(23 s^3+114 s_2 s^2+463 s_2^2 s-16 s_2^3\right) s_1^2 \nn \\
  && \hspace{-2cm}
  -\left(275 s^4+s_2 \left(1181 s^3+s_2 \left(3 s_2 \left(93 s+94 s_2\right)-1441 s^2\right)\right)\right)s_1^4
  +\left(s-s_2\right) \big(103 s^4+767 s_2 s^3 \nn \\
  && \hspace{-2cm}
  -79 s_2^2 s^2-563 s_2^3 s-36 s_2^4\big) s_1^3
  \bigg)
  \bigg], \nn \\ \\
  %
  %
  \underline{ \bf \underline{F_8 (s;\,s_1,\,s_2,\,m^2)}} &=&
- \frac{  e ^2 m^2}{6 \pi^2 s \sigma ^2}
\bigg[ 3 s^2-2 \left(s_1+s_2\right)  s - \left(s_1-s_2\right){}^2\bigg] \nn \\
&& \hspace{-3cm}
- \frac{e^2}{3 \pi^2 s \, \gamma  \sigma ^2} \bigg( \mathcal B_0 (s,m^2)\,m^2 - \mathcal A_0 (m^2)\bigg)
\bigg[4 s^3 - 7 \left(s_1+s_2\right) s^2+2  \left(s_1^2+s_2 s_1+s_2^2\right) s \nn \\
&&\hspace{-2cm}
+\left(s_1-s_2\right){}^2 \left(s_1+s_2\right)\bigg]   
 - \frac{ e ^2 }{12 \pi^2 \gamma  \sigma ^3} \bigg[ 4 s_2^5+\left(14 s_1-11 s\right) s_2^4  +2 \left(s-s_1\right) \left(2 s+9 s_1\right) s_2^3 \nn \\
 && \hspace{-2cm}
   +2 \left(7
   s^3-43 s_1 s^2+33 s_1^2 s-9 s_1^3\right) s_2^2 
   -2 \left(s-s_1\right){}^2 \left(8 s^2-21 s_1 s-7 s_1^2\right)   s_2 \nn \\
 && \hspace{-2cm}   
 +\left(s-s_1\right){}^4 \left(5 s+4 s_1\right)\bigg]  
 - \, e^2 \mathcal D_2 (s, s_2,m^2) \, \bigg[
 \frac{ m^2}{3 \pi^2  s \gamma  \sigma ^3}  \bigg(
 -2 s_2^5+\left(3  s_1-10 s\right) s_2^4 \nn  \\
 && \hspace{-2cm}
 +\left(39 s^2-33 s_1 s+2 s_1^2\right)  s_2^3
  + 7 s \left(s-s_1\right){}^2 \left(s+5 s_1\right)  s_2 \nn \\
&&  \hspace{-2cm}
+\left(-37 s^3+20 s_1 s^2+9 s_1^2 s-4 s_1^3\right)  s_2^2
 +\left(s-s_1\right){}^4 \left(3 s+s_1\right)
 \bigg)
 +\frac{1 }{24 \pi^2 \gamma  \sigma ^4} \bigg( -5 s_2^7  \nn \\
 && \hspace{-2cm}
 +3 \left(s-13  s_1\right) s_2^6 + \left(57 s^2-128 s_1 s+43 s_1^2\right)   s_2^5
 +\left(-155 s^3 + 567 s_1 s^2 - 341 s_1^2 s + 121 s_1^3\right)  s_2^4  \nn \\
 && \hspace{-2cm}
 + 3 \left(55 s^4-176 s_1 s^3+86 s_1^2 s^2+56 s_1^3 s-53  s_1^4\right) s_2^3
 + 3 \left(s-s_1\right){}^6  \left(s+s_1\right) \nn \\
 && \hspace{-2cm}
 - \left(s-s_1\right){}^2 \left(75 s^3+103  s_1 s^2-311 s_1^2 s-11 s_1^3\right)   s_2^2
 +\left(s-s_1\right){}^4 \left(7 s^2+124 s_1 s+25  s_1^2\right) s_2
 \bigg)  \bigg] \nn \\
   && \hspace{-3cm} 
   - \,  e^2 \mathcal  D_1 (s, s_1,m^2) \,
   \bigg[
   \frac{ m^2}{3 \pi^2 s \gamma  \sigma  ^3}
   \bigg(    s_2^5-s s_2^4 +\left(-6 s^2+35 s_1 s-4  s_1^2\right) s_2^3
   +\left(14 s^3 - 63 s_1 s^2+9 s_1^2 s+2 s_1^3\right) s_2^2 \nn \\
  && \hspace{-2cm}
   +\left(-11 s^4+21 s_1 s^3+20 s_1^2 s^2-33  s_1^3 s+3 s_1^4\right) s_2
   +\left(s-s_1\right){}^3 \left(3 s^2+16 s_1 s+2 s_1^2\right)\bigg) \nn \\
   && \hspace{-2cm}
   +\frac{1 }{24 \pi^2  \gamma  \sigma ^4}
   \bigg(3 s_2^7+5 \left(5 s_1-3 s\right)  s_2^6 + \left(27 s^2+24 s_1 s+11 s_1^2\right) s_2^5
    -\left(15 s^3+339 s_1 s^2 \right. \nn \\
   && \hspace{-2cm}
  \left. -289 s_1^2 s+159 s_1^3\right) s_2^4
  +\left(-15 s^4+616 s_1 s^3-714 s_1^2 s^2+168 s_1^3 s+121 s_1^4\right)  s_2^3  \nn \\
  && \hspace{-2cm}
  +\left(s-s_1\right) \left(27 s^4-402 s_1 s^3+40 s_1^2  s^2 + 298 s_1^3 s-43 s_1^4\right) s_2^2
  -\left(s-s_1\right){}^3 \left(15 s^3-51 s_1 s^2 \right. \nn \\
  && \hspace{-2cm}
  \left.-245 s_1^2 s-39 s_1^3\right)
   s_2+\left(s-s_1\right){}^5 \left(3 s^2+22 s_1 s+5
   s_1^2\right)\bigg)\bigg] \nn \\
  && \hspace{-3cm}
   - \, e^2 \mathcal C_0 (s, s_1,s_2, m^2) \,  \bigg[
   \frac{2  m^4}{3 \pi^2 s \gamma  \sigma ^2}
   \bigg(2 s^3-3 \left(s_1+s_2\right) s^2+10 s_1
   s_2 s+\left(s_1-s_2\right){}^2 \left(s_1+s_2\right)\bigg)
   \nn \\
  && \hspace{-2cm}
  +\frac{ m^2}{6 \pi^2 s \sigma ^3}
  \bigg(11 s^5 -18 \left(s_1+s_2\right) s^4+\left(-11 s_1^2+94 s_2 s_1-11  s_2^2\right) s^3 \nn \\
 && \hspace{-2cm}
  +\left(s_1+s_2\right) \left(31 s_1^2-90 s_2 s_1+31 s_2^2\right) s^2 - 4 \left(s_1-s_2\right){}^2 \left(3
   s_1^2+11 s_2 s_1+3 s_2^2\right) s \nn \\
  && \hspace{-2cm}
  -\left(s_1-s_2\right){}^4
   \left(s_1+s_2\right)\bigg)
   +\frac{1}{4 \pi^2 \gamma  \sigma ^4}
   \bigg(    \left(s_1+s_2\right) s^7  -6 \left(s_1^2-s_2   s_1+s_2^2\right) s^6\nn \\
   && \hspace{-2cm}
    + 3 \left(s_1+s_2\right) \left(5 s_1^2-12  s_2 s_1+5 s_2^2\right) s^5
   +2 \left(-10 s_1^4+3 s_2 s_1^3  \right. \nn \\
   && \hspace{-2cm} \left. +54 s_2^2 s_1^2+3 s_2^3 s_1-10 s_2^4\right)   s^4
   +\left(s_1+s_2\right) \left(15 s_1^4+16 s_2 s_1^3-126  s_2^2 s_1^2+16 s_2^3 s_1+15 s_2^4\right) s^3 \nn \\
   && \hspace{-2cm}
   -6 \left(s_1^6 + 5   s_2 s_1^5-s_2^2 s_1^4-18 s_2^3 s_1^3-s_2^4 s_1^2+5 s_2^5  s_1+s_2^6\right) s^2
   +\left(s_1-s_2\right){}^2   \left(s_1+s_2\right) \nn \\
   && \hspace{-2cm}
   \left(s_1^4+6 s_2 s_1^3+34 s_2^2 s_1^2+6
   s_2^3 s_1+s_2^4\right) s+2 s_1 \left(s_1-s_2\right){}^4 s_2
   \left(s_1+s_2\right){}^2\bigg)
   \bigg], \nn \\ \\
  %
  %
   \underline{ \bf \underline{F_9 (s;\,s_1,\,s_2,\,m^2)}}&=& 
 \underline{ \bf \underline{F_{10} (s;\,s_2,\,s_1,\,m^2)}}  =
-  \frac{2  e ^2 m^2}{3 \pi^2 \sigma   s_1} 
- \frac{e^2}{6 \pi^2 s_1}
 \bigg[ \mathcal B_0(s,m^2) \,m^2 - \mathcal A_0(m^2) \bigg] \left( \frac{3}{\gamma ^2}+\frac{1}{\sigma }\right)
 \nn \\
 && \hspace{-3cm}
 + \frac{  e ^2 }{12 \pi^2 \gamma ^2 \sigma ^2} \bigg[
 \left(s-s_1\right){}^4
  - 4 \left(4
   s+s_1\right) s_2 \left(s-s_1\right){}^2-3 s_2^4+4 \left(s_1-2
   s\right) s_2^3+2 \left(13 s^2-2 s_1 s+s_1^2\right) s_2^2\bigg] \nn \\
 && \hspace{-3cm}
   -\,  e^2 \mathcal C_0 (s,s_1,s_2,m^2) \, 
   \bigg[   \frac{4 s_2   m^4}{\pi^2 \gamma ^2 \sigma }
   + \frac{ m^2}{2 \pi^2 \gamma  \sigma ^2}
   \bigg(\left(s-s_1\right){}^3+\left(7 s+s_1\right) s_2
   \left(s-s_1\right)-3 s_2^3+5 \left(s_1-s\right) s_2^2\bigg) \nn \\
   && \hspace{-2cm}
   +\frac{8 s s_2 }{\gamma ^2 \sigma ^3}
   \bigg(s_2^5+\left(2  s_1-3 s\right) s_2^4
   +2 \left(s-s_1\right) \left(s+2 s_1\right) s_2^3+2 \left(s^3-7 s_1 s^2+3 s_1^2 s-s_1^3\right)  s_2^2 \nn \\
   && \hspace{-2cm}
   -\left(s-3 s_1\right) \left(s-s_1\right){}^2 \left(3
   s+s_1\right) s_2+s \left(s-s_1\right){}^4\bigg )\bigg] \nn \\
   && \hspace{-3cm}
   - \, e^2 \mathcal D_2  (s,s_2,m^2) \, \bigg[
   \frac{2 s_2  m^2}{3 \pi^2 \gamma ^2 \sigma ^2}
   \bigg(8 \left(s-s_1\right){}^2 - 5 s_2^2  -3 \left(s+s_1\right)  s_2\bigg) \nn \\
   && \hspace{-2cm}
   +\frac{ s_2}{12 \pi^2 \gamma ^2 \sigma ^3}
   \bigg( s_2^5-\left(35 s+11 s_1\right) s_2^4+30 \left(3  s^2+s_1^2\right) s_2^3
   +2 \left(-35 s^3+17 s_1 s^2 \right. \nn \\
   && \hspace{-2cm} \left.
   +11 s_1^2 s - 17 s_1^3\right) s_2^2
   +\left(s-s_1\right){}^2 \left(5 s^2+26  s_1 s+17 s_1^2\right) s_2+3 \left(s-s_1\right){}^4 \left(3
   s-s_1\right)\bigg) \bigg] \nn \\
   && \hspace{-3cm} 
   - \, e^2 \mathcal D_1 (s,s_1,m^2) \,
   \bigg[ \frac{2  m^2}{3  \pi^2 \gamma ^2 \sigma ^2 s_1}
   \bigg( -s_2^4+2 \left(2 s+3 s_1\right)   s_2^3
   +\left(-6 s^2-6 s_1 s+s_1^2\right) s_2^2 \nn \\
   && \hspace{-2cm}
   +\left(s-s_1\right)
   \left(4 s^2-2 s_1 s+3 s_1^2\right) s_2 -\left(s-3 s_1\right) \left(s-s_1\right){}^3\bigg)
   +\frac{1 }{12 \pi^2 \gamma ^2 \sigma ^3}   \bigg(-s_2^6+\left(18 s+11
   s_1\right) s_2^5 \nn \\
   && \hspace{-2cm}
  -3 \left(21 s^2-3 s_1 s+10 s_1^2\right)   s_2^4 +2 \left(46 s^3-37 s_1 s^2+2 s_1^2 s+17 s_1^3\right)
   s_2^3 \nn \\
   && \hspace{-2cm}
   -\left(63 s^4-82 s_1 s^3+2 s_1^3 s+17 s_1^4\right)  s_2^2 +3 \left(s-s_1\right){}^3 \left(6 s^2+7 s_1
   s-s_1^2\right) s_2-s \left(s-s_1\right){}^5\bigg)\bigg] ,\nn \\ \\
  %
  %
\underline{ \bf \underline{F_{11} (s;\,s_1,\,s_2,\,m^2)}} &=&
\underline{ \bf \underline{F_{12} (s;\,s_2,\,s_1,\,m^2)}}  =
\frac{2  e ^2 m^2}{3 \pi^2  \sigma  s_2}
+\frac{e^2} {6 \pi^2 \sigma  s_2} [\mathcal B_0(s,m^2) \, m^2 - \mathcal A_0(m^2)] \nn \\
&& \hspace{-3cm}
+\frac{ e ^2 }{12\pi^2  \sigma ^2 s_2}  \biggl[ 2 s^3-\left(5 s_1+2   s_2\right) s^2 
+\left(4 s_1^2+4 s_2 s_1-2 s_2^2\right)
   s-\left(s_1-2 s_2\right) \left(s_1-s_2\right){}^2\biggr]  \nn \\
&&  \hspace{-3cm} 
   - \, e^2 \mathcal C_0 (s,s_1,s_2,m^2) \, \left[\frac{ m^4}{\pi^2 \sigma  s_2} 
   +\frac{m^2}{4 \pi^2 \sigma ^2 s_2}
   \biggl(  3 s^3 - \left(5 s_1+3 s_2\right) s^2 + \left(s_1^2+10 s_2 s_1-3 s_2^2\right) s \biggr. \right.\nn \\
 && \hspace{-2cm}  \biggl. 
 +\left(s_1-s_2\right){}^2 \left(s_1+3   s_2\right)\biggr)
 + \frac{ s}{4 \pi^2 \sigma ^3}
   \biggl( s^4+\left(s_1-4 s_2\right) s^3 -\left(s_1-s_2\right)
   \left(5 s_1+6 s_2\right) s^2 \biggr. \nn \\
 && \biggl. \left. \hspace{-2cm}           +\left(s_1+s_2\right) \left(3
   s_1^2+3 s_2 s_1-4 s_2^2\right) s+\left(s_1-s_2\right){}^2 s_2
   \left(3 s_1+s_2\right)
   \biggr)\right] \nn \\
  && \hspace{-3cm} 
  - \, e^2 \mathcal D_2 (s, s_2,m^2) \,
   \left[
     \frac{m^2}{6 \pi^2 \sigma ^2 s_2} \left(-4 \left(s-s_1\right){}^2+9 s_2^2-5
   \left(s+s_1\right) s_2\right)
   +\frac{1}{24 \pi^2  \sigma ^3}
   \biggl (  -17 s^4+\left(26 s_1+48 s_2\right) s^3   \biggr. \right. \nn \\
 && \hspace{-2cm} \biggl. \left.
 - 42 s_2  \left(s_1+s_2\right) s^2 -2 \left(s_1-s_2\right) \left(5
   s_1^2+17 s_2 s_1+4 s_2^2\right) s+\left(s_1-3 s_2\right)
   \left(s_1-s_2\right){}^3\biggr)
   \right] \nn \\
  && \hspace{-3cm}  
  + \, e^2 \mathcal D_1 (s,s_1,m^2)\,
   \left[
   \frac{m^2}{6 \pi^2  \sigma ^2 s_2}
   \left(4 s^2+5 s_1 s-8 s_2 s-9 s_1^2+4 s_2^2+5 s_1 s_2\right)
   -\frac{2 }{3 \sigma ^3 s_2}
     \biggl (3 s^5
     \right. \nn \\
   &&   \hspace{-2cm} -\left(10 s_1+9 s_2\right) s^4 + \, \, 2 \left(6 s_1^2+26 s_2 s_1+3
   s_2^2\right) s^3-6 \left(s_1^3+4 s_2 s_1^2+14 s_2^2
   s_1-s_2^3\right) s^2 	\nn \\
   &&  \left. +\left(s_1-s_2\right) \left(s_1^3-19 s_2
   s_1^2-43 s_2^2 s_1+9 s_2^3\right) s+\left(s_1-3 s_2\right)
   \left(s_1-s_2\right){}^3 s_2
   \biggr)
   \right], \nn \\ \\
  %
  %
\underline{ \bf \underline{F_{13,R} (s;\,s_1,\,s_2,\,m^2)}} &=&
- \frac{  e ^2 m^2 \left(s_1+s_2\right)}{12 \pi^2 s_1 s_2} 
+\frac{e^2 }{48\pi^2 }  \left(\frac{s \gamma }{\sigma }+\frac{3 s}{\gamma}+1\right)
-\frac{1}{2} \left[\Pi_R(s_1,m^2)+ \Pi_R(s_2,m^2) \right]
 \nn \\
&& \hspace{-2cm}
-  \frac{e^2}{12\pi^2 } \, [\mathcal B_0(s,m^2) \, m^2 - \mathcal A_0(m^2)] \, \bigg( \frac{1}{s_1}+\frac{1}{s_2}+\frac{3}{\gamma }\bigg) \nn \\
&& \hspace{-2cm}
 + \, e^2 \mathcal C_0 (s,s_1,s_2,m^2)\, \bigg[\frac{ m^4}{2 \pi^2 \gamma }
 +\frac{m^2 s \gamma }{4 \pi^2 \sigma } 
 +\frac{ s^2 s_1 s_2 \left(s^2-2 \left(s_1+s_2\right)
  s+s_1^2+s_2^2\right)}{4 \pi^2 \gamma  \, \sigma ^2}\bigg] 
  \nn \\
&& \hspace{-3cm}
 - \, e^2 \mathcal D_1(s,s_1,m^2)\, \bigg[\frac{1}{24\pi^2 } m^2 \left(-\frac{5 \left(s+s_1-s_2\right)}{\sigma }
 -\frac{2}{s_1}-\frac{3}{\gamma }\right) \nn \\
 &&
+ \frac{1}{24 \pi^2  \gamma  \sigma ^2} \bigg(\left(s-s_1\right) \left(5 s^3+s_1^2 s-4 s_1^3\right) s_2 
+\big(10 s^2+5 s_1 s +7 s_1^2\big)
  s_2^3
\nn \\
&& \hspace{-2cm}
 - \left(s-s_1\right){}^3 \left(s^2+2 s_1 s-s_1^2\right) +\left(-10 s^3+3 s_1 s^2-7 s_1^3\right) s_2^2 +s_2^5-\left(5 s+4
  s_1\right) s_2^4\bigg)\bigg] \nn \\
  && \hspace{-2cm}
  \nn \\
&& \hspace{-3cm}
- e^2  \, \mathcal D_2 (s, s_2,m^2) \,  
\bigg[ \frac{1}{24 \pi^2 } m^2 \left(-\frac{5 \left(s-s_1+s_2\right)}{\sigma }-\frac{2}{s_2} 
- \frac{3}{\gamma }\right)
+\frac{1}{24 \pi^2  \gamma  \sigma^2} \bigg(\left(4 s^3+s_1 s^2+7 s_1^3\right) s_2^2  \nn \\
&& \hspace{-2.5cm}
-\left(8 s^2+5 s_1 s+7 s_1^2\right) s_2^3
-\left(s-s_1\right){}^5+\left(s-4
  s_1\right) \left(s+s_1\right) s_2 \left(s-s_1\right){}^2-s_2^5+\left(5 s+4 s_1\right) s_2^4\bigg)
  \bigg], \nn \\
\eea
where as previously done the master integrals are collected in Appendix \ref{scalars}. 
These expressions have been analyzed in the text in various kinematical limits to show 
the appearance of anomaly poles and of all the other poles in the off-shell formulation. 

Notice that $F_{13}$ contains two vacuum polarization diagrams with different momenta on the external lines and has been renormalized by a subtraction at zero momentum
\bea
\Pi_R (s,m^2) \equiv \Pi (s,m^2)- \Pi(0,m^2) = \frac{e^2}{36 \, \pi^2 } 
\bigg[ \left(3 +  \frac{6 \, m^2}{s} \right) a_3 \, \log \frac{a_3 +1}{a_3 - 1} \, -  \frac{12 \, m^2}{s} -5 \bigg],
\eea
where $\Pi (s,m^2)$ is defined in Eq. (\ref{vacuumpol}), $a_3 = \sqrt{1-4m^2/s}$ and 
\bea
\Pi(0,m^2) = - \frac{e^2}{12 \, \pi^2 } \, \mathcal B_0 (0,m^2) = 
 - \frac{e^2}{12 \, \pi^2 }  \left [ \frac{1}{\bar \eps} - \log \left( \frac{m^2}{\mu^2} \right) \right]
\eea
with $1/\bar\eps$ defined in (\ref{bareps}).

\section{Appendix. The massless invariant amplitudes}
\label{Fimassless}
We present here the expressions of the invariant amplitudes in the massless limit.
We obtain
%
 %
 \beqa
 \underline{ \bf \underline{F_{1} (s;\,s_1,\,s_2,\,0)}}&=&
- \frac{e^2 }{18 \pi^2 s}, \\
  %
 \underline{ \bf \underline{F_{2} (s;\,s_1,\,s_2,\,0)}} &=& 0, \\ 
 \underline{ \bf \underline{F_{3} (s;\,s_1,\,s_2,\,0)}}  &=&
  \underline{ \bf \underline{F_{5} (s;\,s_2,\,s_1,\,0)}} =
-\frac{e^2} {144 \pi^2 s  \sigma ^3}
\biggl[ s^6-3 \left(s_1-4 s_2\right) s^5+6 \left(3
   s_1-7 s_2\right) s_2 s^4   \nn \\
   && \hspace{-2cm} +2 \left(5 s_1^3-69 s_2 s_1^2+117
   s_2^2 s_1+23 s_2^3\right) s^3   -3 \left(5 s_1^4-62 s_2
   s_1^3+72 s_2^2 s_1^2+50 s_2^3 s_1+7 s_2^4\right) s^2 \biggr. \nn \\
   && \hspace{-2cm}  +3
   \left(s_1-s_2\right){}^2 \left(3 s_1^3-24 s_2 s_1^2-33 s_2^2
   s_1+2 s_2^3\right) s -2 \left(s_1-s_2\right){}^6\biggr]    \nn \\
  && \hspace{-3cm} - \frac{e^2 \, s_1 } {48 \pi^2 \, \si^4 }  \,\mathcal D_1 (s,s_1,0) \,
  \Big[ \left(s-s_1\right){}^6+2 \left(14 s+11
   s_1\right) s_2 \left(s-s_1\right){}^4-\left(23 s^2-214 s_1
   s+19 s_1^2\right) s_2^2 \left(s-s_1\right){}^2   \Big. \nn \\
   && \hspace{-2.5cm} \Big. -21 s_2^6 +2
   \left(5 s_1-2 s\right) s_2^5+\left(107 s^2-318 s_1 s+71
   s_1^2\right) s_2^4+8 \left(-11 s^3+18 s_1 s^2+17 s_1^2 s-8
   s_1^3\right) s_2^3\Big]  \nn \\
   &&  \hspace{-3cm} -  \frac{e^2 \, s_2 } {48 \pi^2 \si^4 }  \,\mathcal D_2 (s, s_2,0) \,
   \Big[ s_2^6-2 \left(s-14 s_1\right) s_2^5 + \left(s^2+120 s_1
   s-37 s_1^2\right) s_2^4  \nn \\
   && \hspace{-2cm}
   - 4 \left(s^3+49 s_1 s^2-69 s_1^2 s+13  s_1^3\right) s_2^3
   +\left(s-s_1\right) \left(11 s^3-69 s_1
   s^2+309 s_1^2 s-83 s_1^3\right) s_2^2
   \nn \\
   && \hspace{-2cm}
   -2 \left(s-s_1\right){}^3 \left(5 s^2-49 s_1 s-4 s_1^2\right) s_2
   +3 \left(s-s_1\right){}^5 \left(s+5 s_1\right)\Big]  \nn \\
   && \hspace{- 3cm} -  \frac{e^2}{16 \pi^2} \mathcal C_0 (s,s_1,s_2,0) \, \bigg[
   \frac {2 s_1 s_2} {\sigma ^4}
   \left[ 2 s^6+3 \left(s_2-3 s_1\right) s^5+\left(15 s_1^2+6  s_2 s_1-13 s_2^2\right) s^4  \right. \nn \\
  && \hspace{-2cm}
   +2 \left(-5 s_1^3-19 s_2 s_1^2+29
   s_2^2 s_1+s_2^3\right) s^3+12 s_2 \left(4 s_1^3-4 s_2
   s_1^2-3 s_2^2 s_1+s_2^3\right) s^2  \nn \\
    && \bigg. \left.  \hspace{-2cm}
   +\left(s_1-s_2\right){}^2
   \left(3 s_1^3-15 s_2 s_1^2-31 s_2^2 s_1-5 s_2^3\right)
   s-\left(s_1-s_2\right){}^4
   \left(s_1+s_2\right){}^2\right]\bigg], \nn \\ \\
  %
  %
  \underline{ \bf \underline{F_{4} (s;\,s_1,\,s_2,\,0)}} &=&
 \underline{ \bf \underline{F_{6} (s;\,s_2,\,s_1,\,0)}} = \nn \\
  && \hspace{-3cm} \frac{ e ^2 }{24 \pi^2 \sigma ^3 s_1}
   \bigg[
      -s_2^5+\left(6 s+11 s_1\right)  s_2^4-\left(14 s^2+s_1 s+5 s_1^2\right) s_2^3
      +\left(16 s^3-35 s_1 s^2+46 s_1^2 s-15 s_1^3\right)   s_2^2   \bigg. \nn \\
   && \hspace{-2cm} \bigg.   -\left(s-s_1\right){}^2 \left(9 s^2-11 s_1 s-6
   s_1^2\right) s_2+2 \left(s-s_1\right){}^4 \left(s+2 s_1\right)
   \bigg] \nn \\
  && \hspace{-3cm} -  \frac{e^2}{16 \pi^2} \mathcal D_2 (s, s_2, 0) \,
   \bigg[  - \frac{1}{3 \sigma ^4 s_1} \bigg(
   3 \left(s+s_1\right)  \left(s-s_1\right){}^6-4 \left(4 s^2-14 s_1 s-5 s_1^2\right)
   s_2 \left(s-s_1\right){}^4 \bigg. \nn \\
   &&  \hspace{-2cm} +\left(35 s^3-119 s_1 s^2+169 s_1^2
   s-13 s_1^3\right) s_2^2 \left(s-s_1\right){}^2+\left(s-3
   s_1\right) s_2^6 - 8 \left(s^2+9 s_1 s+7 s_1^2\right)   s_2^5 \nn \\
   &&  \hspace{-2cm}  \bigg. +\left(25 s^3+159 s_1 s^2-197 s_1^2 s+157 s_1^3\right)
   s_2^4+4 \left(-10 s^4+21 s_1^2 s^2+28 s_1^3 s-27 s_1^4\right)
   s_2^3 \bigg) \bigg] \nn \\
    &&  \hspace{-3cm} -  \frac{e^2}{16 \pi^2} \mathcal D_1 (s, s_1, 0) \,  \bigg[
    -\frac{1}{3 \sigma ^4}
   \bigg(
   s_2^6+4 \left(6 s+11 s_1\right)  s_2^5
   +\left(-87 s^2+106 s_1 s-91 s_1^2\right) s_2^4 \nn \\
   && \hspace{-2cm}
   +4 \left(22 s^3-69 s_1 s^2+40 s_1^2 s+s_1^3\right)  s_2^3
   +\left(s-s_1\right) \left(3 s^3-29 s_1 s^2+209 s_1^2 s-79   s_1^3\right) s_2^2 \nn \\
   && \hspace{-2cm}
   -8 \left(s-s_1\right){}^3 \left(6 s^2-13 s_1 s - 4 s_1^2\right) s_2+\left(s-s_1\right){}^5
   \left(19 s+5  s_1\right)\bigg) \bigg]  \nn \\
   && \hspace{-3cm} -  \frac{e^2}{16 \pi^2} \mathcal C_0  (s, s_1, s_2, 0 ) \, \bigg[
   \frac{1}{\sigma ^4}
   \bigg(-2 s  \left(s-s_1\right){}^6-2 \left(s^2+7 s_1 s+2 s_1^2\right) s_2 \left(s-s_1\right){}^4
    \nn \\
   && \hspace{-2cm}
   + 2 \left(7 s^3+6 s_1 s^2+11 s_1^2 s-4 s_1^3\right) s_2^4
   +12 \left(2 s^3-3 s_1 s^2-2 s_1^2 s+s_1^3\right) s_2^2 \left(s-s_1\right){}^2
    \nn \\
    && \hspace{-2cm}
    -4 \left(s+s_1\right) s_2^6
    + 6 \left(s^2-5 s_1 s+2 s_1^2\right)   s_2^5
   - 4 \left(9 s^4-25 s_1 s^3+33 s_1^2 s^2-15 s_1^3 s+2 s_1^4\right) s_2^3 \bigg)
   \bigg],  \nn \\
 %
 %
 \underline{ \bf \underline{F_{7} (s;\,s_1,\,s_2,\,0)}} &=&
 \frac{e^2}{72 \pi^2}  \bigg[
  \frac{840 s\, s_1^2}{\sigma ^3}  \left( 2 \left(s+s_1\right) s_2-\left(s-s_1\right){}^2\right)  \nn \\
  && \hspace{-2cm}
  + \, \frac{6 s_1}{\sigma ^2}  \left(-13 s^2+166 s_1 s-13
  s_1^2+39 \left(s+s_1\right) s_2\right)
  +\frac{3 \left(-s+27 s_1+s_2\right)}{\sigma}
  + \frac{2}{s} + \frac{9 s}{\gamma^2} -\frac{6}{\gamma}
  \bigg] \nn \\
  && \hspace{-3cm}
  -  \frac{e^2}{16 \pi^2} \mathcal C_0 (s, s_1, s_2,0) \,  \bigg\{
  -\frac{4 s_1 s_2}{\gamma ^2 \sigma ^4}
  \bigg[ \left(-9 s^2+22 s_2 s-4
  s_2^2\right) s_1^6 + 2 s \left(20 s^2-39 s_2 s+21 s_2^2\right) s_1^5  \nn \\
  && \hspace{0.5cm}
  + \left(s-s_2\right){}^6 s_2 \left(4 s+s_2\right)
  + 2 s \left(s-s_2\right){}^4 \left(2 s^2+5 s_2 s+11 s_2^2\right) s_1
   \nn \\
  && \hspace{0.5cm}
  +\, \left(-65 s^4+96 s_2 s^3+33 s_2^2 s^2-62 s_2^3 s+6 s_2^4\right) s_1^4 \nn \\
  && \hspace{0.5cm}
  + \, 2 s \left(27 s^4-22 s_2 s^3-108 s_2^2 s^2+102 s_2^3 s-31 s_2^4\right) s_1^3
 \nn \\
  && \hspace{0.5cm}
  -\left(s-s_2\right){}^2 \big( 23 s^4+40 s_2 s^3-105 s_2^2 s^2
  -34 s_2^3 s+4 s_2^4 \big) s_1^2+ s_1^8-2 s s_1^7 \bigg] \bigg\} \nn \\
  && \hspace{-3cm} - \frac{e^2}{16 \pi^2} \mathcal D_1 (s, s_1, 0) \,
  \bigg\{
  \frac{ 2 s_1}{3 \gamma ^2 \sigma ^4}
  \bigg[-\left(317 s^2+227 s_1 s+64 s_1^2\right)
  s_2^6+\left(s_1-s\right){}^5 \left(-7 s^2+39 s_1 s+32 s_1^2\right)s_2 \nn \\
  && \hspace{-2.5cm}
  +\left(397 s^3+846 s_1 s^2-539 s_1^2 s+312 s_1^3\right)
  s_2^5-\left(s-s_1\right){}^3 \left(23 s^3+114 s_1 s^2+463 s_1^2 s-16 s_1^3\right) s_2^2 \nn \\
  && \hspace{-2.5cm}
  -\left(275 s^4+s_1 \left(1181 s^3+s_1 \left(3 s_1
  \left(93 s+94 s_1\right)-1441 s^2\right)\right)\right)
  s_2^4+\left(s-s_1\right) \big(103 s^4+767 s_1 s^3 \nn \\
  && \hspace{-2.5cm}
  - \, 79 s_1^2 s^2-563 s_1^3
  s-36 s_1^4\big) s_2^3-23 s_2^8+\left(133 s+4 s_1\right)
  s_2^7+\left(s_1-s\right){}^7 \left(2 s+s_1\right) \bigg]
  \bigg\}
  \nn \\
  && \hspace{-3cm}
  -  \frac{e^2}{16 \pi^2} \mathcal D_2 (s, s_2, 0) \, \bigg\{
  \frac{2 s_2}{3 \gamma ^2 \sigma ^4} \bigg[
  -23 s_1^8+\left(133 s+4 s_2\right) s_1^7  -\left(317 s^2+227 s_2 s+64 s_2^2\right)s_1^6 \nn \\
  && \hspace{-2.5cm}
  +\left(s-s_2\right){}^5 \left(7 s^2-39 s_2 s-32 s_2^2\right)s_1 	
  -\left(s-s_2\right){}^7 \left(2 s+s_2\right) \nn \\
  && \hspace{-2.5cm}
  +\left(397 s^3+846 s_2 s^2-539 s_2^2 s+312 s_2^3\right)s_1^5
  -\left(s-s_2\right){}^3 \left(23 s^3+114 s_2 s^2+463 s_2^2 s-16 s_2^3\right) s_1^2 \nn \\
  && \hspace{-2.5cm}
  -\left(275 s^4+s_2 \left(1181 s^3+s_2 \left(3 s_2 \left(93 s+94 s_2\right)-1441 s^2\right)\right)\right)s_1^4
  +\left(s-s_2\right) \big(103 s^4+767 s_2 s^3 \nn \\
  && \hspace{-2.5cm}
  -79 s_2^2 s^2-563 s_2^3 s-36 s_2^4\big) s_1^3
  \bigg]
  \bigg\},  \nn \\ \\
  %
  %
  \underline{ \bf \underline{F_{8} (s;\,s_1,\,s_2,\,0)}}&=&
 -\frac{ e ^2 }{12 \pi^2 \gamma  \sigma ^3} \bigg[ 4 s_2^5+\left(14 s_1-11 s\right) s_2^4+2
   \left(s-s_1\right) \left(2 s+9 s_1\right) s_2^3 \nn \\
 && \hspace{0.5cm}
 +2 \left(7  s^3-43 s_1 s^2+33 s_1^2 s-9 s_1^3\right) s_2^2
  -2 \left(s-s_1\right){}^2 \left(8 s^2-21 s_1 s-7 s_1^2\right)   s_2 \nn \\
  &&\hspace{0.5cm}
 +\left(s-s_1\right){}^4 \left(5 s+4 s_1\right)\bigg]  \nn \\
 && \hspace{-3cm}
 -  \frac{e^2}{16 \pi^2} \mathcal D_2 (s, s_2, 0) \,  \bigg\{
 \frac{2 }{3 \gamma  \sigma ^4}
 \bigg[ -5 s_2^7  +3 \left(s-13  s_1\right) s_2^6 + \left(57 s^2-128 s_1 s+43 s_1^2\right)   s_2^5
 \nn \\
 && \hspace{-1cm}
 +\left(-155 s^3 + 567 s_1 s^2 - 341 s_1^2 s + 121 s_1^3\right)  s_2^4  \nn \\
 && \hspace{-1cm}
 + 3 \left(55 s^4-176 s_1 s^3+86 s_1^2 s^2+56 s_1^3 s-53  s_1^4\right) s_2^3
 + 3 \left(s-s_1\right){}^6  \left(s+s_1\right) \nn \\
 && \hspace{-2cm}
 - \left(s-s_1\right){}^2 \left(75 s^3+103  s_1 s^2-311 s_1^2 s-11 s_1^3\right)   s_2^2
 +\left(s-s_1\right){}^4 \left(7 s^2+124 s_1 s+25  s_1^2\right) s_2
 \bigg]  \bigg\} \nn \\
   && \hspace{-3cm} -  \frac{e^2}{16 \pi^2} \mathcal D_1 (s, s_1, 0) \,
   \bigg\{
 \frac{2 }{3 \gamma  \sigma ^4}
   \bigg[ 3 s_2^7+5 \left(5 s_1-3 s\right)  s_2^6 + \left(27 s^2+24 s_1 s+11 s_1^2\right) s_2^5
    -\left(15 s^3+339 s_1 s^2 \right. \nn \\
   && \hspace{-2cm}
  \left. -289 s_1^2 s+159 s_1^3\right) s_2^4
  +\left(-15 s^4+616 s_1 s^3-714 s_1^2 s^2+168 s_1^3 s+121 s_1^4\right)  s_2^3  \nn \\
  && \hspace{-2cm}
  +\left(s-s_1\right) \left(27 s^4-402 s_1 s^3+40 s_1^2  s^2 + 298 s_1^3 s-43 s_1^4\right) s_2^2
   \nn \\
  && \hspace{-2cm}
  -\left(s-s_1\right){}^3 \left(15 s^3-51 s_1 s^2 -245 s_1^2 s-39 s_1^3\right)
   s_2+\left(s-s_1\right){}^5 \left(3 s^2+22 s_1 s+5
   s_1^2\right)\bigg]\bigg\} \nn \\
  && \hspace{-3cm}
   -  \frac{e^2}{16 \pi^2} \mathcal C_0 (s, s_1, s_2, 0) \,  \bigg\{
  \frac{4}{\gamma  \sigma ^4}
   \bigg[    \left(s_1+s_2\right) s^7  -6 \left(s_1^2-s_2   s_1+s_2^2\right) s^6
   \nn \\
   && \hspace{-2cm}
   + \, 3 \left(s_1+s_2\right) \left(5 s_1^2-12  s_2 s_1+5 s_2^2\right) s^5
   +2 \left(-10 s_1^4+3 s_2 s_1^3  +54 s_2^2 s_1^2+3 s_2^3 s_1-10 s_2^4\right)   s^4  \nn \\
   && \hspace{-2cm}
   + \, \left(s_1+s_2\right) \left(15 s_1^4+16 s_2 s_1^3-126  s_2^2 s_1^2+16 s_2^3 s_1+15 s_2^4\right) s^3 \nn \\
   && \hspace{-2cm}
   - \, 6 \left(s_1^6 + 5   s_2 s_1^5-s_2^2 s_1^4-18 s_2^3 s_1^3-s_2^4 s_1^2+5 s_2^5  s_1+s_2^6\right) s^2
    \nn \\
   && \hspace{-2cm}
   + \, \left(s_1-s_2\right){}^2   \left(s_1+s_2\right)
   \left(s_1^4+6 s_2 s_1^3+34 s_2^2 s_1^2+6  s_2^3 s_1+s_2^4\right)  \, s  \nn \\
   && \hspace{-2cm}
   +\, 2 s_1 \left(s_1-s_2\right){}^4 s_2
   \left(s_1+s_2\right){}^2\bigg]
   \bigg\} ,\nn \\ \\
  %
  %
  \underline{ \bf \underline{F_{9} (s;\,s_1,\,s_2,\,0)}}&=&
  \underline{ \bf \underline{F_{10} (s;\,s_2,\,s_1,\,0)}}  =
 \frac{e^2 }{12 \pi^2 \gamma ^2 \sigma ^2}
 \bigg[    \left(s-s_1\right){}^4
  - 4 \left(4  s+s_1\right) s_2 \left(s-s_1\right){}^2
 \nn \\
 && \hspace{1cm}
   - \, 3 s_2^4  +4 \left(s_1-2   s\right) s_2^3+2 \left(13 s^2-2 s_1 s+s_1^2\right) s_2^2\bigg] \nn \\
&& \hspace{-3cm}
   -  \frac{e^2}{16 \pi^2} \mathcal C_0(s, s_1, s_2, 0)\,  \bigg[  \frac{8 s s_2 }{\gamma ^2 \sigma ^3}
   \bigg(s_2^5+\left(2  s_1-3 s\right) s_2^4
   + \,2 \left(s-s_1\right) \left(s+2 s_1\right) s_2^3 \nn \\
    && \hspace{-2cm}
   + \,2 \left(s^3-7 s_1 s^2+3 s_1^2 s-s_1^3\right)  s_2^2
   -\left(s-3 s_1\right) \left(s-s_1\right){}^2 \left(3
   s+s_1\right) s_2+s \left(s-s_1\right){}^4\bigg )\bigg] \nn \\
    && \hspace{-3cm}
   -  \frac{e^2}{16 \pi^2} \mathcal D_2(s, s_2, 0) \,  \bigg[
   \frac{4 s_2}{3 \gamma ^2 \sigma ^3}
   \bigg( s_2^5-\left(35 s+11 s_1\right) s_2^4+30 \left(3  s^2+s_1^2\right) s_2^3
   +2 \left(-35 s^3+17 s_1 s^2 \right. \nn \\
   && \hspace{-2cm} \left.
   +\, 11 s_1^2 s - 17 s_1^3\right) s_2^2
   +\left(s-s_1\right){}^2 \left(5 s^2+26  s_1 s+17 s_1^2\right) s_2+3 \left(s-s_1\right){}^4 \left(3
   s-s_1\right)\bigg) \bigg] \nn \\
   && \hspace{-3cm} -  \frac{e^2}{16 \pi^2} \mathcal D_1 (s, s_1, 0) \,
   \bigg[
  \frac{4 }{3 \gamma ^2 \sigma ^3}   \bigg(-s_2^6+\left(18 s+11
   s_1\right) s_2^5-3 \left(21 s^2-3 s_1 s+10 s_1^2\right)   s_2^4 \nn \\
   && \hspace{-2cm}
   + \, 2  \left(46 s^3-37 s_1 s^2+2 s_1^2 s+17 s_1^3\right)
   s_2^3-\left(63 s^4-82 s_1 s^3+2 s_1^3 s+17 s_1^4\right)  s_2^2 \nn \\
   && \hspace{-2cm}
   +3 \left(s-s_1\right){}^3 \left(6 s^2+7 s_1
   s-s_1^2\right) s_2-s \left(s-s_1\right){}^5\bigg)\bigg],   \nn \\ \\
  %
  %
  \underline{ \bf \underline{F_{11} (s;\,s_1,\,s_2,\,0)}} &=&
 \underline{ \bf \underline{F_{12} (s;\,s_2,\,s_1,\,0)}}= \, \,
  \nn \\
&& \hspace{-2cm} 
\frac{e ^2 }{12 \pi^2  \sigma ^2 s_2}  \biggl[ 2 s^3-\left(5 s_1+2   s_2\right) s^2
+\left(4 s_1^2+4 s_2 s_1-2 s_2^2\right)  \, s
-\left(s_1-2 s_2\right) \left(s_1-s_2\right){}^2\biggr] \nn \\
&& \hspace{-2cm}
   -  \frac{e^2}{16 \pi^2} \mathcal C_0 (s,s_1,s_2,0) \, \left[
 -\frac{4 s}{\sigma ^3}
   \biggl(s^4+\left(s_1-4 s_2\right) s^3   -\left(s_1-s_2\right)
   \left(5 s_1+6 s_2\right) s^2 \right. \nn \\
  && \hspace{-1cm}
  \left.  +\left(s_1+s_2\right) \left(3
   s_1^2+3 s_2 s_1-4 s_2^2\right) s+\left(s_1-s_2\right){}^2 s_2
   \left(3 s_1+s_2\right)
   \biggr)\right]  \nn \\
  && \hspace{-2cm}-  \frac{e^2}{16 \pi^2} \mathcal D_2 (s, s_2, 0) \,
   \left[
  \frac{2}{3 \sigma ^3}
   \biggl (
         -17 s^4+\left(26 s_1+48 s_2\right) s^3
          - 42 s_2  \left(s_1+s_2\right) s^2   \right. \nn \\
 && \hspace{-1cm}  \left.
 -2 \left(s_1-s_2\right) \left(5 s_1^2+17 s_2 s_1+4 s_2^2\right) s
 +\left(s_1-3 s_2\right)
   \left(s_1-s_2\right){}^3\biggr)
   \right] \nn \\
  && \hspace{-2cm}
  -  \frac{e^2}{16 \pi^2} \mathcal D_1 (s, s_1, 0) \,
   \left[
   - \, \frac{2 }{3 \sigma ^3 s_2}
     \biggl (   3 s^5-\left(10 s_1+9 s_2\right) s^4  \right. \nn \\
   &&   \hspace{-0.5cm}
   + \, \, 2 \left(6 s_1^2+26 s_2 s_1+3
   s_2^2\right) s^3-6 \left(s_1^3+4 s_2 s_1^2+14 s_2^2
   s_1-s_2^3\right) s^2 	\nn \\
   &&  \left. \hspace{-0.5cm}
   +\left(s_1-s_2\right) \left(s_1^3-19 s_2
   s_1^2-43 s_2^2 s_1+9 s_2^3\right) s+\left(s_1-3 s_2\right)
   \left(s_1-s_2\right){}^3 s_2
   \biggr)
   \right],  \nn \\ \\
  %
  %
   \underline{ \bf \underline{F_{13,R} (s;\,s_1,\,s_2,\,0)}} &=&
  -\frac{1}{2} \left[ \Pi_R(s_1,0)+ \Pi_R(s_2,0) \right]
 + \frac{e^2}{48 \pi^2}  \left(\frac{s \gamma }{\sigma }+\frac{3 s}{\gamma}+1\right)\nn \\
&& \hspace{-1cm}
 +  \frac{e^2}{16 \pi^2} \mathcal C_0 (s,s_1,s_2,0) \,  \bigg[\frac{4 s^2 s_1 s_2 \left(s^2-2 \left(s_1+s_2\right)
  s+s_1^2+s_2^2\right)}{\gamma  \sigma ^2}\bigg] \nn \\
&& \hspace{-3cm}
-  \frac{e^2}{16 \pi^2}  \mathcal D_1 (s,s_1,0) \,  \bigg[
\frac{2}{3 \gamma  \sigma ^2} \bigg(\left(s-s_1\right) \left(5 s^3+s_1^2 s-4 s_1^3\right) s_2
+\big(10 s^2+5 s_1 s +7 s_1^2\big)\,  s_2^3\nn \\
&& \hspace{-2cm}
 -\left(s-s_1\right){}^3 \left(s^2+2 s_1 s-s_1^2\right)+\left(-10 s^3+3 s_1 s^2-7 s_1^3\right) s_2^2+s_2^5-\left(5 s+4
  s_1\right) s_2^4\bigg)\bigg] \nn \\
&& \hspace{-3cm}
-  \frac{e^2}{16 \pi^2} \mathcal D_2 (s, s_2, 0)\, \bigg[ \frac{2}{3 \gamma  \sigma^2} \bigg(\left(4 s^3+s_1 s^2+7 s_1^3\right) s_2^2 -\left(8 s^2+5 s_1 s+7 s_1^2\right) s_2^3 \nn \\
&& \hspace{-2cm}
-\left(s-s_1\right){}^5+\left(s-4
  s_1\right) \left(s+s_1\right) s_2 \left(s-s_1\right){}^2-s_2^5+\left(5 s+4 s_1\right) s_2^4\bigg)
\bigg];
 \eea
as already noticed above for the case of the massive form factors the last one, i.e. $F_{13,R} (s;\,s_1,\,s_2,\,0)$, has been affected by the renormalization procedure for which the one-loop transverse photon propagator with a virtual pair of massless fermions is given by
 \bea
 \Pi_R (s,0) = - \frac{e^2}{12 \,  \pi^2 } \, \left[\frac{5}{3} - \log \left(-\frac{s}{\mu^2} \right)\right],
 \eea
  where the dependence on the renormalization scale $\mu$ remains explicit.

\section{Appendix. The asymptotic behavior of the off-shell massless $\lag TJJ \rag $ correlator}
We present here the asymptotic expression of the form factor in the high energy limit. The leading contributions to the expansion in each expression come from the pole singularities (conformal or anomalous) except for 
$F_{13}$  which has a constant asymptotic term.
\bea
F_{1} (s, s_1, s_2, 0) &=& - \frac{e^2}{18 \pi^2 s},  \nn \\
F_{2} (s, s_1, s_2, 0) &=& 0,  \nn \\
F_{3} (s, s_1, s_2, 0) &=& 
- \frac{ e ^2}{144 \pi^2 s}
- \frac{ e ^2 }{48 \pi^2 s^2} \left[s_1+6 s_2+s_1 \log \left(\frac{s_1}{s}\right)+3 s_2 \log \left(\frac{s_2}{s}\right)\right] +O\left(\frac{1}{s^3}\right), \nn \\
F_{4} (s, s_1, s_2, 0) &=& \frac{ e ^2}{48 \pi^2  s_1 s} \left[3 \log \left(\frac{s_2}{s}\right)+4\right]
+\frac{e ^2 }{48 \pi^2 s_1 s^2} \bigg[2 \pi ^2 s_1+16 s_1+6 s_2+19 s_1 \log \left(\frac{s_1}{s}\right) \nn \\
&& + \log \left(\frac{s_2}{s}\right) \left(9 s_1+8 s_2+6 s_1 \log \left(\frac{s_1}{s}\right)\right)\bigg] +O\left(\frac{1}{s^3}\right), \nn \\
F_{7} (s, s_1, s_2, 0) &=& 
\frac{e^2}{36 \pi^2 s}
+ \frac{ e^2}{24 \pi^2 s^2} \left[3 \left(s_1+s_2\right)+2 s_1 \log \left(\frac{s_1}{s}\right)+2 s_2 \log \left(\frac{s_2}{s}\right)\right] +O\left(\frac{1}{s^3}\right), \nn \\
F_{8} (s, s_1, s_2, 0) &=& - \frac{ e^2}{24 \pi^2  s^2} \left[3 \log \left(\frac{s_1}{s}\right)+3 \log \left(\frac{s_2}{s}\right)+10\right] + O\left(\frac{1}{s^3}\right), \nn \\
F_{9} (s, s_1, s_2, 0) &=& \frac{ e^2}{12 \pi^2 s^2} \left[\log \left(\frac{s_1}{s}\right)+1\right]+ O\left(\frac{1}{s^3}\right), \nn \\
F_{11} (s, s_1, s_2, 0) &=& \frac{ e^2 }{24 \pi^2 s_2 s}\left[3 \log \left(\frac{s_1}{s}\right)+4\right]
+ \frac{ e^2 }{24 \pi^2 s_2 s^2}\bigg[6 s_1+2 \pi ^2 s_2+12 s_2+17
  s_2 \log \left(\frac{s_2}{s}\right) \nn \\
&& + \log \left(\frac{s_1}{s}\right) \left(8 s_1+9 s_2+6 s_2 \log \left(\frac{s_2}{s}\right)\right)\bigg] +O\left(\frac{1}{s^3}\right), \nn \\
F_{13} (s, s_1, s_2, 0) &=& 
- \frac{1}{2} \left[ \Pi_R(s_1,0) + \Pi_R (s_2,0) \right]
+ \frac{e^2}{24 \pi^2}  \left[ \log \left(\frac{s_1}{s}\right) + \log \left(\frac{s_2}{s}\right) + \frac{5}{2} \right] \nn \\
%
&& + \frac{  e^2}{12 \pi^2 s} \bigg[s_1+s_2+2 s_1 \log \left(\frac{s_1}{s}\right)+2 s_2 \log \left(\frac{s_2}{s}\right)\bigg] \nn \\
&& +\frac{  e^2}{24 \pi^2 s^2} \bigg[ 2 \left(s_1^2+\left(3+\pi ^2\right) s_2 s_1+s_2^2\right) 
+s_2 \left(13 s_1+6 s_2\right) \log \left(\frac{s_2}{s}\right) \nn \\
&& +s_1 \log \left(\frac{s_1}{s}\right) \left(6 s_1+13 s_2+6 s_2 \log \left(\frac{s_2}{s}\right)\right)\bigg] +O\left(\frac{1}{s^3}\right). \nn \\
\eea
\section{Appendix. The asymptotic behavior of the on-shell massive $\lag TJJ \rag $ correlator}
This appendix contains the asymptotic expansion of the relevant on-shell massive form factors, that is their dominant contributions as $s\rightarrow \infty$ with $s>0$ after taking into account the suitable analytic continuation. They result
\bea
F_{1} (s, 0, 0, m^2) &=& 
- \frac{  e ^2}{18 \pi^2 s}
+\frac{e^2 m^2}{12 \pi^2 s^2} 
\left[ 4 - \log ^2\left(\frac{m^2}{s}\right) - 2 \, i \pi \log\left(\frac{m^2}{s}\right) + \pi^2 \right] +O\left(\frac{1}{s^3}\right), \nn \\ \\
F_{3} (s, 0, 0, m^2) &=& F_{5} (s, 0, 0, m^2) = - \frac{e ^2}{144 \pi^2  s}
-\frac{e^2 m^2}{24 \pi^2 s^2}\bigg[-\log ^2\left(\frac{m^2}{s}\right)-(6+2 i \pi ) \log
  \left(\frac{m^2}{s}\right)+\pi ^2 \nn \\
&& -6 i \pi -14\bigg] +O \left(\frac{1}{s^3}\right),   \\
F_{7} (s, 0, 0, m^2) &=& -4 F_{3} (s, 0, 0, m^2),   \\
F_{13, R} (s, 0, 0, m^2) &=& \frac{ e^2}{144 \,  \pi ^2} \left[12  \log \left(\frac{m^2}{s}\right)+ 35 + 12\, i \, \pi \right]
+\frac{ e^2 \, m^2 }{8 \, \pi ^2 \, s} \bigg[\log ^2\left(\frac{m^2}{s}\right) +10 - \pi ^2+2 \, i \,  \pi  \nn \\
&& +(\,  2 + 2 i \, \pi \, ) \log \left(\frac{m^2}{s}\right)\bigg] 
- \frac{e ^2 m^4}{4 \, \pi^2 \, s^2} \bigg[-\log ^2\left(\frac{m^2}{s}\right)+(2-2 i \pi ) \log \left(\frac{m^2}{s}\right)+\pi^2 \nn \\
&& +2 i \pi -3\bigg] +O\left(\frac{1}{s^3}\right)
\eea
%

\end{appendix}
\bibliographystyle{h-elsevier3}

\end{document}